\documentclass[12pt]{article}
\usepackage[utf8]{inputenc}
\usepackage{amssymb}
\usepackage{latexsym}
\usepackage{amsmath}
\usepackage{epsfig}
\usepackage{subfigure}
\usepackage{natbib}
\usepackage{xcolor}
\usepackage{array}
\usepackage{physics}
\usepackage{graphicx}
\usepackage{caption}
\usepackage{float}
\usepackage{multirow}
\usepackage{array}
\usepackage{tikz}
\newcolumntype{P}[1]{>{\centering\arraybackslash}p{#1}}
\usepackage{comment}
\usepackage[export]{adjustbox}
\usepackage{hyperref}
\newcommand{\beq}{\begin{equation}}
\newcommand{\eeq}{\end{equation}}
\newcommand{\beqs}{\begin{eqnarray}}
\newcommand{\eeqs}{\end{eqnarray}}

\newcommand{\calD}{{\cal D}}
\newcommand{\calE}{{\cal E}}

\newcommand{\calK}{{\cal K}}
\newcommand{\calL}{{\cal L}}

\newcommand{\calN}{{\cal N}}

\newcommand{\calP}{{\cal P}}

\newcommand{\calS}{{\cal S}}

\newcommand{\rmath}{\mathbb{R}}

\newcommand{\addr}{\textcolor{red}}

\newcommand{\R}{\mathbb{R}}    

\newcommand{\Lko}{\lfloor Lk_0 \rceil}
\newcommand{\rsm}[1]{\textcolor{cyan}{#1}}

\newcommand{\rs}[1]{\textcolor{blue}{#1}}
\begin{document}
\begin{center}
{\LARGE $cgNA+min$: computation of sequence-dependent dsDNA energy-minimising minicircle configurations}\\
\normalsize
\vspace{0.2in}
\vspace{2mm}
Raushan Singh$^1$, Jaroslaw Glowacki$^2$, Marius Beaud$^2$, Federica Padovano$^2$, Robert S. Manning$^3$, John H. Maddocks$^2$\\
$^1$Department of Mechanical Engineering, IIT Madras, Chennai 600036, India\\
$^2$Institute of Mathematics, {\'E}cole Polytechnique F{\'e}d{\'e}rale de Lausanne, Lausanne 1015, Switzerland\\
$^3$Haverford College, Department of Mathematics and Statistics,
Haverford, PA, 19041, USA\\
Correspondence: raushan@iitm.ac.in\\
(Dated: November 08, 2024)\\
\end{center}
\vspace{0.2in}
\normalsize
\noindent {\bf Abstract:} DNA minicircles are closed double-stranded DNA (dsDNA) fragments that have been demonstrated to be an important experimental tool to understand supercoiled, or stressed,  DNA mechanics, such as nucleosome positioning and DNA–protein interactions. Specific minicircles can be simulated using Molecular Dynamics (MD) simulation. However, the enormous sequence space makes it unfeasible to exhaustively explore the sequence-dependent mechanics of DNA minicircles using either experiment or MD. For linear fragments, the $cgNA+$ model, a computationally efficient sequence-dependent coarse-grained model using enhanced Curves+ internal coordinates (rigid base plus rigid phosphate) of double-stranded nucleic acids (dsNAs), predicts highly accurate nonlocal sequence-dependent equilibrium distributions for an arbitrary sequence when compared with MD simulations. This article addresses the problem of modeling sequence-dependent topologically closed and, therefore, stressed fragments of dsDNA. We introduce $cgNA+min$, a computational approach within the $cgNA+$ framework, which extends the $cgNA+$ model applicability to compute the sequence-dependent energy minimising configurations of covalently closed dsDNA minicircles of various lengths and linking numbers ($Lk$). The main computational idea is to derive the appropriate chain rule to express the $cgNA+$ energy in absolute coordinates involving quaternions where the closure condition is simple to handle. We also present a semi-analytic method for efficiently computing sequence-dependent initial minicircles having arbitrary $Lk$ and length. For different classes and lengths of sequences, we demonstrate that the dsDNA minicircle energies computed using $cgNA+min$ agree well with the energies approximated from experimentally measured $J$-factor values. Utilizing computational efficiency of $cgNA+min$, finally, we present minicircle shape, energy, and multiplicity of $Lk$ (and multiplicity of shapes/energies for fixed $Lk$) for more than $120K$ random DNA sequences of different lengths (ranging between $92-$ to $106-$base-pairs).  
\section{Introduction}
A long-standing problem in the field of DNA modeling is to learn how basepair sequence 
affects local mechanical properties such as intrinsic shape and stiffness \citep{Calladine1882,Olson1998,Rief1999,Gonzalez2013,Young2022,Sharma2023,Farré-Gil2024,Li2024}. One experimental measurement that is used to understand these properties is the cyclization $J$-factor \citep{FlorySuterMutter1976,LeveneCrothers1986,ZhangCrothers03,CzaplaSwigonOlson06,Kahn_Crothers1992,Cloutier_Widom2004,Basu2021r,Basu2021}. There are a variety of experimental protocols to measure $J$-factors, but roughly the idea is to alter the DNA to have ``sticky ends'' and then track the progress of two reactions: dimerisation (in which two molecules attach to form a double-length molecule) and cyclisation (in which one molecule has its two ends attach to each other).  The $J$-factor is the ratio of the equilibrium constants of these two reactions.  We note in particular two recent advances in this field: \cite{Basu2021r} developed a high-throughput tool to measure 
J-factors (note also the related work in \citep{Basu2021} and \citep{Basu2022}), 
and \cite{Li2022} 
trained a deep-learning model on experimental data to make predictions of $J$.

Computationally, the ability to compute closed-loop (or minicircle) configurations that locally minimize the energy is a key ingredient in modeling the $J$-factor \citep{Jacobson_Stockmayer_1950,Crothers1992,Manning1996,Cotta-Ramusino_Maddocks_2010}. Those who study cyclisation (either experimentally or computationally) typically categorise closed loop configurations according to their {\it linking number} $Lk$, which counts the number of complete turns experienced by the double helix. In principle, for {\it any} value of $Lk$, the DNA will have an energy minimiser of that link (since energy is a non-negative number). For most values of $Lk$ (those representing considerable overwinding or underwinding), these minimisers will involve relatively high energies and DNA self-contact.  In contrast, for a few values of $Lk$ close to $N/10.5$ (where $N$ is the number of base pairs and $10.5$ is roughly the number of basepairs per helix turn), we can expect an energy minimiser with lower energy and no self-contact.  We focus here only on this latter case, i.e., we will
ignore DNA self-contact. For molecules of length 100-300 bp, the self-contact configurations that are ignored 
should be of considerably higher energy and hence not play a significant role in the $J$-factor.


Specific minicircles can be simulated using Molecular Dynamics (MD) simulations 
\citep{Lankas2006,Pasi2017,Mitchell_Harris_2013,Curuksu2023,Kim2022}. However, the enormous sequence space of even short 
fragments makes it unfeasible to explore exhaustively the sequence-dependent mechanics of DNA minicircles using 
either experiment or MD. Here, we instead use a coarse-grained model, specifically a recently 
introduced sequence-dependent model (applicable to several varieties of double-stranded nucleic acids) in 
the ``family of $cgDNA$ models" called $cgNA+$ \citep{Sharma2023,Patelli2019,Sharma2023t}, which treats each base 
and each phosphate group as a rigid body. For linear fragments, the $cgNA+$ model 
predicts highly accurate non-local sequence-dependent equilibrium distributions for an arbitrary sequence 
when compared with MD simulations. In this article, we introduce a formulation within the $cgNA+$ model framework 
that allows modelling of sequence-dependent DNA minicircles of arbitrary sequence and linking number. 

In the continuum setting, \cite{Glowacki2016} describes an algorithm (named $bBDNA$) that aims to compute non-self-contact energy-minimizers (of different links) for a DNA molecule, using a {\it birod} model \cite{MoakherMaddocks2005} that is a continuum limit (developed with DNA in mind) of a sequence-dependent {\it rigid base} model of DNA called $cgDNA$ \citep{Petkeviciute2012, Gonzalez2013, Petkeviciute2014}, the predecessor model of $cgNA+$. This is in contrast to a fairly substantial literature of {\it rod} models, which can be viewed as a continuum limit of some {\it rigid base pair} model of DNA. The $cgDNA$ model arose from the MD runs performed by the Ascona B-DNA Consortium \cite{Beveridge2004}.  Analysis of this data revealed a relatively poor fit when using a rigid basepair model, but a relatively good fit when using rigid bases, leading to the development of $cgDNA$ in \cite{Gonzalez2013, Petkeviciute2014}. 
An advantage of passing to a continuum model is the ability to connect with some mathematical tools not present for a large-dimensional minimization problem.  In the case of energy minimisers of a birod, one can look for critical points of an energy functional within the calculus of variations. For the birod model \cite{Glowacki2016} and \cite{Grandchamp2016} show that the equilibrium equations satisfied by these critical points can be put in Hamiltonian form and find within those equations the familiar elastic rod equations as a special case. The known solution set for that special case, mapped out in full in \cite{DichmannLiMaddocks1996,ManningMaddocks1999}, thus serves as a useful starting point for birod computations. Furthermore, they exploit previous work on symmetry-breaking when starting on this known solution set \cite{ManningMaddocks1999} to develop an automated algorithm that finds with fairly high confidence all low-energy branches of solutions.

For $cgNA+$ minicircles in its native Curves+ coordinates, the closure constraints are 
completely nonlocal, posing a substantial computational challenge. 
We show here a remarkably tractable, local change of variable from the Curves+ relative coordinates to absolute coordinates (with Cartesian translations and rotations expressed in quaternions) of all rigid base and phosphate groups. In these coordinates the energy is no longer quadratic (but simpler than the closure constraints in Curves+ coordinates) and the minicircle closure condition is very simple, making the overall problem numerically tractable. Another central element of our algorithm is the design of suitable initial guesses for the energy minimisation algorithm.  These initial guesses are generated by first computing uniform helical configurations that achieve a chosen number of turns (eventually becoming $Lk$) in a way that incorporates the $cgNA+$ energy, and then wrapping these helices into a loop, so that the base pair origins lie on a circular torus.  This second step involves a choice of {\it register}, i.e., which direction to bend the helix to create a loop.

The paper is organized as follows. In Section 2, we give a brief description of 
the $cgNA+$ model, including a mild extension that computes a ``periodic'' stiffness matrix and linear 
ground-state suitable for modeling a covalently closed minicircle. 
In Section 3, we present the change of coordinates and our approach to computing link. 
In Section 4, we present the minimization problem in the new variables and our algorithmic approach to
solving it, including designing an ensemble of initial guesses and approaches to determining if the 
algorithm has converged and whether the results from two different initial guesses are categorized as
distinct. 
Section 5 present the results, where we show a ``typical case'' where we see exactly two energy minimisers
with adjacent values of $Lk$, and then cases that exhibit more than two links and/or more than one 
distinct minimisers at a given link.  Later in Section 5, we show the correlation between $cgNA+min$ energies 
and energies computed from experimentally measured $J$-factors. Finally, we show the statistics of $cgNA+min$ minicircles 
for $120K$ sequences of different lengths. Section 6 concludes our paper.
\section{The $cgNA+$ model}
\label{sec:cgnaplus}
In this Section we summarize the aspects of the prior $cgNA+$ model that provide our starting point for modelling sequence-dependent, energy minimising, minicircle configurations.
\subsection{The $cgNA+$ model of linear fragments}
\label{cgNA+linear}
For linear dsNA fragments of any sequence, the $cgNA+$ model \citep{Sharma2023} (with full detail available in the theses \cite{Patelli2019} and \cite{Sharma2023t}) provides a computationally highly efficient prediction of sequence-dependent equilibrium distributions expressed in enhanced Curves+ internal configuration coordinates.
More precisely, given a sequence $\calS$ (along a designated reading strand) of length $N$ bp and a parameter set $\calP$ 
(sets are available for dsDNA in both standard and epigenetically modified alphabets, for dsRNA, and for DNA-RNA hybrid fragments), 
the $cgNA+$ model predicts a Gaussian, or multivariate normal, probability density function (or pdf) 
\begin{equation}
\rho(\Omega; \calS,\calP) = \frac{1}{Z} e^{-U(\Omega;\calS,\calP)} ~~~~  \text{with energy } ~~~~ U(\Omega; \calS,\calP) = \frac{1}{2} \big[\Omega-\hat{\Omega} \big]^T \calK \big[\Omega-\hat{\Omega}\big], 
\label{cgNA+pdf}
\end{equation}
where the components of the vector $\Omega \in \rmath^{24N - 18}$ are the configuration space coordinates,
$\calK (\calS) \in \R^{(24N-18) \times (24N-18)}$ is the  positive-definite, banded {\em stiffness} (or inverse covariance, or precision) matrix, $\hat{\Omega}$ are the coordinates of the linear ground state (or intrinsic or minimum energy) configuration, $Z(\calS,\calP)$ is the (of course explicitly known) normalization constant (or partition function), and we refer to the shifted quadratic form $U$ as the $cgNA+$ energy of the configuration $\Omega$. This predictive $cgNA+$ model pdf has consistently been shown to be highly accurate for a wide variety of test sequences $\calS_i$ of varying lengths in the sense that the differences between the pdf mean $\hat{\Omega}(\calS_i)$ and covariance $\calK^{-1}(\calS_i)$, and the first and second moments estimated from the appropriate time averages over fully atomistic MD trajectories, are very small when compared to variations of $\hat{\Omega}(\calS_i)$ and $\calK^{-1}(\calS_i)$ as a function of the sequences $\calS_i$. In this presentation we are concerned only with finding configurations that minimise (a modified periodic version of) the $cgNA+$ energy $U(\Omega)$, i.e.\ finding configurations that provide peaks of the pdf, when the configuration coordinate $\Omega$ is constrained to express a cyclisation condition, so that the absolute energy minimiser achieved by the linear ground state $\hat{\Omega}$ is not accessible.

\begin{figure}[ht]
\centering
\includegraphics[width=1.9in]{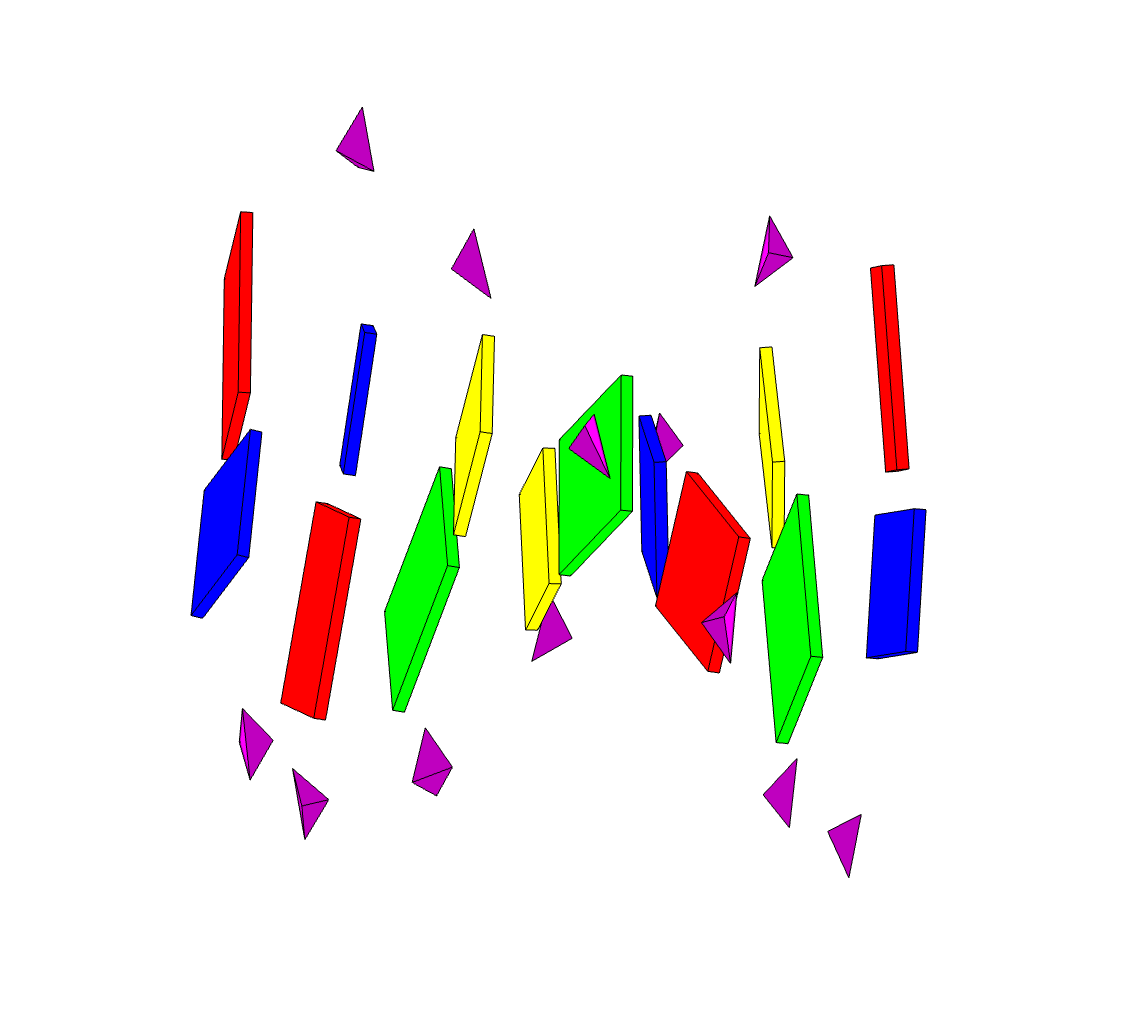}
\qquad
\includegraphics[width=3.2in]{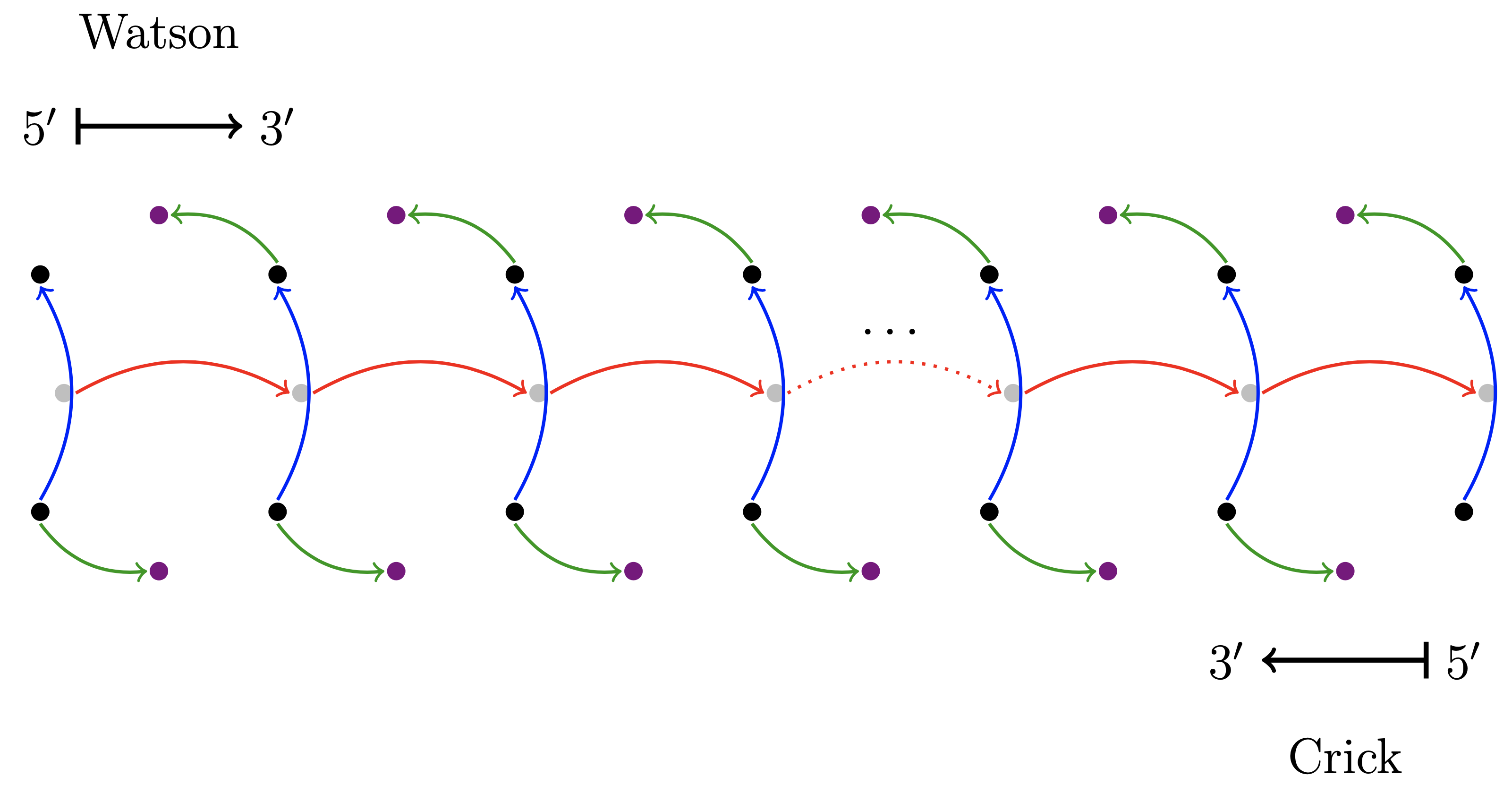}

\caption{Left: A cartoon representation of the level of coarse-graining in the $cgNA+$ model. Each base and phosphate group is approximated as a rigid body, or frame. The oligomer is blunt-ended so that the two $5'$ end phosphates are absent. Right: Schematic of the tree structure of the $cgNA+$ relative rigid body displacement coordinates: dots represent frames (black for base, gray for base-pair, purple for phosphate) and arrows represent coordinates (red for inter, blue for intra, green for phosphate)
expressed in their associated frame.}
\label{fig:cgNA+_schematic}
\end{figure}

We will have need of more detail concerning the $cgNA+$ model coordinates. The level of coarse graining in the $cgNA+$ model is indicated in the left panel of Fig.~\ref{fig:cgNA+_schematic}. The oligomer shown has $7$ base pairs, with blunt ends, i.e.\ the two final $5'$ end phosphate groups are absent. 
Each of the two nucleotides in an interior base pair level is described at the coarse grain level as a pair of rigid bodies, one embedded in the atoms of the base and one embedded in the atoms of the phosphate group. Consequently the $cgNA+$ coarse grain configuration of each interior base pair level consists of four rigid bodies. The first and last base pair levels have only three rigid bodies because of the two missing phosphate groups.  Of course in the atomistic MD simulations used to train the model parameter set $\calP$ all the water and ions in the solvent, along with the other atoms of the nucleic acid, including the sugar groups, are treated explicitly, but they only enter the $cgNA+$ model implicitly via $\calP$. Effectively the $cgNA+$ model pdf represents a marginal over all of the degrees of freedom omitted from the chosen coarse grain level description of the configuration.

More abstractly any rigid body configuration can be described as a {\em frame}, or Cartesian reference point $o$ combined with a proper rotation (or orientation or direction cosine) matrix $R$, both regarded as absolute coordinates with respect to some fixed reference (or laboratory) frame. The $cgNA+$ model coordinates are derived from {\em relative} rigid body displacements, i.e.\ relative 3D rotations and relative 3D translations, between pairs of frames that are designated by the model as adjacent (as we now describe).  We introduce $N$ additional {\em base pair} frames, defined as an appropriate average of the two base frames at each base pair level.  We then show in the right hand panel of Figure \ref{fig:cgNA+_schematic} the $4N-3$ relative rigid body displacements that are used to define the $cgNA+$ coordinates ($N$ base-to-base, $N-1$ base pair to base pair, and $2N-2$ base-to-phosphate).  
Prescribing these $4N-3$ relative displacements prescribes the shape of the oligomer, and the absolute configuration of the oligomer is prescribed if the absolute location and orientation of any one frame is also prescribed.
The choice of relative displacements to define $cgNA+$ coordinates is convenient because the pdf (\ref{cgNA+pdf}) and its energy $U(\Omega; \calS,\calP)$ model a dsNA fragment in the absence of any external field, and so have no dependence on an overall translation and rotation. 

Another important detail involves the use of junction frames (an appropriate average of two adjacent base pair frames) in order to define the $cgNA+$ coordinates corresponding to the base pair-to base pair relative rigid body displacement.
The 
use of base pair and junction frames is a standard way to guarantee a simple transformation rule for the relative rigid body translations  under the inherent symmetry of dsDNA in which either of the two Crick and Watson anti-parallel backbones could be chosen as the reference 
strand for both the coordinates $\Omega$ and the sequence $\calS$.

To be even more explicit, each of the $4N-3$ relative rigid body displacement is described by three rotation and three translation scalar coordinates in a relatively standard way. The relative displacements between the base frames are parametrized by 
 twelve familiar and standard coordinates namely (a Curves+ \citep{Lavery2009} implementation of the Tsukuba convention \citep{Olson2001} for) six {\it intra} base pair coordinates (buckle, propeller, opening, shear, stretch, stagger) and six {\it inter} base pair (or junction) coordinates (tilt, roll, twist, shift, slide, rise). The relative displacements to the phosphate frames are then given in terms of less familiar, but analogous, sets of three relative rotation and three translation coordinates serving to locate each phosphate group, one in each backbone, with respect to the base within its nucleotide 
 As a consequence of these choices, the $cgNA+$ internal coordinate vector $\Omega$ of a linear, blunt-ended dsNA fragment of length $N$ base pairs is finally written in the form
\begin{equation}
\Omega = \{ x_1, z_1^{-}, y_1, z_2^{+}, x_2, z_2^{-}, y_2,..,z_i^{+},x_i,z_i^{-},y_i,.., y_{n-1},z_N^{+}, x_N \} \in \rmath^{24N-18}. 
\end{equation}
Here $x_i = (\eta_i, \xi_i) \in \rmath^6$ are intras, $z_i^{-} = (\tau_i^{-}, \zeta_i^{-})\in \rmath^6$ are coordinates of Crick phosphates, $z_i^{+} = (\tau_i^{+}, \zeta_i^{+})\in \rmath^6$ are coordinates of Watson phosphates, and $y_i = (u_i,v_i)\in \rmath^6$ are inters. The first base pair level has only a Crick phosphate, and the last base pair level has only a Watson phosphate, so $i$ runs from $1$ to $N - 1$ for Crick phosphates and inters, $2$ to $N$ for Watson phosphates, and $1$ to $N$ for intras, making the total length of $\Omega$ be $6(N-1)+6(N-1)+6(N-1)+6N=24N-18$. Each symbol in $\{\eta_i, \xi_i, \tau_i^{\pm}, \zeta_i^{\pm}, u_i, v_i \}$ represents a 3-tuple of coordinates of a vector in an associated frame, with $i$ being an index along the dsNA chain:

\noindent $\bullet$ each $\eta_i$ is a Cayley vector encoding a relative intra base pair rotation;

\noindent $\bullet$ each $\xi_i$ is an intra base pair translation expressed in the base pair frame; 

\noindent $\bullet$ each $\tau_i^{-}$ or $\tau_i^{+}$ is a Cayley vector encoding a relative base-to-phosphate rotation;

\noindent $\bullet$ each $\zeta_i^{-}$ or $\zeta_i^{+}$ is a base-to-phosphate translation expressed in the associated base frame;

\noindent $\bullet$ each $u_i$ is a Cayley vector encoding a relative inter base pair rotation; and

\noindent $\bullet$ each $v_i$ is an inter base pair translation expressed in the junction frame. 

\noindent We refer to Appendix \ref{Annexe: Cayley vector} for details on the (standard) Cayley vector representation of rotations. We do however note that the Cayley vector parameterisation is only valid for rotations through less than $\pi$ radians (or $180$ degrees). As the Cayley vector is used within the $cgNA+$ model to parametrise relative rotations this restriction is effectively of no consequence. This point is discussed further, along with the choices of units and scaling within the $cgNA+$ model, in Appendix \ref{Annexe: quaternion}.

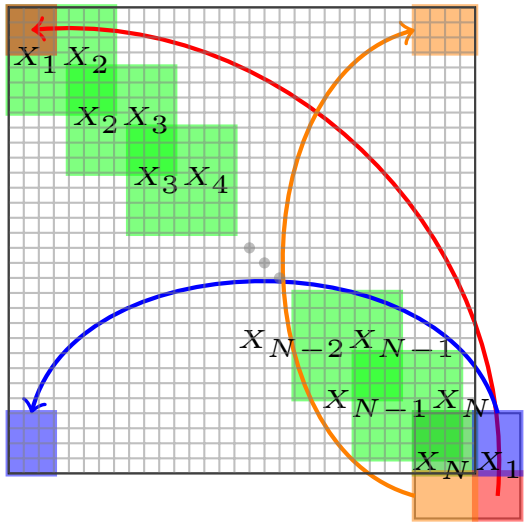
\begin{figure}
\centering
\scalebox{2.0}{
\begin{tikzpicture}
    \filldraw[green,very thick,opacity=.5] (0,3.1) rectangle (0.7,2.4);
    \filldraw[green,very thick,opacity=.5] (0.4,2.7) rectangle (1.1,2.0);
    \filldraw[green,very thick,opacity=.5] (0.8,2.3) rectangle (1.5,1.6);
    \filldraw[green,very thick,opacity=.5] (1.9,1.2) rectangle (2.6,0.5);
    \filldraw[green,very thick,opacity=.5] (2.3,0.8) rectangle (3.0,0.1);
    \draw[gray] (2.7,0.4) rectangle (3.4,-0.3);
    \filldraw[black!25!green,very thick,opacity=.5] (2.7,0.4) rectangle (3.1,0);
    \filldraw[red,very thick,opacity=.5] (0,3.1) rectangle (0.3,2.8);
    \filldraw[red,very thick,opacity=.5] (3.1,0) rectangle (3.4,-0.3);
    \draw[red,->,thick] (3.25,-0.15) to [bend right=50](0.15,2.95);
    \filldraw[blue,very thick,opacity=.5] (0,0) rectangle (0.3,0.4);
    \filldraw[blue,very thick,opacity=.5] (3.1,0.4) rectangle (3.4,0);
    \draw[blue,->,thick] (3.25,0.4) to [bend right=75](0.15,0.4);
    \filldraw[orange,very thick,opacity=.5] (2.7,2.8) rectangle (3.1,3.1);
    \filldraw[orange,very thick,opacity=.5] (2.7,0) rectangle (3.1,-0.3);
    \draw[orange,->,thick] (2.7,-0.15) to [bend left=75](2.7,2.95);
    \filldraw[gray,opacity=.5] (1.6,1.5) circle (0.03);
    \filldraw[gray,opacity=.5] (1.7,1.4) circle (0.03);
    \filldraw[gray,opacity=.5] (1.8,1.3) circle (0.03);
    \fontsize{2pt}{12pt}\selectfont
    \node[] at (0.35,2.75) {$X_1X_2$};
    \node[] at (0.75,2.35) {$X_2X_3$};
    \node[] at (1.15,1.95) {$X_3X_4$};
    \node[] at (2.25,0.85) {$X_{N-2}X_{N-1}$};
    \node[] at (2.65,0.45) {$X_{N-1}X_N$};
    \node[] at (3.05,0.05) {$X_N X_1$};
    \draw[black] (0,0) rectangle (3.1,3.1);
    \draw[gray,opacity=.5,step=0.10] (0,0) grid (3.1,3.1);
\end{tikzpicture}
}
\quad
\caption{Schematic of the sparsity pattern of the periodic $cgNA+$ stiffness matrix $\calK_p(\calS)$ for a sequence $\calS = X_1X_2..X_N$ of length $N$. Each small square in the grid is $6\times6$ and the overlaid green squares are $42\times 42$, each one modelling the physical interactions between adjacent base pair levels  $X_iX_{i+1}$. For covalently closed minicircles there is an interaction between first and last base pair levels $X_NX_1$ with the purple, red, and orange sub-blocks indicating the placement of the associated coefficients, modifying the sparsity pattern from the strictly banded $cgNA+$ stiffness matrix $\calK(\calS)$ that arises for linear fragments.}
\label{fig: periodic cgDNA+ stiffness}
\end{figure}

\subsection{The periodic $cgNA+$ model}
\label{subsec:stiffness_matrix}
Our goal is to compute the sequence-dependent minimal energy shapes of covalently closed dsNA minicircle configurations.  There are then no blunt ends and no missing phosphate groups. And there are the same number of base pair junctions as base pair levels, because the first and last base pair levels interact in just the same way as any other two adjacent base pair levels. Consequently the choice of coordinates must be modified to introduce $cgNA+$ periodic internal coordinates $\Omega_p$ and corresponding periodic ground state $\hat{\Omega}_p$:
\begin{equation}\label{periodic_gr_variables}
\Omega_p = \{z_1^{+}, x_1, z_1^{-}, y_1, z_2^{+}, x_2, z_2^{-}, y_2,..,z_i^{+},x_i,z_i^{-},y_i,.., y_{N-1}, z_N^{+}, x_N, z_N^{-}, y_N \} \in \rmath^{24N} 
\end{equation}
Here $i$ now runs from $1$ to $N$ for all of $\{x_i, z_i^{\pm}, y_i\}$, in particular $\Omega_p$ includes $\{z_1^{+},  z_N^{-}, y_N\}$ which are absent from $\Omega$. 

A suitably modified periodic stiffness matrix $\calK_p$ also needs to be introduced to reflect the interaction of base pair levels $N$ and $1$. The way to do this was first described in \citep{Glowacki2016, Grandchamp2016}, albeit working in the context of a precursor linear fragment model called $cgDNA$ \citep{Petkeviciute2012, Gonzalez2013, Petkeviciute2014} which did not explicitly treat the phosphate groups, and with the different motivation of modelling tandem repeat dsDNA sequences. 

The construction of the periodic stiffness matrix $\calK_p$ for a basal base sequence $\calS= X_1X_2..X_N$ of length $N$ is illustrated in Figure \ref{fig: periodic cgDNA+ stiffness}. In the examples presented here we restrict ourselves to the case of dsDNA minicircles, and moreover without epigenetic base modifications, so that each $X_i\in\{\text{A,T,C,G}\}$, but this restriction could be easily removed as more general parameter sets $\calP$ are already available.
The periodic stiffness matrix $\calK_p$ is then a $24N \times 24N$ matrix whose non-zero entries lie within $42 \times 42$ overlapping blocks, each centered on the inter, or junction, variable of the $(i,i+1)$ (or by convention $i$th) junction step, and with $18 \times 18$ overlaps centered on the base pair level variables $i$ common to the junction steps $(i-1,i)$ and $(i,i+1)$. If the dimer sequence at the $(i,i+1)$ dinucleotide step (read along the reading or Watson strand) is for example $X_iX_{i+1} = \text{AG}$ then the entries in the corresponding stiffness block are read from a symmetric positive-definite matrix $K^{AG}$ appearing in the $cgNA+$ parameter set $\calP$, with no dependence on the index $i$ other than the dimer sequence context. Similarly at the $X_NX_1$ junction the entries of the stiffness block just depend on the dimer sequence context (in the order $X_N$ followed by $X_1$), but the sub-blocks of the $42\times 42$ stiffness block have to be placed as indicated in Figure \ref{fig: periodic cgDNA+ stiffness} reflecting the fact that by periodicity the $N$th intra, phosphate, and inter variables are coupled to the $1$st base pair level intra and phosphate coordinates.

For all detail of how the $cgNA+$ parameter set $\calP$ is estimated we defer to \citep{Sharma2023, Patelli2019, Sharma2023t}. Here we merely remark that the parameter set blocks are estimated from a suitable small library of fully atomistic MD simulations of short linear dsNA fragments. The dimension $42$ of the stiffness blocks arises because they model all nearest-neighbour interactions between all eight rigid bodies making up two successive base pair levels,  expressed as a function of the associated seven relative rigid body displacements, each with six scalar coordinates, as illustrated in the tree structure in Figure~\ref{fig:cgNA+_schematic}. The absence of any beyond nearest neighbour interactions, and the purely dimer sequence-dependence of the stiffness blocks are both shown to be highly accurate approximations as part of the estimation of the parameter set $\calP$. We do however remark that, in contrast, the entries in the ground state vector $\Omega_p$ have a significant long-range sequence-dependence well beyond the dimer sequence context, which is accurately captured within the $cgNA+$ model.

In fact most of the challenge in computing minicircle energy minimisers is related to the closure condition which involves only the inter coordinates $y_i$. And we address these difficulties by introducing changes of variable that involve only the $y_i$, and which leave the intra base pair $x_i$ and phosphate coordinates $z_i^{\pm}$ unaltered. For this reason it is notationally convenient to introduce $w_i:= \{z_i^+, x_i, z_i^-\} \in \rmath^{18}$ which we will refer to as {\em  base pair level coordinates}. Then we arrive at the periodic $cgNA+$ energy
\begin{equation}\label{eq: total_energy_periodic}
U_p(\Omega_p; \calS,\calP) = \frac{1}{2}(\Omega_p - \hat{\Omega}_p)^T \calK_p (\Omega_p - \hat{\Omega}_p). 
\end{equation}
where 
\begin{equation}
\label{base_pair_level_coordinates}
\Omega_p = \{w_1, y_1, w_2, y_2, \dots , w_N, y_N \} \in \rmath^{24N}. 
\end{equation}
In the rest of the article we discuss only the periodic case, so we drop the subscripts $p$, and $\Omega\in \rmath^{24N}$, $\hat\Omega(\calS)\in \rmath^{24N}$ and $\calK(\calS)\in \rmath^{24N}\times \rmath^{24N}$ will hereafter be understood to refer to the periodic versions of the configuration, ground state, and stiffness matrix.

\section{Absolute base pair frame quaternion coordinates, and the minicircle closure condition}
\label{sec:cyclization_prob_stat}

It is not the case that all sets of periodic coordinates $\Omega$ correspond to closed minicircle configurations. In particular the ground state $\hat\Omega(\calS)$ of the periodic $cgNA+$ energy (\ref{eq: total_energy_periodic}) will usually not correspond to a closed loop configuration, which is why the computation of energy minimising  minicircle configurations is nontrivial. In this Section we explain the constraints on $\Omega$ that express closure of the minicircle, and introduce a change of variable that renders the search for minicircle energy minimisers computationally tractable.

\subsection{Absolute base pair frame coordinates and loop closure}
\label{sec:homogeneous_absolute}

We now introduce a  standard (often called homogeneous) coordinate representation of a rigid body configuration 
to describe the absolute orientation and location of each base pair frame by defining
\begin{equation}
\label{Ds}
D_i := \begin{bmatrix}
    R_i & o_i \\
    {\bf 0}^{T} & 1 
\end{bmatrix}  \in \R^{4\times 4}, \quad i= 1, \dots , N,
\end{equation}
where $R_i \in SO(3)$ (the group of $3\times 3$ proper rotation matrices) and $o_i \in \R^3$ are respectively the orientation (or direction cosine) matrix and Cartesian origin coordinates of the $i$th base pair frame with respect to a fixed (lab) reference frame. (Here the notation is $({\bf 0}^{T}, 1) = (0,0,0,1)$, so the bottom row of $D_i$ is constant independent of $i$.)
The set of all such $D_i$ is often denoted by $SE(3)$.

An overall translation and rotation of the dsDNA fragment  configuration can be eliminated by prescribing the first base pair frame $D_1$ to take any value in $SE(3)$, and often it is convenient to choose $D_1=I_4$ (the $4\times 4$ identity matrix). The convenience of the coordinate representation (\ref{Ds}) lies in the fact that the remaining base pair frames can then be determined by a matrix recursion involving only the inter variables $y_i = (u_i,v_i)$ as coefficients
\begin{equation}
D_1= I_4, \qquad D_{i+1} = D_i \begin{bmatrix}
    P(u_i) & \sqrt{P(u_i)} v_i \\
    {\bf 0}^{T} & 1 
\end{bmatrix},  \qquad i= 1, \dots , N,
\label{eqn:base_pair_frame_construction}
\end{equation}
where $P(u)\in SO(3)$ is
\begin{equation}
P(u) := 
\frac{10^2-\|u\|^2}{10^2+\|u\|^2} I_3 + \frac{20}{10^2+\|u\|^2} [u \times] + \frac{2}{10^2+\|u\|^2} u u^T,
\label{eqn:P}
\end{equation}
$[u \times]$ denotes the skew-symmetric matrix corresponding to the vector cross product with the Cayley vector $u$ ($u=u_i$ for the specific junction Cayley vectors), and  ${\small\sqrt{P}}$ is the principal square root of any rotation matrix $P$  (i.e.\ the rotation about the same axis as $P$ but through half the angle). The formula in (\ref{eqn:P}) is the Euler-Rodrigues formula for $P$ in terms of $u$, see Appendix \ref{Annexe: Cayley vector}. The factors of $10$ arise due to the choice of scaling of the Cayley vectors $u_i$ that is natural within the $cgNA+$ model of dsDNA, see Appendix 
\ref{Annexe: quaternion}.
The $\sqrt{P(u_i)}$ matrix appears in the $(1,2)$ block of the recursion coefficient matrix because within the $cgNA+$ model the inter translation coordinates $v_i$ are expressed in the midway, or junction,  frame (which complicates life here, but, as previously remarked, facilitates the way that the Crick-Watson symmetry of switching the choice of reading strand is manifested in the model).

For $i=1, \dots , N-1$ the initial condition and recursion (\ref{eqn:base_pair_frame_construction}) give the absolute homogeneous coordinates of all of the dsDNA base pair frames $D_1, \dots D_N$ in terms of the $N-1$ inter variables $y_1, \dots y_{N-1}$. But the recursion also holds in the case $i=N$ to further define a frame $D_{N+1}$. And the minicircle closure condition in absolute coordinates is precisely $D_{N+1}=D_1$, i.e.\ the $(N+1)$st $D_i$ frame coincides with the first. In turn, from the recursion relation (\ref{eqn:base_pair_frame_construction}) we can see that closure is satisfied for any choice of $D_1$ precisely if the inter variables $y_i$, $i=1, \dots , N$ satisfy the $SE(3)$ matrix equality constraint 
\begin{equation}
\prod_{i=1}^N \begin{bmatrix}
    P(u_i) & \sqrt{P(u_i)} v_i \\
    {\bf 0}^{T} & 1 
\end{bmatrix} = I_4.
\label{SE3_constraint}
\end{equation}
Because $SE(3)$ is a group, the matrix product on the left hand side of (\ref {SE3_constraint}) lies in $SE(3)$. Consequently there are only six independent scalar constraints in the matrix equation (\ref{SE3_constraint}). 
In fact because each factor in the matrix product in (\ref{SE3_constraint}) is a function of only one set of inter variables $y_i\in \R^6$, any given set of inter variables $y_i$ can be explicitly eliminated in favour of the other $N-1$ sets. For example to eliminate $y_N$, (\ref{SE3_constraint}) can be rewritten in the form
\begin{equation}
\begin{bmatrix}
    P(u_N) & \sqrt{P(u_N)} v_N \\
    {\bf 0}^{T} & 1 
\end{bmatrix} = \prod_{i=(N-1), \dots ,1} \begin{bmatrix}
    P^T(u_i) & -P^T(u_i)\sqrt{P(u_i)} v_i \\
    {\bf 0}^{T} & 1 
\end{bmatrix},
\label{elimination_constraint}
\end{equation}
where on the right hand side the explicit form of the inverse of a $SE(3)$ matrix has been used, and the reverse order of the product should be remarked. The $(1,1)$ block that arises after the product on the right is evaluated via block multiplication retains the property that each factor depends on only one $y_i$, in fact only on the inter Cayley vector $u_i$. The $(1,1)$ block equality can be used to extract the Cayley vector $u_N$ (provided that issues concerning rotations close to $\pi$ are resolved). But the analogous $(1,2)$ block is nonlinear and nonlocal, coupling all of the inter variables $(y_1, \dots , y_{N-1}) \in \R^{6(N-1)}$.

The problem of computing minicircle energy minimising configurations can now be given its first complete formulation: for a given sequence $\calS$ (and implicitly a given parameter set $\calP$) minimise the periodic $cgNA+$ energy (\ref{eq: total_energy_periodic}) over all periodic configuration coordinates (\ref{base_pair_level_coordinates}) for which the inter components $(y_1, \dots ,y_N)\in \R^{6N}$ satisfy constraints (\ref{SE3_constraint}). That is a well-posed problem which in principle could be fed to a nonlinear optimisation code that treats equality constraints. 
However we are unaware of any successful numerical treatment using this approach.
The difficulty is apparently that while the objective function (\ref{eq: total_energy_periodic}) is as simple as could be desired, just a quadratic well in the given coordinates, the constraints (\ref{SE3_constraint}) are highly nonlinear and nonlocal.  Similarly one set of inters could be explicitly eliminated via (\ref{elimination_constraint}) leading to a modified periodic $cgNA+$ energy (\ref{eq: total_energy_periodic}) in which one (and only one) inter argument has a nonlinear dependence on all of the other inters. However that approach appears to be too imbalanced in the necessary modifications to the objective function to lead to a robust numerical unconstrained minimisation algorithm.

\subsection{The change of variable to absolute base pair frame coordinates}
\label{homogeneous_absolute}
We now introduce more balanced changes of variable involving all of the inter coordinates $y_i$, $i=1, \dots , N$. The objective function for minimisation is then no longer quadratic, but in a more symmetric way than directly exploiting (\ref{elimination_constraint}), and the new set of independent variables satisfy the closure constraint automatically. Throughout, the base pair level coordinates $(w_1, \dots ,w_N)\in \R^{18N}$ are left untouched. The first change of variable of this type is to the coordinates
 \begin{equation}
 D_1 \left(=D_{N+1}\right)\  \text{prescribed.}\   D_i\in SE(3), \  i=2, \dots , N: 
 \label{independent_base_frames}
 \end{equation}
 \begin{equation}
 \det(D_i+D_{i+1}) > 0,  \; \forall i =1, \dots , N
 \label{d_det_condition}
 \end{equation}
where the independent unknowns are now the $N-1$ base pair frames $D_2$ through $D_N$. Prescribing $D_1$ is just eliminating the overall rotation and translation, and setting $D_{N+1} = D_1$ is the closure constraint. Both $D_1$ and $D_{N+1}$ need to be prescribed because in the (across junction) nearest-neighbour determinant condition $\det(D_i+D_{i+1}) > 0$ 
on the unknowns $D_2, \dots ,D_N$ the prescribed value of $D_1$ enters for $i=1$ and the prescribed value of $D_{N+1}$ is needed when $i=N$. The origin of this determinant condition is just that it is a compact mathematical notation for the condition that each of the $N$ relative junction rotations $R_i^TR_{i+1}$ do not have an eigenvalue $-1$, i.e.\ they are not a rotation through $\pi$, so that the associated junction Cayley vector is defined. In fact it is easily computed that $\det(D_i+D_{i+1})= 2\, \det(R_i+R_{i+1}) = 2\,\det (I_3 +R^T_iR_{i+1}) = 8(1+\cos\Theta_i)$, where $0\le\Theta_i\le \pi$ is the rotation angle in the $i$th junction. Consequently the determinant is always non-negative, and vanishes only when $\Theta_i = \pi$, i.e.\ when the junction Cayley vector $u_i$ is not defined.  

There is an invertible mapping (or change of coordinates) between the base pair frame variables $D_i$ defined in (\ref{independent_base_frames},\ref{d_det_condition}) and the set of inter variables $y_i$ defined in
\begin{equation}
\label{independent_inters}
D_1 \left(=D_{N+1}\right)\  \text{prescribed.}\  \text{Inters}\ y_i, \  i=1, \dots , N\ \text{satisfying constraints (\ref{SE3_constraint})}.
\end{equation}
The mapping from (\ref{independent_inters}) to (\ref{independent_base_frames},\ref{d_det_condition}) is just the recursion (\ref{eqn:base_pair_frame_construction}). The mapping from (\ref{independent_base_frames},\ref{d_det_condition}) to (\ref{independent_inters}) is
\begin{equation}
\label{Ds_to_inters}
\left[\frac{u_i}{10}\times\right] = \frac{4(R_i^TR_{i+1} - R_{i+1}^TR_i)}{\det(D_j+D_{j+1})}, \qquad v_i = \sqrt{R_i^TR_{i+1}}^TR_i^T (o_{i+1} - o_i),\quad i=1, \dots ,N.
\end{equation}
where, for any $u \in \R^3$, the symbol $[u\times]$ denotes the skew symmetric matrix 
such that the cross product $u \times v$ equals $[u\times] v$ for all $v$, and $R_i, o_i$ are the matrix and origin vector appearing in the blocks of the base pair frames $D_i$ (\ref{Ds}), including the two cases $i=1,N+1$ that are prescribed. As discussed in Appendix \ref{Annexe: Cayley vector} the first formula in (\ref{Ds_to_inters}) is the standard way to extract the Cayley vector parametrising the junction rotation matrix $R_i^TR_{i+1}$, with the factor of $10$ again appearing because of the particular choice of scaling in the $cgNA+$ model, as discussed in Appendix \ref{Annexe: quaternion}.

Composition with the change of coordinates (\ref{Ds_to_inters}) allows the periodic $cgNA+$ energy (\ref{eq: total_energy_periodic}) to be expressed as a function of the base pair level coordinates $w_1,\dots , w_N$ (which are uninvolved in the change of coordinates) and the set of independent unknowns $(R_2, o_2, \dots , R_N, o_N)$, taken along with all four of $R_1=R_{N+1}$ and $o_1=o_{N+1}$ whose values are prescribed. Consequently an unconstrained iterative minimisation algorithm could in principle be applied with calculus carried out on the rotation group $SO(3)$ for the $R_i$ variables. We do not pursue this approach because a further change of variables to quaternion coordinates yields an even simpler formulation, with, moreover, a little more information. 

\subsection{The change of variable to quaternion coordinates of absolute base pair frames}
\label{quaternion_absolute}

We now introduce the (standard) quaternion parameterisation of matrices in $SO(3)$. For any $0\neq q \in \R^4$ we follow the convention that $(q^1,q^2,q^3)$ denotes the vector part of the quaternion, while $q^4$ is the scalar part. Then, as described with more detail in Appendix \ref{Annexe: Cayley vector}, for rotations through $\Theta < \pi$ where the Cayley vector $u$ (in the $cgNA+$ scaling) for the same rotation matrix is defined, there is an invertible relation between $u$ and {\em unit} quaternions with $(q^1)^2 + (q^2)^2 + (q^3)^2 + (q^4)^2=1$, and $q^4 >0$:
\begin{equation}
\label{junction_u_to_p}
u/10 = (q^1,q^2,q^3)/q^4, \qquad q = (u/10,1)/\sqrt{1+\norm{u/10}^2},
\end{equation}
or 
\begin{equation}
\label{quat_axis}
(q^1,q^2,q^3) =  \sin(\Theta/2)\, \mathbf{n}, \qquad q^4 = \cos(\Theta/2),  
\end{equation}
where $\mathbf{n}$ is the unit rotation axis vector (oriented so that $\Theta$ is a right handed rotation) and we have used the fact that $\norm{u/10}=\tan(\Theta/2)$ to compute that $\cos(\Theta/2) = +1/\sqrt{1+\norm{u_i/10}^2}$. It is easily verified that substitution of the expression (\ref{junction_u_to_p}) for $u/10$ in the Cayley vector Euler-Rodrigues formula (\ref{eqn:P}) yields the quaternion  Euler-Rodrigues reconstruction formula for $R(q)$: 
\begin{equation}
    R(q) = \scriptstyle{\frac{1}{(q^1)^2 + (q^2)^2 + (q^3)^2 + (q^4)^2}}
    \begin{bmatrix} 
     \scriptstyle{(q^1)^2 - (q^2)^2 - (q^3)^2 + (q^4)^2} & 
     \scriptstyle{2(q^1q^2 - q^3q^4)} & \scriptstyle{2(q^1q^3 + q^2q^4)} \\
     \scriptstyle{2(q^1q^2 + q^3q^4)} 
     & \scriptstyle{-(q^1)^2 + (q^2)^2 - (q^3)^2 + (q^4)^2} 
     & \scriptstyle{2(q^2q^3 - q^1q^4)} \\
     \scriptstyle{2(q^1q^3 - q^2q^4)} & \scriptstyle{2(q^2q^3 + q^1q^4)} &
     \scriptstyle{-(q^1)^2 - (q^2)^2 + (q^3)^2 + (q^4)^2}
    \end{bmatrix}. 
    \label{eqn:quaternion_defn}
\end{equation}
Here the explicit normalisation in the pre-factor means that (\ref{eqn:quaternion_defn}) is not limited to unit quaternions, and, moreover, that (\ref{eqn:quaternion_defn}) is a double covering in the sense that $R(q)=R(-q)$. However the real utility of the quaternion coordinates is that (\ref{eqn:quaternion_defn}) is also valid for $q^4=0$ which is the case of rotations through $\Theta = \pi$, in which the Cayley vector is not defined. In fact (\ref{quat_axis}) and (\ref{eqn:quaternion_defn}) remain valid for rotation angles $\Theta$ taken to lie in the range $0\le \Theta \le 2\pi$. The implied double covering of $SO(3)$ removes the coordinate singularity of Cayley vectors close to rotations through $\pi$. For quaternion coordinates it is always true that for two close by rotation matrices $R_1$ and $R_2 \in SO(3)$ (independent of whether or not they are close to rotations through $\Theta = \pi$) there are two close by quaternions $q_1$ and $q_2$ with $q_2^Tq_1 > 0$ and $R_1 = R(q_1)$ and $R_2 = R(q_2)$. It is for this reason that in our energy minimisation we use quaternions to track the absolute orientations of the sequence of base pair frames $\{R_i\}$ around a minicircle configuration. In contrast $cgNA+$ adopts Cayley vector coordinates for the junction rotations $\{R_i^TR_{i+1}\}$ between two adjacent base pair frames, because in that context rotations close to $\pi$ can be avoided, and the Cayley vector parametrisation has other desirable features, for example it has the minimal possible dimension, namely three.

We will make use of the remarkable composition rule for quaternions \cite{Rodrigues1840}.
With the notation
\beqs
    B_1 :=
    \begin{bmatrix}
     0 &  0 &  0 &  1 \\
     0 &  0 &  1 &  0 \\
     0 & -1 &  0 &  0 \\
    -1 &  0 &  0 &  0
    \end{bmatrix}, \;\;
    B_2 :=
    \begin{bmatrix}
     0 &  0 & -1 &  0 \\
     0 &  0 &  0 &  1 \\
     1 &  0 &  0 &  0 \\
     0 & -1 &  0 &  0
    \end{bmatrix}, \;\;
    B_3 :=
    \begin{bmatrix}
     0 &  1 &  0 &  0 \\
    -1 &  0 &  0 &  0 \\
     0 &  0 &  0 &  1 \\
     0 &  0 & -1 &  0
    \end{bmatrix},\;\;
    B_4 := I_4,
\eeqs
we note that for any unit quaternion $q$, $\{B_j q,\  j = 1,2,3,4\}$ is an orthonormal basis for $\R^4$. Then if $q$ and $p$ are quaternions for matrices $R(q)$ and $R(p)$, one of the two quaternions $\pm r$ for the matrix product $R(r) = R(q)R(p)$ (where the ordering in the matrix product is of course significant) is given by
\begin{equation}
\label{composition}
r = \left (\sum_{j=1}^4 p^jB_j\right ) q.
\end{equation}
As the right hand side is bilinear, taking the alternative sign for one of $p$ or $q$ delivers the opposite sign for $r$. The relation (\ref{composition}) does not depend on the quaternions being unit, but if $p$ and $q$ are both unit, then so is $r$, as can be seen immediately from orthonormality of the basis $\{B_j q\}$. More generally $\norm{r} = \norm{p}\norm{q}$. Using orthogonality of $\{B_j q\}$ for any $q$, the relation (\ref{composition}) can be inverted in component form
\begin{equation}
\label{composition_components}
p^j = r^T B_j q,\qquad j=1,2,3,4.
\end{equation}

We now apply (\ref{composition_components}) in the specific cases where $q=q_i$ is a quaternion for the absolute base pair frame orientation matrix $R_i$, $r=q_{i+1}$ is a quaternion for the absolute base pair frame orientation matrix $R_{i+1}$, and $p=p_i$ is the $i$th junction quaternion as can be expressed in (\ref{junction_u_to_p}) in terms of the junction Cayley vector $u_i$. The result is a recurrence relation, analogous to (\ref{eqn:base_pair_frame_construction}) but now directly on the quaternion coordinates rather than on base pair frame orientation matrices
\begin{equation}
\label{quaternion_recursion}
q_1 \quad \text{prescribed,}\qquad q_{i+1} = \frac{\left (I_4 + \sum_{j=1}^3 u_i^j\, B_j/10 \right )}{\sqrt{1+\norm{u_i/10}^2}}\quad q_i, \qquad i= 1, \dots , N.
\end{equation}
Notice that if $q_1$ is chosen as a unit quaternion, then, because each junction quaternion specified in (\ref{junction_u_to_p}) is also unit, all subsequent $q_i$ are also unit. Moreover $q_{i+1}^Tq_i > 0$, $i=1, \dots , N$. If the sign of $q_1$ is switched, then the sign of all the subsequent $q_i$ also switches, but with the specific choice for junction quaternion (\ref{junction_u_to_p}) there is no further freedom in switching signs of individual quaternions within the recursion. Effectively the specific choice (\ref{junction_u_to_p}) for the junction quaternion enforces a nearest neighbour continuity in index of the choice of quaternion sign, as expressed in the conditions $q_{i+1}^Tq_i > 0$, $i=1, \dots , N.$

To complete the recursion we need an analogous expression for the absolute base pair frame origins $o_i$. To achieve that we exploit another beautiful feature of the quaternion parameterisation. For
the absolute base pair frame orientations $R_i=R(q_i)$ expressed by (\ref{eqn:quaternion_defn}) as a function of their quaternions $q_i$, which themelves are constructed from recursion (\ref{quaternion_recursion}), then a quaternion of the junction frame can immediately be computed from the arithmetic sum of the two base pair frame quaternions (with the sum scaled to be unit or not)
\begin{equation}
\label{junction_quat}
R_i\sqrt{R_i^TR_{i+1}}= R(q_i)\sqrt{R(q_i)^TR(q_{i+1})}  = R(q_i + q_{i+1}), \quad i=1, \dots , N.
\end{equation}
The validity of (\ref{junction_quat}) does rest on the equal norm condition $\norm{q_i}=\norm{q_{i+1}}$ and the inequality $q_{i+1}^Tq_i > 0$, both of which are satisfied by quaternions generated from the recursion (\ref{quaternion_recursion}). A proof is given in Appendix \ref{Annexe: quaternion}. 

With the quaternion of the absolute junction frame in hand, the recursion for the base pair frame origins is very simple
\begin{equation}
\label{origins_recursion}
o_1 \quad \text{prescribed,}\qquad o_{i+1} = o_i + R(q_i+q_{i+1}) \, v_i \qquad i= 1, \dots , N,
\end{equation}
which is just vector addition after the junction translation coordinates $v_i$ are rotated to be expressed in the absolute or laboratory frame. When implementing the combined recursion (\ref{quaternion_recursion})+(\ref{origins_recursion}), $q_{i+1}$ is first found from (\ref{quaternion_recursion}), after which (\ref{origins_recursion}) is an explicit expression for $o_{i+1}$.

Finally, we are able to introduce the set of coordinates in which we will compute. There is an invertible mapping (or change of coordinates) between the set of inter variables $y_i$ defined in (\ref{independent_inters}) (including satisfying the closure constraint (\ref{SE3_constraint})) and 
\begin{equation}
\label{independent_quat}
(q_1,o_1) \left(=(\pm q_{N+1},o_{N+1})\right)\  \text{prescribed,}\  \  (q_i,o_i)\in \R^4\times \R^3, \  i=2, \dots , N: q_{i+1}^Tq_i > 0,  \forall i =1, \dots , N.
\end{equation}
The mapping from (\ref{independent_inters}) to (\ref{independent_quat}) is just the recursions (\ref{quaternion_recursion})+(\ref{origins_recursion}). The mapping from (\ref{independent_quat}) to (\ref{independent_inters}) has two parts. First for the junction Cayley vectors $u_i$, using orthonormality of $\{B_j q\}$ for any unit $q$, the quaternion recursion (\ref{quaternion_recursion})  can be inverted to arrive at
\begin{equation}
\label{quat_junctions_to_cayley}
q_{i+1}^T q_i = \frac{1}{\sqrt{1+\norm{u_i/10}^2}} \left (= \cos(\Theta_i/2)\right), \qquad  \frac{u_i^j}{10} = \frac{q_{i+1}^T B_j q_i}{q_{i+1}^T q_i},\qquad j=1,2,3, \quad i= 1, \dots ,N.
\end{equation}
For the junction inter translations we merely invert (\ref{origins_recursion}) to obtain
\begin{equation}
\label{origins_recursion_inv}
v_i = [R(q_i+q_{i+1})]^T\, (o_{i+1} - o_i), \qquad i= 1, \dots , N.
\end{equation}

It remains only to clarify the connexion between the conditions
$D_1 \left(=D_{N+1}\right)$ prescribed, appearing in (\ref{independent_base_frames}), and
$(q_1,o_1) \left(=(\pm q_{N+1},o_{N+1})\right)$ prescribed, appearing in (\ref{independent_quat}). The conditions on the origins $o_1=o_{N+1}$ are identical with the conditions on the $(1,2)$ blocks of $D_1$ and $D_{N+1}$, but the conditions on the quaternions $q_1$ and $q_{N+1}$ deserve comment. First for the $(1,1)$ block $R_1$ prescribed there are the two possible choices of sign for the unit quaternion satisfying $R(\pm q_1) = R_1$. But as already remarked, changing this choice for $q_1$, but for the same inter Cayley vectors $u_i$ corresponding to a closed minicircle configuration, just flips the sign in all the $q_i$, and the continuity sign condition $q_{i+1}^Tq_i > 0$ is unaffected. But it is not possible to predict which sign in $q_{N+1}=\pm q_1$ arises, either of which guarantees the closure condition $R_{N+1} = R_1$, until the recursion relation for $q_i$ for given coefficients $u_i$ is solved. Some sets of $u_i$ satisfying closure will lead to $q_{N+1} = + q_1$ and some to $q_{N+1} = - q_1$. As explained in the next Section the two possible signs that can arise are not a degeneracy, rather they are well-defined extra information.

\subsection{Link}
\label{subsec:link}

The Gauss linking number is an integer (including possibly zero and negative values) defined for a pair of closed, pairwise nonintersecting  curves $x$ and $y$ (self intersections of $x$ with itself or $y$ with itself are of no consequence). We take the linking number of a $cgNA+$ minicircle to be the link of the two closed piecewise linear curves that interpolate, respectively, the origins of the phosphate groups on the Crick, and on the Watson, backbones. The link of two disjoint curves can be evaluated in various ways including counting signed crossings of $x$ and $y$ in some nondegenerate projection, or by evaluating the classic Gauss double integral. We evaluate numerically the Gauss double integral, using an algorithm adapted from that developed by \cite{KleninLangowski2000} for computing the analogous double integral expression for the  writhe of a single piecewise linear curve. 

We note that the $cgNA+$ model is a phantom chain approach in the sense that it includes no energy or penalty term included to prohibit strand passage of the two backbones adjacent to base pair level $I$ and the two backbones adjacent to another base pair level $J$, with $J$ very different from $I$. Indeed we frequently observe such events in intermediate computations of configurations generated by our iterative energy minimisation algorithm (prior to its convergence to a minimiser). The important \citep{Bates_Maxwell_2005} 
point for us is that as the (approximately) double helical backbones pass through each other there are two pertinent strand passages: the Crick backbone adjacent to base pair level $I$ crosses the Watson backbone adjacent to base pair level $J$, and the Watson backbone adjacent to base pair level $I$ crosses the Crick backbone adjacent to base pair level $J$. The consequence is that the link changes by $\pm 2$ during such a full double strand passage. Therefore if an initial configuration for energy minimisation is taken with an odd linking number, then all along the sequence of configurations generated by our iterative energy minimisation algorithm,  the link remains odd, although it can
change its value due to strand passage. Similarly if the initial configuration has an even link, then the link remains even.

The sign in the quaternion closure condition $q_1 = \pm q_{N+1}$ corresponds to whether the link is odd or even, with $q_1 =  -q_{N+1}$ corresponding to even links. Explaining this fact in full generality would take us too far afield, but it has been checked on the numerical examples presented below.

The fact that strand passages of dsDNA change link by $\pm 2$ implies that for any sequence of any length in any physically reasonable, phantom chain model, there should be minimisers with at least two distinct links, one odd and one even. Our computations within $cgNA+min$ suggest that for most sequences of around $100$ bp in length there are precisely two distinct links with minimisers, but we have found $94$ bp sequences with minimisers at three and four distinct links. And the expectation is that the longer the sequence, the more likely that there will be  minimisers with more than two distinct links, although we have not investigated this point thoroughly.  
It is also the case that there are sequences with multiple minimisers at the same link, but that is a different phenomenon.

\section{Minicircle energy minimisation algorithm}

\subsection{The objective function}\label{sec: cgDNAmin energy}

Finally we can arrive at our formulation of a numerically tractable, unconstrained minimisation problem that generates minicircle configurations that are local minimisers of the $cgNA+$ energy.
As objective function we take
\begin{equation}
\label{eq: cgDNAmin zvec energy}
   E \left(z, {\mathbb P} \right) := \frac{1}{2}\left(\Omega-\hat \Omega\right)^T \calK \left(\Omega-\hat \Omega\right)
   + p_w \sum_{i=2}^N{(\norm{q_i}^2-1)^2}.
\end{equation}
Here ${\mathbb P}= (\calS,\calP, o_1, q_1, o_{N+1}, q_{N+1}, p_w)$ are parameters, specifically a base sequence $\calS$ of length $N$, a $cgNA+$ model parameter set $\calP$, the absolute frame origins $o_1=o_{N+1}$ and orientations $q_1=\pm q_{N+1}$ simultaneously fixing the location and orientation of the minicircle in space and expressing the closure constraints (with the choice of $\pm$ set by the link of the choice of initial configuration), and $p_w$ a weight parameter in a penalisation term. 
As discussed above, most components of our energy are invariant to rescaling any quaternion $q_i$, but the expression for the inter translation $v_i$ relied on having unit quaternions.  (The general expression for $v_i$ for any set of quaternions can be readily computed, but is more complicated, so we opted to avoid its use in the interest of keeping the computation of the energy gradient and Hessian numerically tractable.)  As a result, we introduce a penalty term into the energy in \eqref{eq: cgDNAmin zvec energy}, so that the minimisation algorithm will force each quaternion to be (quite close to) normalized.
All of the parameters ${\mathbb P}$ are set as an input to a minimisation run and stay fixed during the run. Minimisation is over the unknowns $ \R^{25\,N-7} \ni z =(w_1, o_2, q_2, w_2, o_3, q_3, \dots , o_N, q_N, w_N)$. The $w_i\in \R^{18}$ are the base pair level coordinates introduced in (\ref{base_pair_level_coordinates}); the $(o_i, q_i)\in \R^7$ are the base pair frame absolute origins and quaternions introduced in (\ref{independent_quat}). We discuss the associated inequality constraints in the next paragraph. 

On the right the first term is the $cgNA+$ periodic energy (\ref{eq: total_energy_periodic}), which is given as an explicit quadratic function of the $cgNA+$ periodic variables $\Omega_p$ (\ref{base_pair_level_coordinates}). The base pair level coordinates $w_i$ appear explicitly on both sides of the expression (\ref{eq: cgDNAmin zvec energy}) for the objective function.
The right hand side can be written as an explicit composition function depending only the problem unknowns $z$ and the parameters $(o_1, q_1, o_{N+1}, q_{N+1})$, after the $cgNA+$ inter variables $y_i= (u_i,v_i)$,   $i=1, \dots , N$ are eliminated using the inverse recursions (\ref{quat_junctions_to_cayley}) and (\ref{origins_recursion_inv}). The key point is that the values of $y_i= (u_i,v_i)$ obtained in this way lie in the set (\ref{independent_inters}), and specifically they satisfy the minicircle closure constraint (\ref{SE3_constraint}) automatically.  For the formulas (\ref{quat_junctions_to_cayley}) to be valid we need that the (sharp) inequality constraints $q_{i+1}^T q_i > 0$ (appearing in (\ref{independent_quat})) hold for all  $i=1, \dots ,N$. These inequalities express the condition that the inter rotation angle $\Theta_i$ in each junction remains strictly less than $\pi$. In the next Section we describe how to construct initial conditions for which these inequalities are satisfied. Then the conditions can be anticipated to remain satisfied throughout the iterative minimisation algorithm, because as the $i$th constraint boundary is approached, the inter rotation angle $\Theta_i$ in the junction approaches $\pi$, which means that the norm of the associated Cayley vector grows without bound,  and if even one component $u^j_i$ of the corresponding Cayley vector  approaches infinity, then the $cgNA+$ energy will approach infinity, which should not arise during a minimisation algorithm. These expectations are indeed borne out in all of our numerical examples, so that an unconstrained minimisation routine can be effectively applied to find minimisers of (\ref{eq: cgDNAmin zvec energy}).


It is possible to compute closed form expressions for both the gradient and (the sparse) Hessian of the objective functional (\ref{eq: cgDNAmin zvec energy}) with respect to the unknown $z$. The necessary computations are outlined in Appendix \ref{sec:grad_and_hess}. 
Having these explicit formulas available substantially improves the efficiency of our energy-minimisation algorithm. 

\subsection{Generating initial configurations}\label{sec: minicircle_initial_guess}
\begin{figure}
\centering
\includegraphics[width=3.6in]{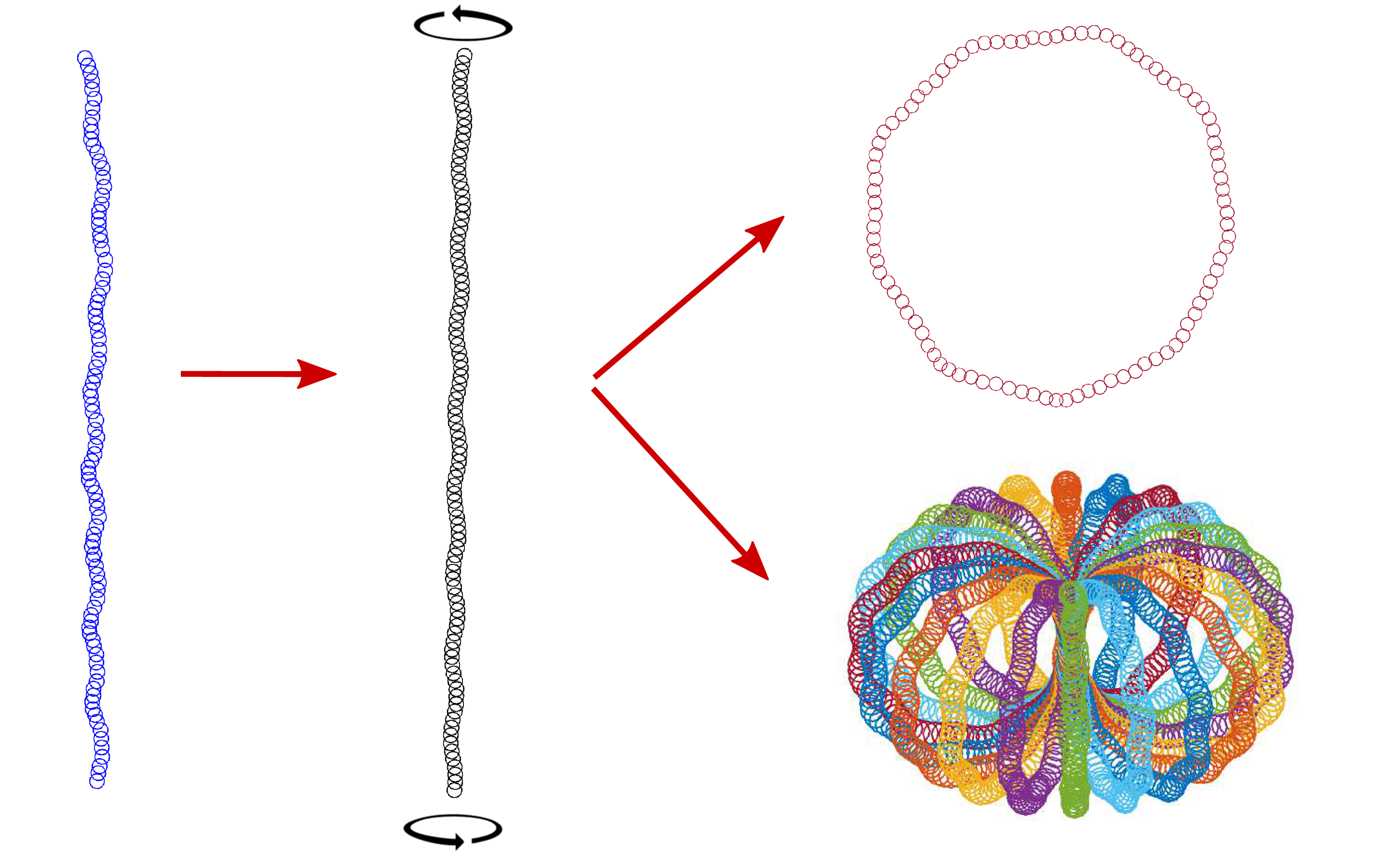} 
\caption{Illustration of procedure for determining initial-guess minicircles.  Left: periodic ground-state (i.e., non-uniform helical configuration) for a 94-bp DNA. Middle: ideal helix with a prescribed integer number of turns obtained from ground state by computing appropriate uniform inter coordinates.
Top right: bending of the ideal helix 
into a toroid of prescribed link (which equals the number of turns in the ideal helix). Bottom right: rosette of minicircle configurations corresponding to 20 distinct register angles. Our computations generate all $cgNA+$ coordinates for each configuration shown, but for clarity we plot just bp origins ($o_i^h$ for black, $o_i^t$ for red, $o_i^f$ for rosette).}\label{fig: method}
\end{figure}
We generate initial conditions with a range of links and registers through a two step preliminary computation (see below for a discussion of the notion of register). The associated detailed computations  are provided in Appendix \ref{initial}. We here just outline the central ideas of the approach using Fig.\ \ref{fig: method}. Any sequence has a periodic $cgNA+$ linear ground state $\hat \Omega$ which is typically a modulated double helix after the absolute coordinates are reconstructed (blue left panel). The standard theory of screw (or helical) deformations then states that if all of the inter variables $(u_i,v_i)$ are constrained to take index independent, but generally unknown, values $(u,v)$, then the absolute base pair origins lie exactly on a circular helix, whose radius and rise are explicitly known as simple functions of $(u,v)$. Moreover the Cayley vector $u$ is exactly aligned with the straight centreline of the helix, with the absolute orientation of the base pair frame being an index independent rotation from the centreline tangent and the radial vector. In  particular there is a closed form expression available for the origins and absolute quaternions $(o_i,q_i)$ for each of the base pair frames. With that geometry in mind it is possible to formulate the constrained quadratic minimisation problem
\beq
\label{helix_min}
\min_{\{\R^{18N+6}\ni \Omega: v_i=v\in \R^6\}} \frac{1}{2}\left(\Omega-\hat \Omega\right)^T \calK \left(\Omega-\hat \Omega\right) -\lambda \norm{u}^2,
\eeq
as a function of the scalar parameter $\lambda$. Each minimisation of the quadratic function (\ref{helix_min}) for a given $\lambda$ involves a single linear (sparse) solve in $\R^{18N+6}$, where the inters all take a single common, but unknown, set of values $(u,v)$ and the intra base pair level coordinates $w_i$ are all to be computed individually. A typical solution is shown after reconstruction to absolute coordinates in the second panel (black) of Fig.\ \ref{fig: method}. Values of the multiplier $\lambda$ that give rise to an arbitrary integer number of turns after $N$ bp steps (with the $N+1$st base pair composition matching the first bp composition) can be computed numerically. This integer will give rise to the Link of our initial minicircle configuration. With an exactly  helical configuration of known pitch and radius and an integer number of turns in hand, a particular plane containing the helix centreline can be picked. Then the straight central line segment can be wrapped onto a circle of known radius lying in that plane, and the base pair origins will lie on the surface of a circular torus with both radii known, and the first and $(N+1)$st bp frames exactly overlap. The intra base pair level variables $w_i$ are left unchanged from their helical sequence and index dependent computed values, while the inters are recomputed to index independent (but sequence dependent) values corresponding to a uniform wrapping around the torus. This is our first initial configuration of prescribed link, shown in red in the top right panel of Fig.\ \ref{fig: method}. Finally any of a one parameter family of planes can be selected for the deformation into a torus, leading to the rosette of initial configurations all of the same link shown in the bottom right panel of Fig.\ \ref{fig: method}. The different configurations in the rosette correspond to different registers, i.e.\ in its initial configuration the major groove face of the first bp could be oriented toward the centre of the torus, or away from the centre of the torus, or anywhere in between. 


\subsection{Determining if minimisers are distinct}\label{sec:distinct}

We follow the classic strategy in nonlinear optimisation of trying to find the maximum possible number of local minima by taking a large number of judiciously chosen initial configurations. (Of course we have no guarantee of finding all local minima.) We adopt quite rigorous stopping criteria for each iterative minimisation run, as described in Sec.\ \ref{sec: method} and Appendix \ref{app:opt_details}. However we are still left with establishing criteria to determine whether or not two configurations that have both satisfied the iterative stopping criteria are numerical approximations to the same local minimiser or to two distinct local mimimisers.

The first filter is to compute the link of the stopped configuration. Different link certainly implies different minimisers.

As a second filter we made the decision that we are primarily interested in differences in shape. For that reason we decided to construct a metric on the $cgNA+$ coordinates $\Omega$ in order to identify distinct minima, not on absolute coordinates. And because $cgNA+$ has coordinates of a distinctly different physical character, including different dimensions and scalings, we decided to use Mahalonobis distance, which is a weighted norm using the positive-definite, symmetric, sequence-dependent $cgNA+$ stiffness matrix $\calK(\calS)$ that is closely related to energy difference. Specifically given two $cgNA+$ configurations $\Omega_1$ and $\Omega_2$ for a common sequence $\calS$ with $N$ base pairs we define the scaled Mahalonobis distance as 
\beq
24N d^n(\Omega_1,\Omega_2;\calS) := (\Omega_1-\Omega_2)^T {\calK(\calS)} (\Omega_1-\Omega_2)/2
\eeq
The pre-factor $24 N$ on the left is so that $d^n$ is scaled per degree of freedom, so that a single tolerance can be shared over sequences of different lengths.

We will also consider sequences of length $N$ that are either periodic, or nearly periodic, with period less than $N$. For such sequences we can expect multiplicity of minimisers that are the same shape after a shift in bp index. To handle that case, we define a ``shifted Mahalanobis'' $d^{sn}$ as follows:
 \[ d^{sn}(\Omega_1, \Omega_2) \equiv \min_{k \in \{0,1,\cdots,N-1\}} d^n(\Omega_1,\Omega_{2,k}), \]
where $\Omega_{2,k}$ is the coordinate-vector $\Omega_2$ shifted periodically by $k$ bp, i.e., the subvector $y_i$ of $\Omega_{2,k}$ is the subvector $y_{(i+k)(mod N)}$ of $\Omega_2$, and similarly for subvectors $x_i$,
$z_i^+$, and $z_i^-$.  Intuitively, for a given $\Omega_1 \in \mathbb{R}^{24N}$, if $\Omega_2 \in \mathbb{R}^{24N}$ would be very close to $\Omega_1$ if we ``started reading'' at position $24K+1$ rather than position 1, the value of $d^{sn}(\Omega_1,\Omega_2)$ would be small even though $d^n(\Omega_1,\Omega_2)$ is relatively large.

\subsection{Methodology and Numerics}\label{sec: method}


Given a sequence $\calS$ with $N$ bp, we use the following procedure 
to compute all the distinct minicircle shapes and associated energies of the sequence. 

Following standard notation, we let $Lk_0 = N/10.5$ and use the symbol
$\lfloor Lk_0 \rceil$ to denote the integer closest to $Lk_0$. 
We consider 100 different toroidal initial guesses (cf. Sec.\ \ref{sec: minicircle_initial_guess}) 
by taking all combinations of:

\medskip

\noindent $\bullet$ one of the 5 values of $Lk$ from $\{\Lko-2, \Lko-1, \Lko, \Lko+1, \Lko+2 \}$; and

\noindent $\bullet$ one of the $20$ register angles $\gamma = 2 i \pi/20$ ($0 \le i \le 19$).

\medskip

\noindent For each of these 100 initial guesses, we run MATLAB's unconstrained optimization algorithm 
{\tt fminunc} seeking a local minimum of the energy $E$ from Eq.\ \eqref{eq: cgDNAmin zvec energy}.
We use the penalty weight $p_w = 100$ corresponding to the unit-quaternion constraint; with this value,
we find that any converged solution has each $\| q_i \|$ within $10^{-3}$ of 1.  
Appendix \ref{app:opt_details} contains the input parameters we feed to {\tt fminunc} relating to its 
algorithm and stopping condition.  

Given a configuration output by {\tt fminunc} as a possible energy minimiser, 
we accept it as a minimiser if it passes all of the following criteria:

\medskip

\noindent $\bullet$ {\tt fminunc} returns the output message ``Local minimum found''.

\noindent $\bullet$ The gradient vector of the energy (cf.\ Sec.\ \ref{gradient_subsection})
is less than $10^{-6}$ in absolute value in every slot.

\noindent $\bullet$ The Hessian matrix of the energy (cf.\ Sec.\ \ref{hessian_subsection})
has all positive eigenvalues.


\medskip

\noindent If a minimiser passes these tests, its link $Lk$ is computed as described in Sec.\ \ref{subsec:link}.

After all 100 runs are complete, 
among all accepted minimisers of a given $Lk$, we calculate the Mahalanobis distance
per degree of freedom ($d^n$) between each pair of minimisers, and if $d^n > 0.01$, we declare the
two minimisers to be distinct.

\section{Results}\label{sec: results}
We applied our method to 120K random DNA of lengths 92-106 bp.  Before reporting in Sec.\ \ref{subsec:random} summary data from this ensemble of runs, we present a typical case in Sec.\ \ref{subsec:typical} and some exceptional cases in Sec.\ \ref{subsec:exceptional}.  We also make comparisons to experimental results in Sec.\ \ref{subsec:experimental}.
\subsection{Typical case} \label{subsec:typical}
\begin{figure}
\centering
\includegraphics[width=6.0in]{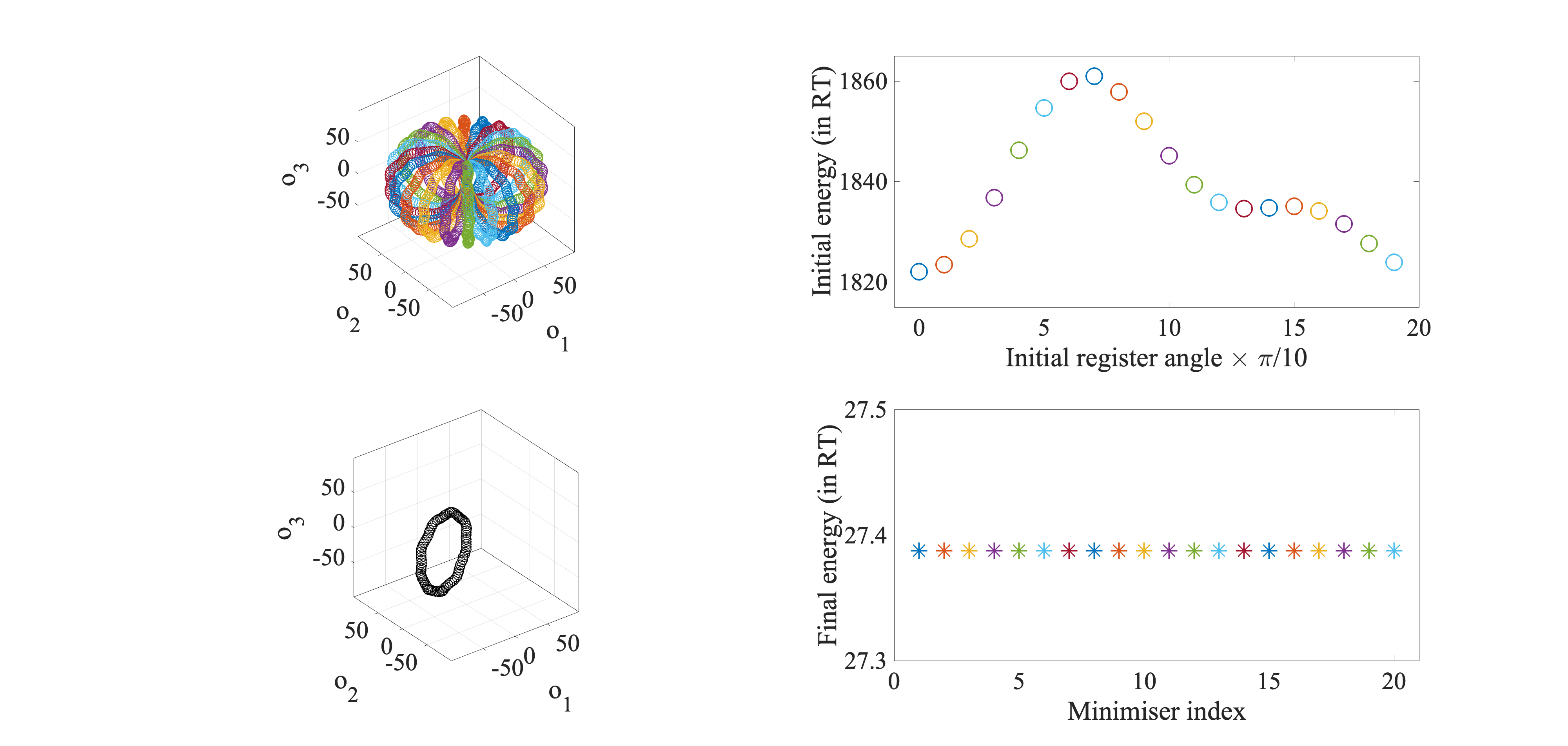}
\caption{Visualization of computation of minicircle equilibria for a random 94 bp long sequence from twenty different initial guesses all with initial $Lk$ 9 and with register angles sampled uniformly from $0$ to $2\pi$.   Top left rosette: bp origins (in $\AA$) of the initial minicircle configurations (i.e., guesses input to $cgNA+min$). Top right: energies of initial guesses. Bottom right: final minicircle energies (output by $cgNA+min$) after the twenty minimisation runs (and note different enrgy scales on the two ordinates).  All runs converge to the same energy value and same minicircle configuration, whose bp origins are shown in the bottom left panel.
}\label{fig: ex1f1}
\end{figure}

We show in Fig.\ \ref{fig: ex1f1} some key results for a 94 bp DNA whose behavior matches qualitatively the majority of cases in the ensemble of 120K random sequences we studied.
As described in Sec.\ \ref{sec: method}, we ran our minimisation algorithm for 100 different initial helicoidal guesses: 20 each (with register angles $\gamma = 2 i \pi/20$ for $0 \le i \le 19$) for 5 different values of $Lk$. In Fig.\ \ref{fig: ex1f1}, we focus on the 20 runs whose initial guess have $Lk = \Lko = 9$, with the top two images showing the shapes and energies of the initial guesses.  The bottom two images show that the 20 different runs appear to all find the same minimiser.

Considering all 100 runs, the typical outcome involves finding two minimisers, one for $Lk = \Lko$ and the other for $Lk = \Lko - 1$ or $\Lko+1$.  [As we will see in Sec.\ \ref{subsec:random}, 
the majority of random sequences of lengths 92-106 bp have exactly two minimisers, but specifically for 94 bp, about 28\% of molecules have this property.] For the molecule considered in this section, the two minimisers are the $Lk$ 9 configuration with energy $\approx 27.38 RT$ shown in Fig.\ \ref{fig: ex1f1} and an $Lk$ 8 configuration with energy $\approx 63.65 RT$ that is found from runs with initial $Lk$ different from the initial-$Lk$ 9 runs shown in Fig.\ \ref{fig: ex1f1}. For this molecule, the 60 runs whose initial guesses had $Lk \notin \{ 8, 9 \}$ either found the $Lk$ 8 or $Lk$ 9 solution---which means that strand passage occurred during the minimisation, resulting in a $\Delta Lk$ of $\pm 2$---or they failed to converge according to our acceptance criteria (cf.\ Appendix \ref{app:opt_details}).

Having two minimisers with adjacent $Lk$ is more than just ``typical'': it occurs in every single case we studied. Since strand-passage during a minimisation run will change $Lk$ by $\pm 2$, the odd-$Lk$ and even-$Lk$ runs must generate distinct minimisers, and it stands to reason that the lowest-energy odd-$Lk$ and even-$Lk$ minimisers would be of adjacent $Lk$. In other words, any exceptional cases only add minimisers to the consistent baseline of two minimisers with adjacent $Lk$.
\begin{figure}
\centering
\includegraphics[width=7.0in]{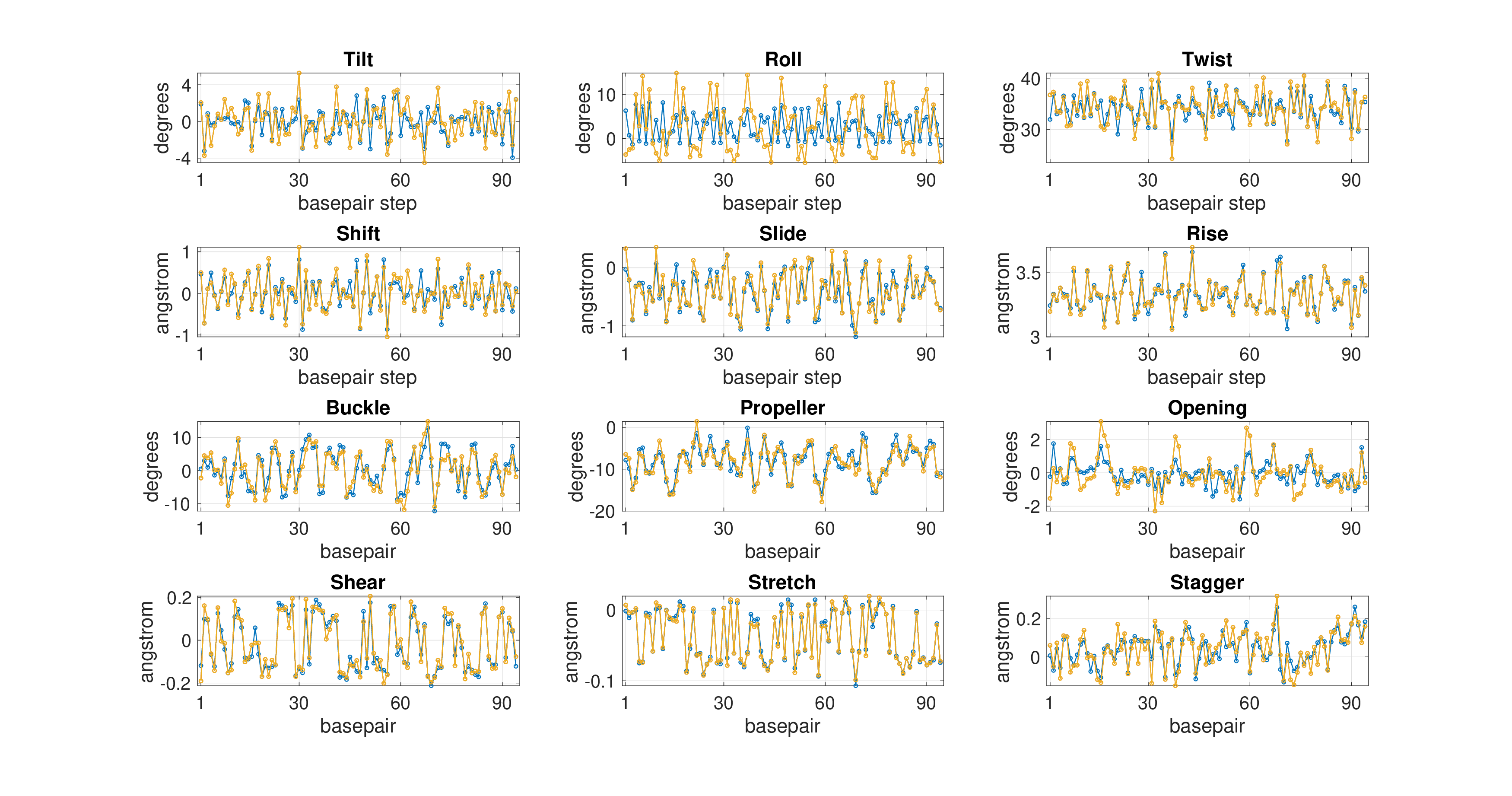} 
\includegraphics[width=7.0in]{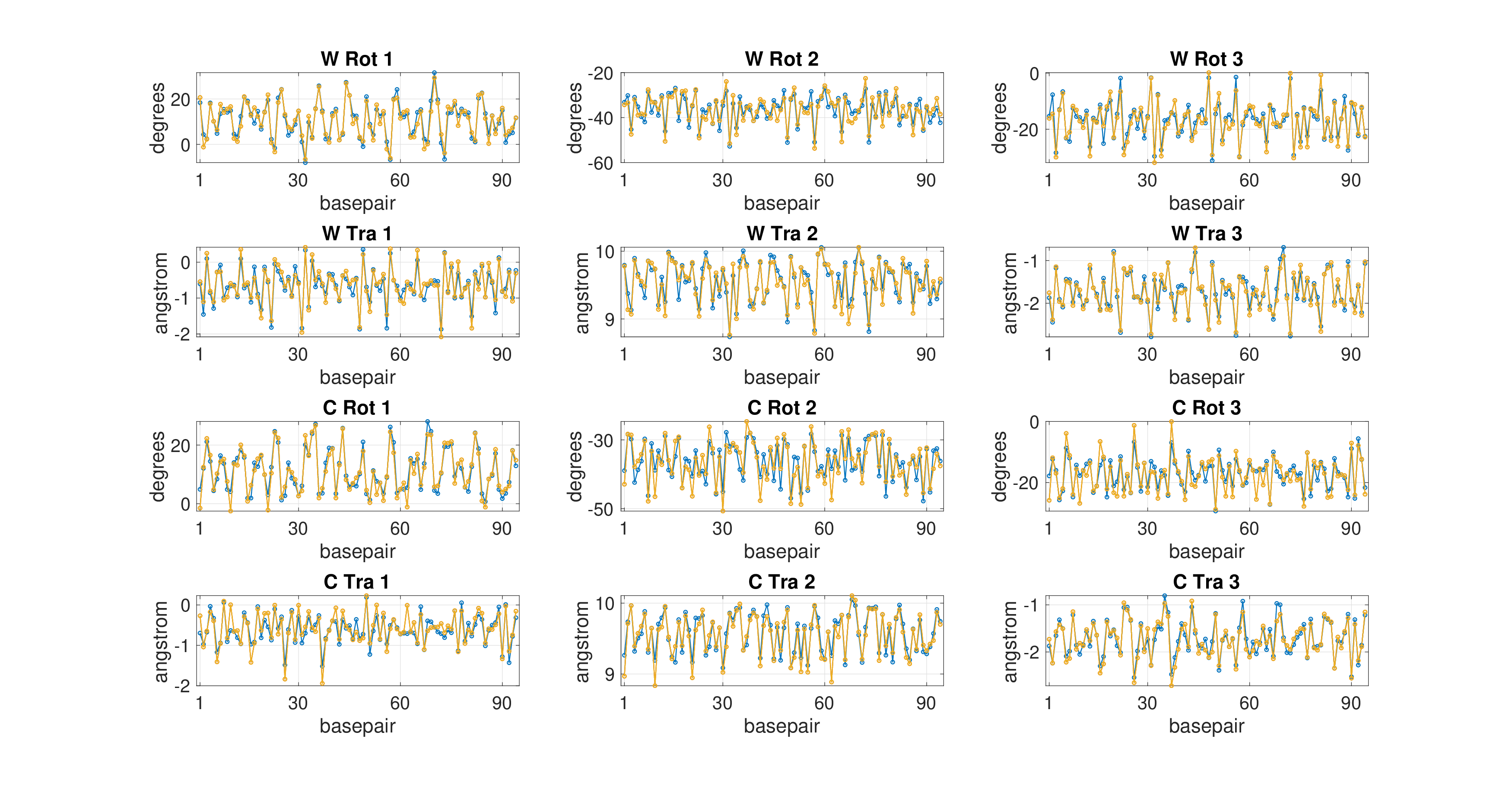}
\caption{For the 94 bp sequence considered in Fig.\ \ref{fig: ex1f1}, $cgNA+$ coordinates of the ground state of the linear fragment (blue) and global energy-minimising minicircle (i.e., the $Lk$ 9 minimum energy configuration)  (yellow).
Top panel: inter and intra coordinates; bottom panel: phosphate coordinates. For many coordinates, the linear-versus-cyclised difference is smaller than the sequence variation of the coordinate along the oligomer.  Significant differences do occur in Tilt, Roll, and Twist.
}\label{fig: ex1f3}
\end{figure}

We delve more deeply in Fig.\ \ref{fig: ex1f3} into the $Lk$ 9 energy-minimising configuration for the typical case shown in Fig.\ \ref{fig: ex1f1}. The 24 images in Fig.\ \ref{fig: ex1f3} plot the internal coordinates along the length of the DNA, for both the ground state (``linear'' energy minimiser) and the cyclised minimiser.  The top row of Fig.\ \ref{fig: ex1f3} shows linear-versus-cyclised differences (of varying magnitudes) in tilt, roll, and twist.  Since substantial bending is needed to turn a short piece of DNA into a loop, the cyclized tilt and roll are larger than in the ground state; we can see in the figure that changes in roll are more prominent than changes in tilt.  Since cyclisation also involves twist closure, the change in twist from linear to cyclised is also sensible, though that twist change appears to be relatively small, presumably because cyclisation of a 94 bp molecule does not require a lot of excess twist since 94 is roughly an integer multiple of the typical twist-per-bp of DNA. Strikingly, the other 21 images in Fig.\ \ref{fig: ex1f3} show only small differences between the linear and cyclised states. The bottom six rows suggest that the internal structures of a basepair, base, or base-phosphate connection are relatively independent of the bending and twisting needed to achieve cyclisation. The second row indicates that even the basepair-to-basepair translation coordinates are not majorly impacted by the need to adjust the basepair-to-basepair rotation coordinates (tilt, roll, twist) to achieve cyclisation.
\subsection{Exceptional cases} \label{subsec:exceptional}
\subsubsection{Multiplicity of distinct minimisers of the same link}
\label{subsubsec:except_same_link}
\begin{figure}[H]
\centering
\includegraphics[width=7.0in]{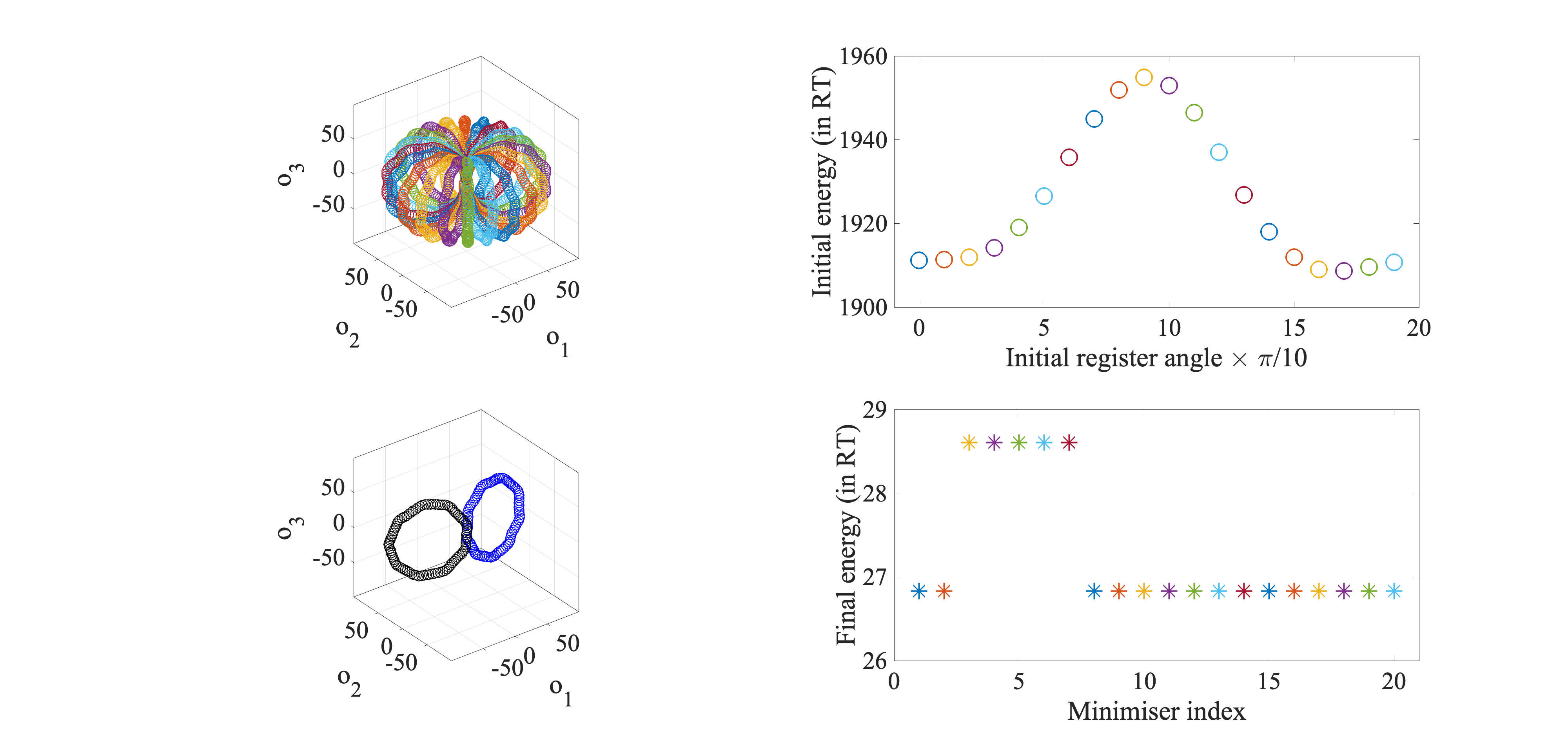} 
\caption{Visualization of computation of minicircle equilibria for a 94 bp sequence that yields two distinct $Lk$ 9 energy-minimisers.  Top left rosette: bp origins (in $\AA$) of the 20 different initial guesses with $Lk$ 9. Top right: energies of initial guesses. Bottom right: final minicircle energies, showing two different energy values.  Bottom left: the two $Lk$ 9 minicircle configurations.}\label{fig: ex2f1}
\end{figure}
In Fig.\ \ref{fig: ex2f1}, we show results for a 94 bp DNA with two distinct $Lk$ 9 local minimisers of energy.  The top panels show the shapes and energies of the 20 initial guesses with $Lk$ 9; these appear similar to the typical case (top half of Fig.\ \ref{fig: ex1f1}).  However, the lower panels show that the 20 runs yield two distinct configurations with different energies; by eye, the two configurations are each roughly circular but with register angles approximately $\pi$ apart.

\begin{figure}
\centering
\includegraphics[width=6.0in]{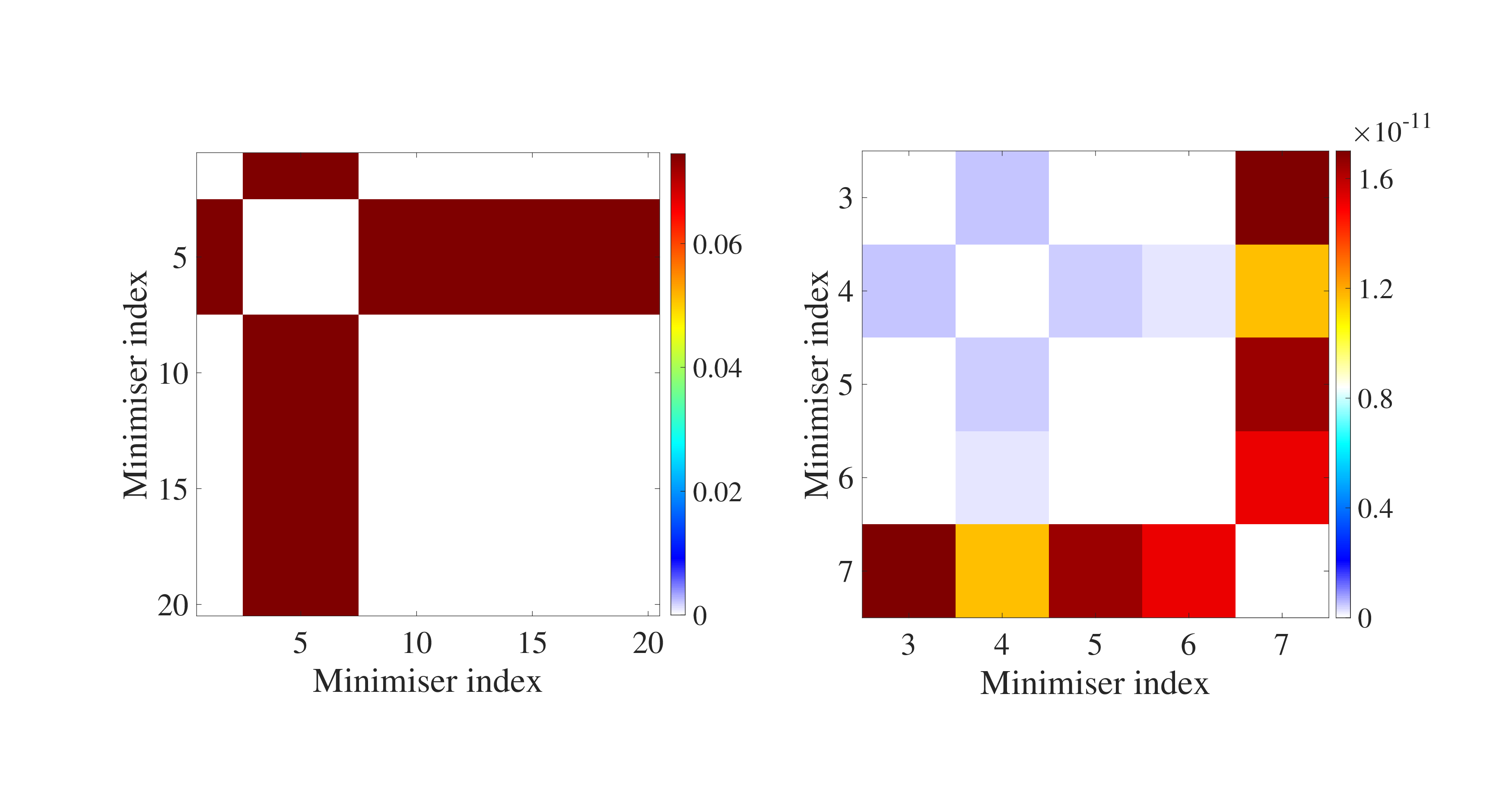} 
\caption{Mahalanobis distance-squared per degree of freedom $d^n$ between pairs of minimisers from Fig.\ \ref{fig: ex2f1}. Left: $d^n$ among all pairs from the twenty computed minimisers; the two distinct clusters at the scale $0.06$ indicate that there are two {\it distinct} minimisers arising from the twenty initial guesses.  
Right: $d^n$ among pairs from only minimisers with index 3-7; the fact that $d^n \sim 10^{-11}$ indicates that all five computed configurations are numerical approximations of a single minimiser. 
}
\label{fig: ex2f3}
\end{figure}

The distinct energies seen in the lower right of Fig.\ \ref{fig: ex2f1} indicate clearly the distinctness of the two minimisers found.  To probe the configurations themselves, we use the Mahalanobis distance $d^n$ as defined in Sec.\ \ref{sec:distinct}.  In Fig.\ \ref{fig: ex2f3}, we visualize $d^n(\Omega_i, \Omega_j)$ for each pair $(\Omega_i,\Omega_j)$ of minimisers. In the left image, the white square in the region $3 \le i,j \le 7$ shows that $d^n$ is very small when comparing $\Omega_i$ and $\Omega_j$ for $i,j$ both in $\{ 3,4,5,6,7\}$, indicating that these five minimisers are all the same configuration.  The right image in Fig.\ \ref{fig: ex2f3} zooms in on this $3 \le i,j \le 7$ square to show that $d^n \sim 10^{-11}$ in all cases; given the very small scale, we do not think the structure within the right image is significant. Similarly, the other four white regions in the left image in Fig.\ \ref{fig: ex2f1} show that $d^n$ is very small when comparing $\Omega_i$ and $\Omega_j$ for $i,j$ both in $\{ 0,1,2,8,9,10,\cdots,19 \}$, indicating that these fifteen minimisers are all the same ($d^n \sim 10^{-11}$ for these cases; data not shown). On the other hand, the four maroon rectangles show that $d^n$ is relatively large when comparing $\Omega_i$ for  $i \in \{ 3,4,5,6,7\}$ and $\Omega_j$ for $j \in \{ 0,1,2,8,9,10,\cdots,19 \}$, or vice versa, indicating that the two minimisers are distinct (consistent with their energies being significantly different).  

\begin{figure} 
\centering
\includegraphics[width=7.0in]{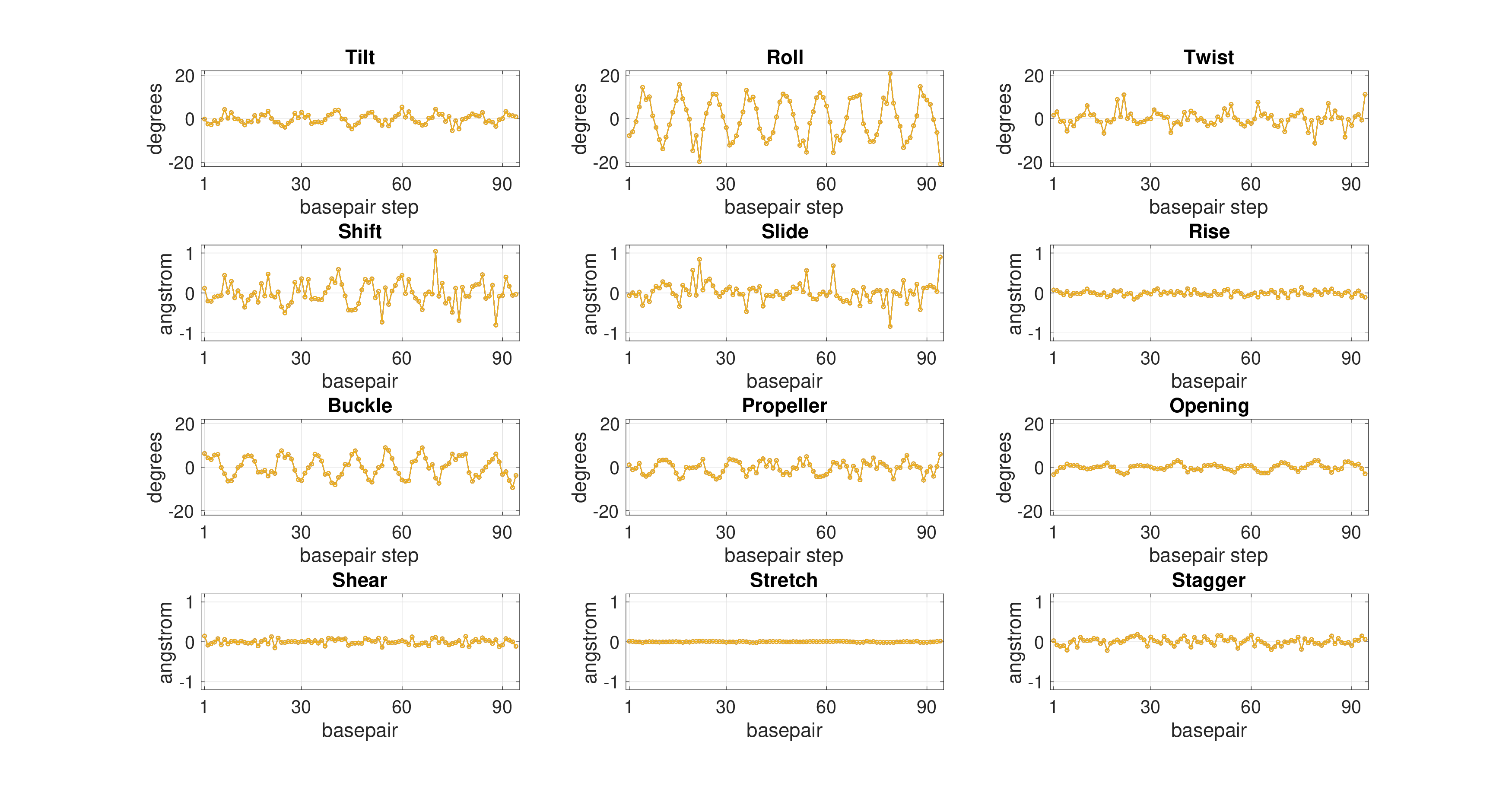} 
\includegraphics[width=7.0in]{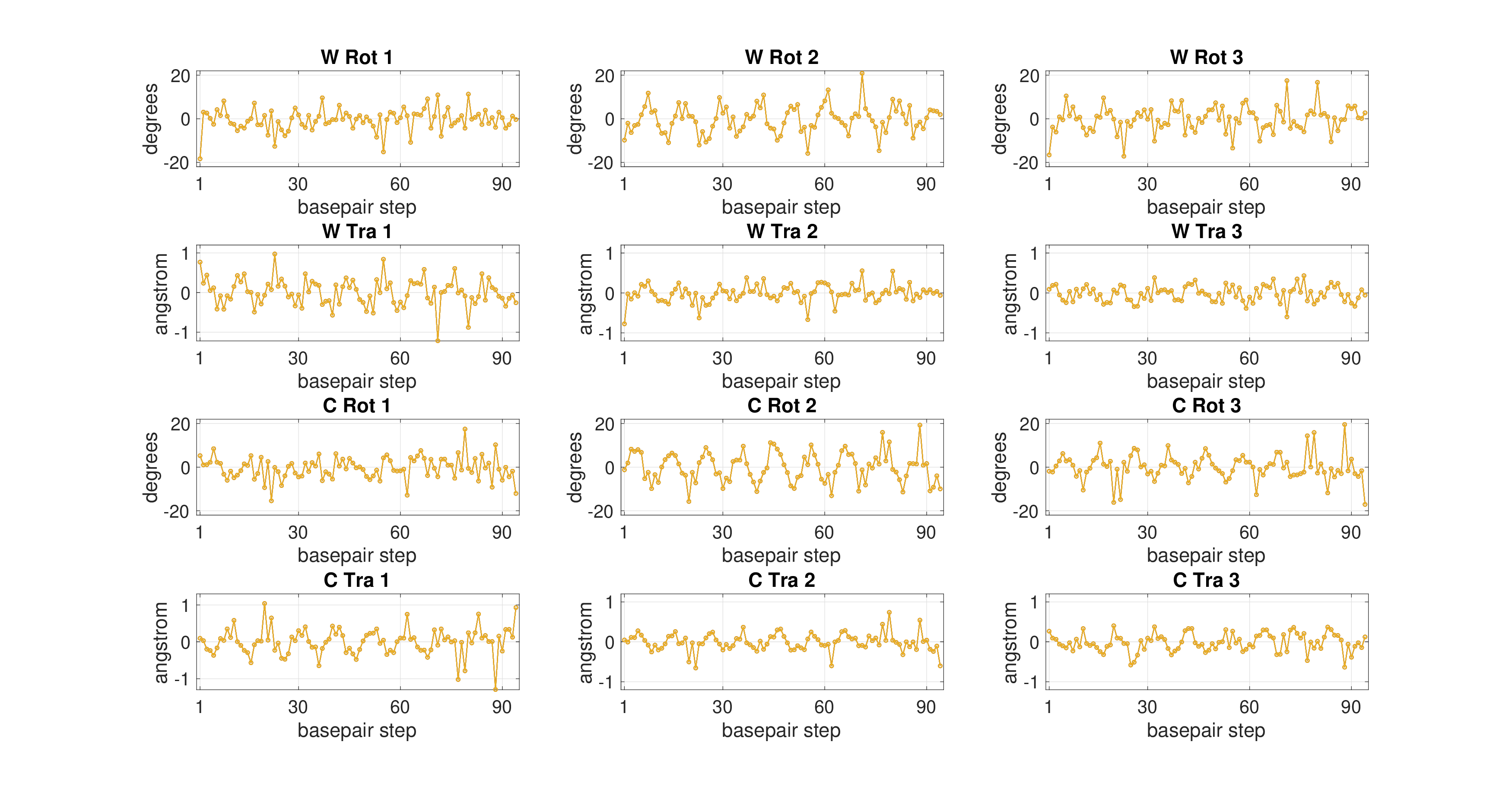} 
\caption{Differences in $cgNA+$ coordinates between the two distinct $Lk$ 9 minicircles from Fig.\ \ref{fig: ex2f1}. Top: differences in inter and intra coordinates.  Bottom: differences in phosphate coordinates. 
}\label{fig: ex2f4}
\end{figure}

We show in Fig.\ \ref{fig: ex2f4} how the internal coordinates of the two $Lk$ 9 minicircles differ; unlike in Fig.\ \ref{fig: ex1f3}, where we plotted {\it values} of internal coordinates (for the ground state linear fragment and the lone minicircle), in Fig.\ \ref{fig: ex2f4} we plot {\it differences} of internal coordinates (between the two minicircles of $Lk$ 9).  Many coordinates show differences on the order of 10$^\circ$ (for angles) or 0.5 Angstroms (for distances), often with a periodicity that suggests a similar overall shape with different register angle.  
\subsubsection{Multiplicity of minimisers of distinct links}
\begin{figure}
\centering
\includegraphics[width=3.6in]{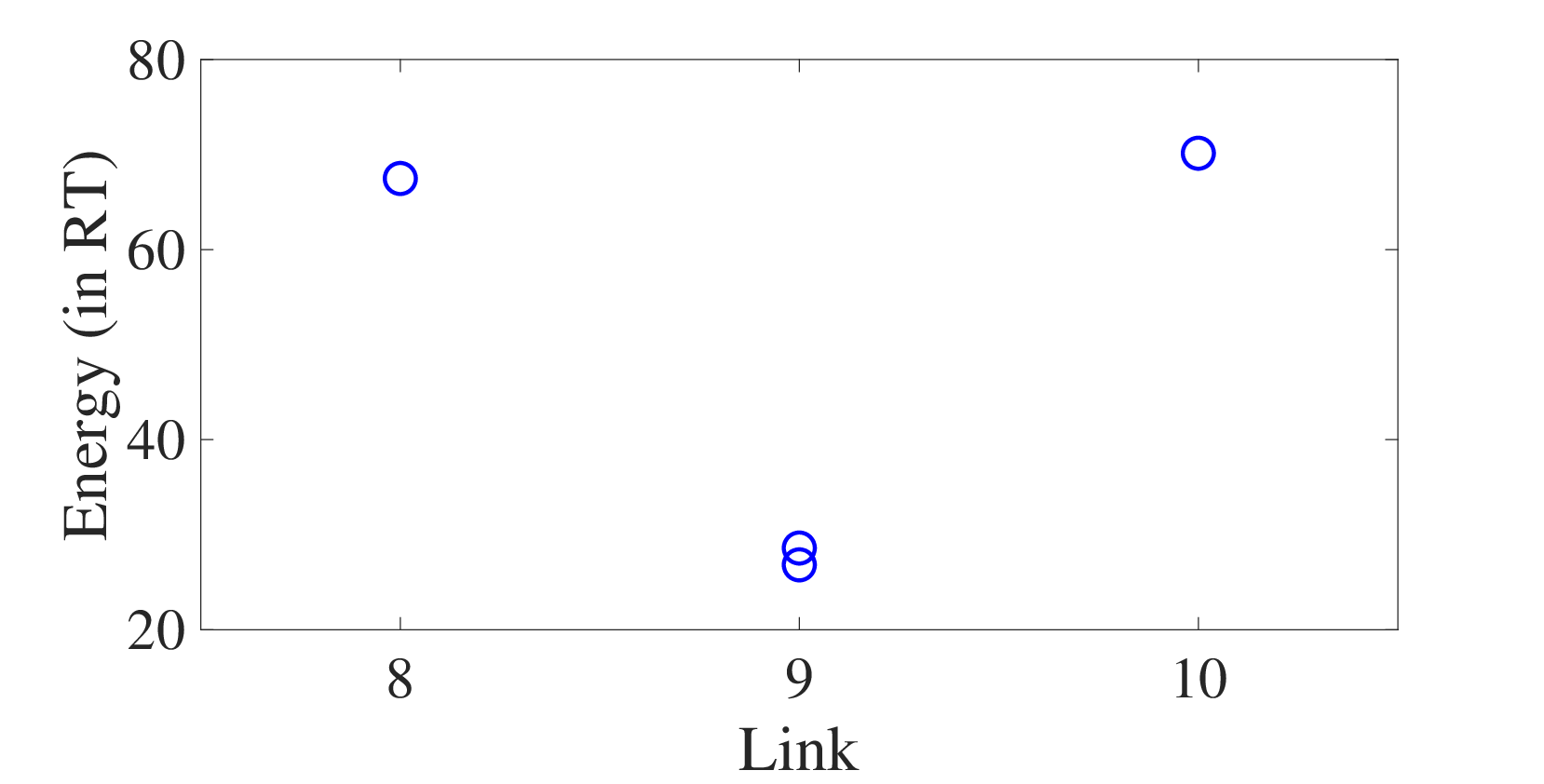} 
\caption{Energies and $Lk$ for distinct minimisers identified for the 94 bp sequence studied in Sec.\ \ref{subsubsec:except_same_link}.  The two dots at $Lk=9$ match those shown in Fig.\ \ref{fig: ex2f1} for the 20 minimisation runs with initial $Lk$ of 9.  The figure also shows $Lk = 8$ and 10 minicircles found by some of the $Lk \ne 9$ minimisation runs, showing that this sequence has minicircles with more than two distinct $Lk$.  
}
\label{fig: ex2f2}
\end{figure}

In addition to the possibility of finding multiple minimisers at a single value of $Lk$, a molecule can  have minicircles at more than two different values of $Lk$.  The molecule studied in Sec.\ \ref{subsubsec:except_same_link} actually exhibits {\it both} phenomena.  (As we will see in Sec.\ \ref{subsec:random}, the phenomena of multiple-minicircles-at-one-$Lk$ and more-than-two-$Lk$ often occur separately; we chose a case exhibiting both phenomena purely for conciseness.)  In Fig.\ \ref{fig: ex2f2}, we show the energies and $Lk$ for four different minimisers---one with $Lk=8$, two with $Lk=9$, and one with $Lk=10$---identified as distinct from the 100 different minimisation runs for this molecule.  
\subsubsection{Six distinct poly (dinucleotide) sequences} 
\label{sec: polys}
Since random-sequence DNA shows significant qualitative differences in results, one might expect DNA with very uniform sequences to exhibit exceptional behavior.  Indeed, in this section, we show quite distinctive results for sequences that repeat single dinucleotide. 

\begin{figure}
\centering
\includegraphics[width=4.80in]{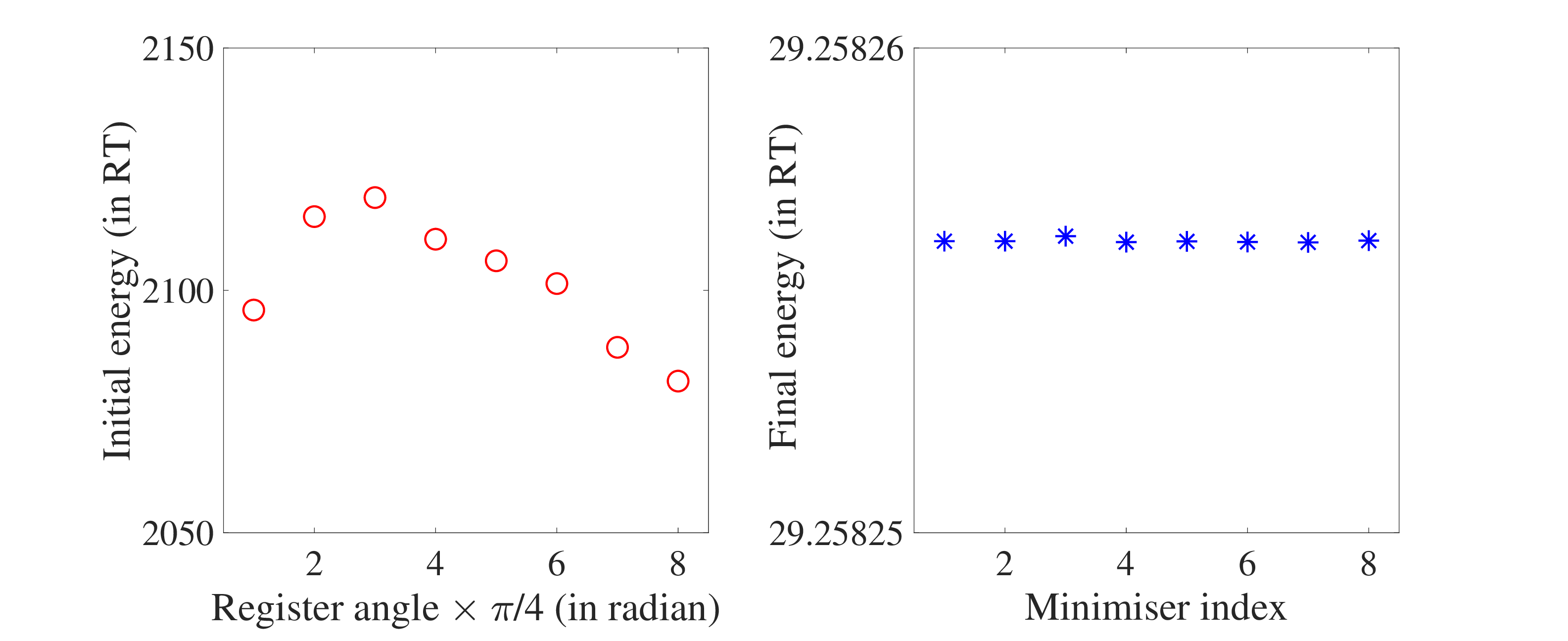} 
\includegraphics[width=2.00in]{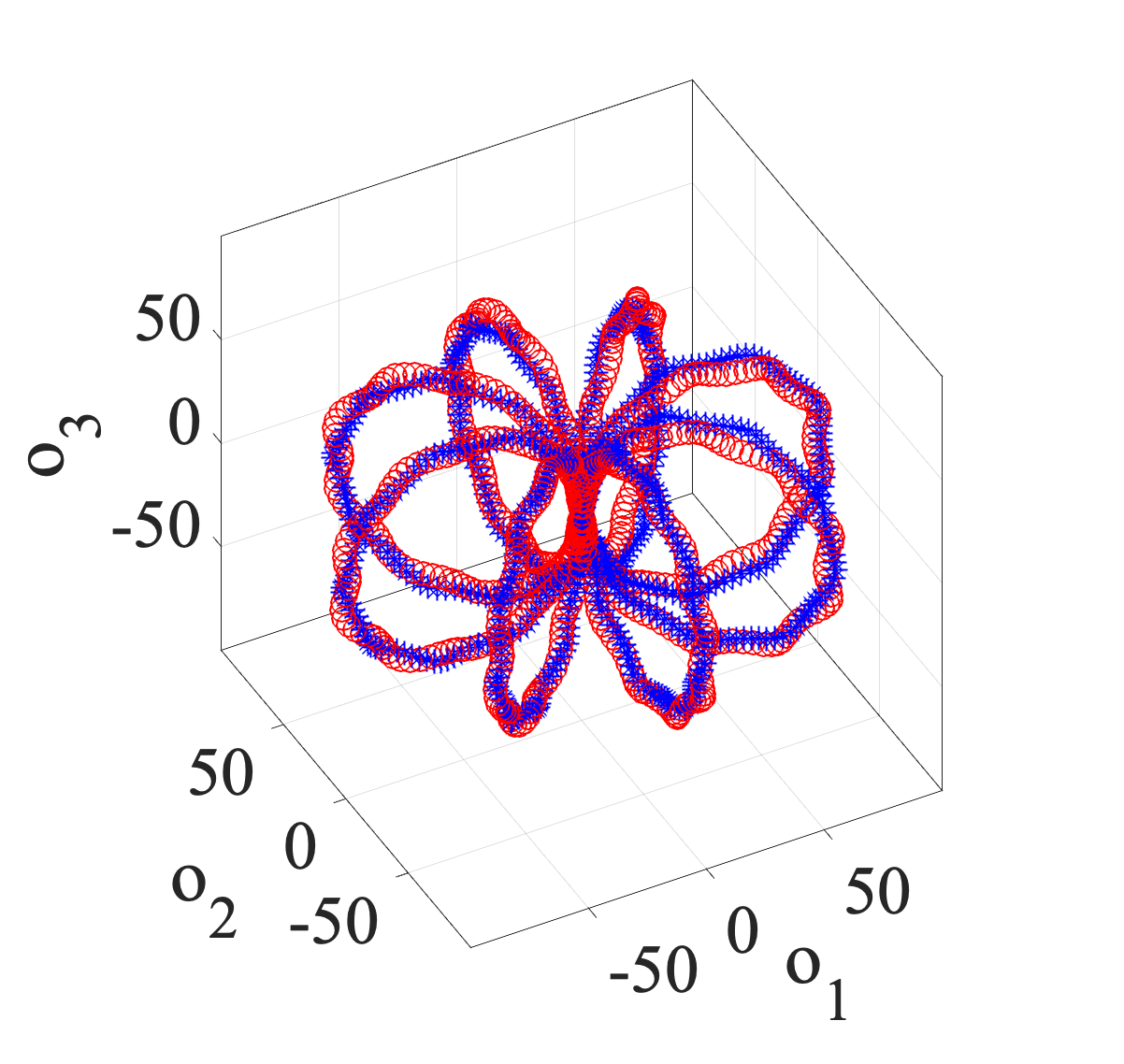}
\caption{Computing minicircle equilibria for $(GG)_{47}$ starting from eight different $Lk$ 9 initial guesses with different register angles.  Left: energies of initial guesses. Middle: final minicircle energies. Right: bp origins (in $\AA$) of initial (red) and minimum-energy (blue) configurations.  Among the eight cases, the final energies are equal (to 5 decimal places) but the configurations are distinct.
}\label{fig: poly_gg_multi}
\end{figure}

\begin{figure}
\centering
\includegraphics[width=7.2in]{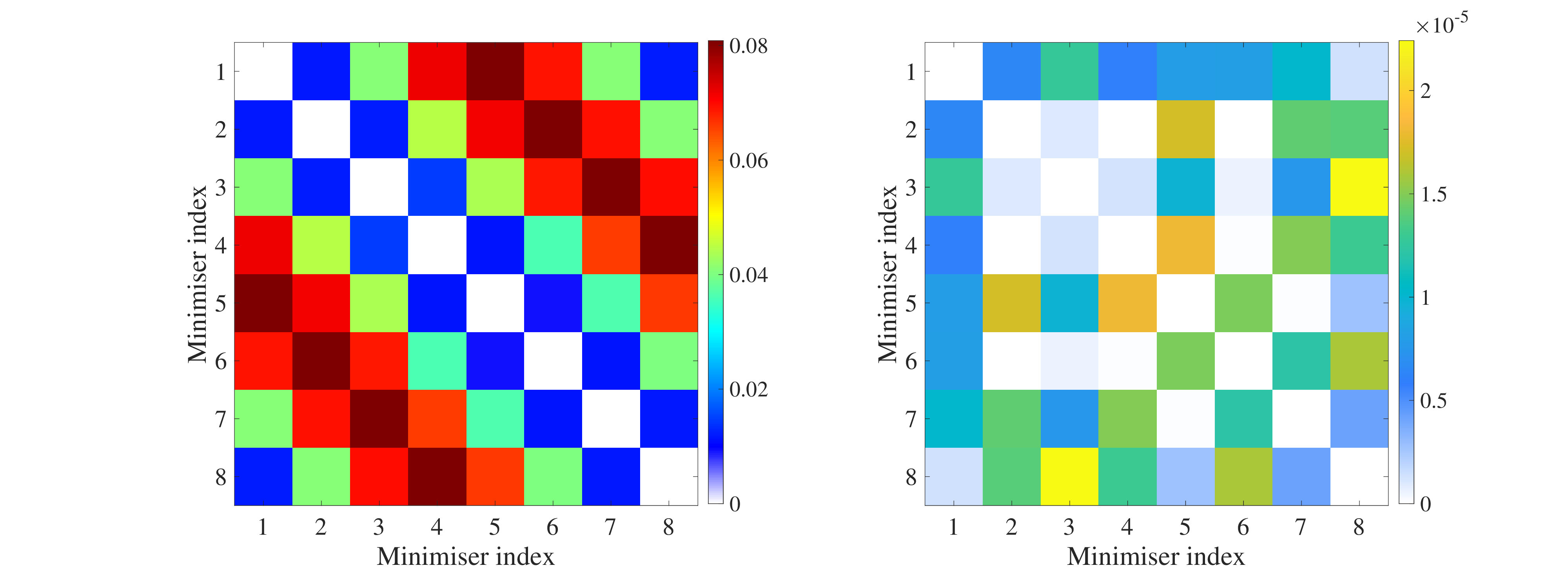} 
\caption{Left: standard Mahalanobis distance-squared per dof $d^n$ between all eight $Lk$ 9 minimisers from Fig.\ \ref{fig: poly_gg_multi} for $(GG)_{47}$, revealing significant differences of shape. Right: a modified Mahalanobis measure $d^{sn}$ (which minimises over a shift in bp index), showing that all eight minimisers are very close to the same shape under the symmetry of translation, or relabelling bp index.}
\label{fig: ex4f2}
\end{figure}

We begin with the $94$-bp DNA $(GG)_{47}$ that repeats $GG$ $47$ times.  Fig.\ \ref{fig: poly_gg_multi} shows results for initial guesses with $Lk$ 9; in contrast to earlier results, we use 8 rather than 20 different registers, in order to convey the results more clearly.  The left image in Fig.\ \ref{fig: poly_gg_multi} shows the energies for the 8 initial guesses, and the center and right images show the energies and configurations for the minimisers found by our procedure.  The energies are all quite close to each other (note the scale of the center image), but the configurations are all distinct (note the contrast with Fig.\ \ref{fig: ex1f1}, where the $20$ minimisers had the same energy {\it and} configuration).  
The distinctness of the minimisers that is apparent in the right image of Fig.\ \ref{fig: poly_gg_multi} is confirmed by visualizing the values of $d^n$ in the left image of Fig.\ \ref{fig: ex4f2}: we can see from the non-white squares that any pair of solutions $\Omega_i$ and $\Omega_j$ with $i \ne j$ has $d^n > 0.01$. This visualization of $d^n$ has a diagonal structure not seen in our previous $d^n$ visualization in Fig.\ \ref{fig: ex2f3}, suggesting that an additional symmetry for this uniform sequence, namely translation along the DNA, plays a key role.   

In the right image of Fig.\ \ref{fig: ex4f2}, we can see that $d^{sn}$ is uniformly small ($< 3 \times 10^{-5}$) for all pairs of minimisers found for $(GG)_{47}$, suggesting they would all be quite similar to each other if we allowed for the translation-along-molecule operation.  For a continuous uniform elastic rod, it is known \cite{ManningMaddocks1999} that any minicircle is a member of an infinite family of equal-energy configurations related to each other by the translation-along-rod symmetry.  The discrete setting here presumably breaks this perfect symmetry, but Figs.\ \ref{fig: poly_gg_multi} and \ref{fig: ex4f2} show that the qualitative feature is largely preserved: we have a large number of minimisers with nearly equal energy that would be very close to each other if we applied the operation of translating along the molecule.  This large multiplicity of minimisers could presumably have a significant impact on the molecule's $J$-factor, but that issue is beyond the scope of our study.

We show in Table \ref{table: polys} the energies and $Lk$ for minimisers for all six independent poly $(XY)$ sequences.  All of them exhibit the multiplicty-of-minimisers issue explored above for $(GG)_{47}$, but we have suppressed that issue in Table \ref{table: polys}, reporting minimisers as distinct only if their $d^{sn}$ values are relatively large.  This leads to a single minimiser per value of $Lk$, though we sometimes get two $Lk$ values (for $TG$, $AG$, or $CG$) and sometimes get three (for $AA$, $GG$, and $TA$).  
\begin{table}
\caption{Values of $Lk$ and energy for $cgNA+$ minicircles for poly(XY) sequences of length 94 bp. 
}
\begin{center}
\begin{tabular}{ |p{2.0cm}|p{6.4cm}|  }
\hline
Sequence  & $Lk$ (energy)     \\ \hline
$\tt (TG)_{47}$  & 8 (66.078), 9 (27.876)             \\ \hline 
$\tt (AA)_{47}$  & 8 (82.804), 9 (31.760), 10 (83.147)         \\ \hline 
$\tt (GG)_{47}$  & 8 (44.047), 9 (29.258), 10 (82.194)         \\ \hline 
$\tt (TA)_{47}$  & 8 (66.234), 9 (24.173), 10 (68.446)         \\ \hline 
$\tt (AG)_{47}$  & \quad\quad\quad\quad\quad 9 (31.296), 10 (65.420)             \\ \hline
$\tt (CG)_{47}$  & \quad\quad\quad\quad\quad 9 (28.825), 10 (68.022)             \\ \hline
\end{tabular}
\label{table: polys}
\end{center}
\end{table}
\subsection{Experimentally studied sequences} \label{subsec:experimental}
Direct experimental measurement of minicircle energies is not currently possible, but it is relatively common to measure a ``cyclization $J$-factor'' \cite{Jacobson_Stockmayer_1950,FlorySuterMutter1976} that is thought to be at least partially related to such energies.  There is, in fact, not a unique notion of $J$-factor, since somewhat different experimental protocols are used to measure it, but one standard approach follows \cite{ShoreBaldwin1983} in seeking to measure $J = K_c/K_d$ (for $K_c$ the equilibrium coefficient of the cyclization reaction and $K_d$ the equilibrium coefficient of the dimerization reaction).  Subject to some set of approximations, one finds that $J$ is proportional to $\exp[-(\Delta G_c-\Delta G_d)/(RT)]$ for $\Delta G_c$ the free-energy change in cyclization and $\Delta G_d$ the free-energy change in dimerization.  Since the enthalpic piece of $\Delta G_c$ is the minimum energy $E_c$ of a minicircle minus zero (presuming the linear form of the molecule is in its ground state), one might hope that $\exp[-E_c/(RT)]$ would show some correlation to the $J$-factor.

Accordingly, we compare in this section the global minima of energies from our procedure to energies extracted from experimentally measured $J$ factors, cognizant of many issues that complicate
this comparison, including: (1) the assumptions inherent in the Shore and Baldwin theory, (2) the entropic portion of free energy not accounted for in $E_c$, (3) the role of local minimisers other than the global minimum, (4) the fact that our experimental results come from a few different researchers, and (5) the fact that the $cgNA+$ model can only be as accurate a model of DNA as the MD simulations from which it is derived. To extract corresponding energies from the $J$-factors, our procedure is as follows:
\medskip

\noindent $\bullet$ For CAP sequences, we used the energies (in $RT$) from Table 3 of \cite{Manning1996} (which had been converted from experimental $J$ factors from Table 1 in \cite{Kahn_Crothers_1998}). 

\noindent $\bullet$ For TATA-box and TACA-box sequences, experimental $J$ factors (in $nM$) from Tables 1 and 2 in \cite{Davis1999} were converted to energies (in $RT$) via the formula $-\left( ln(J) - ln(10^9) \right)$.

\noindent $\bullet$ For Cloutier and Widom sequences, we first extracted $J$ values from Fig. 3(a) in \cite{Cloutier_Widom_2005} using the web interface WebPlotDigitizer \citep{Rohatgi2017}. Then these values of $J$ (in $M$ in a logarithmic scale) were converted to energies (in $RT$) via the formula $-ln(10^J)$. 

\subsubsection{Correlation with experimental data}
\begin{figure}
\centering
\includegraphics[width=3.6in]{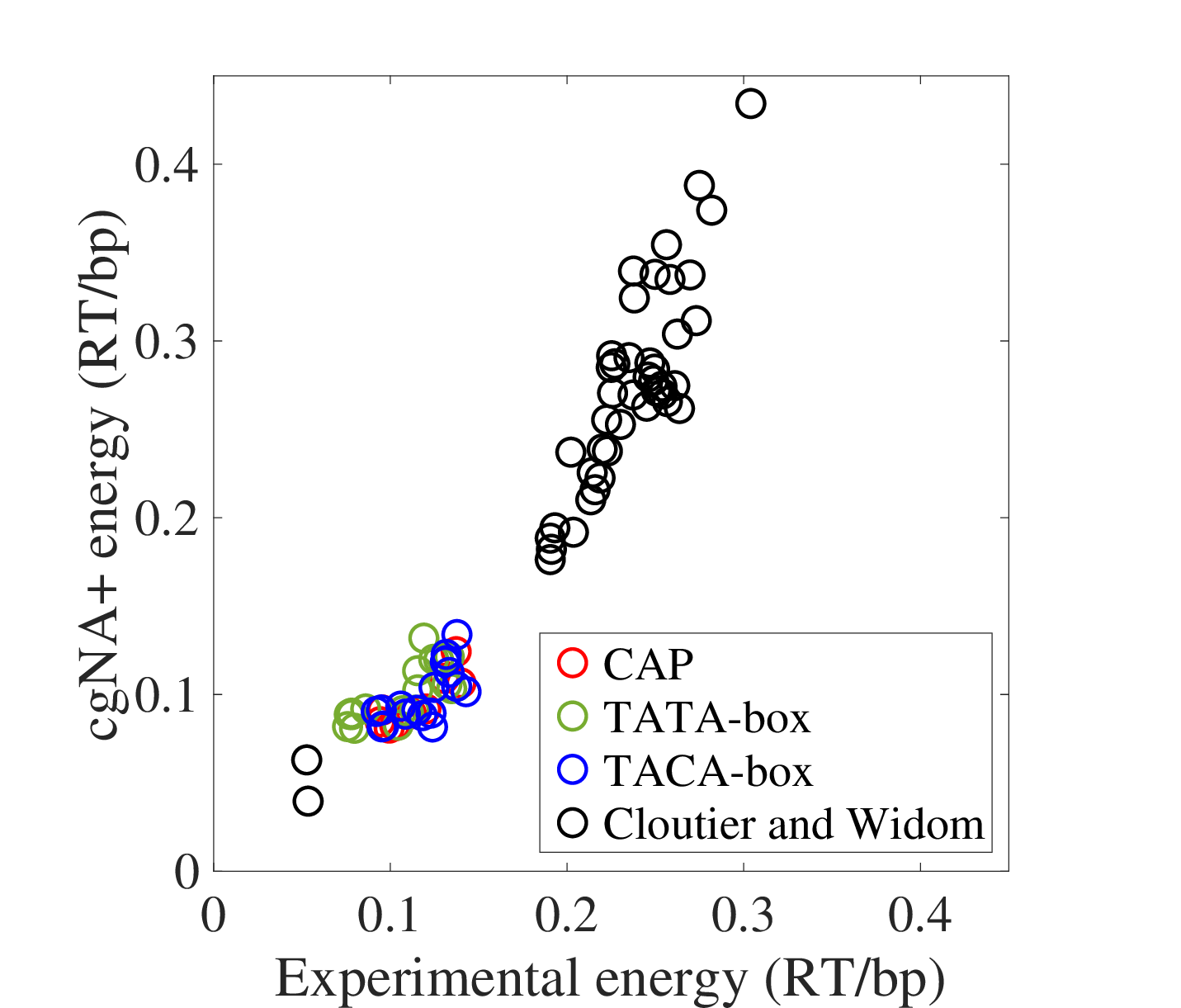}
\caption{Scatterplot of $cgNA+min$ energies (i.e., the energy of the global energy-minimising shape over all possible $Lk$) versus energies extracted from experimentally measured $J$-factors (see text for details), for several families of experimentally studied sequence fragments.}
\label{fig: exp_plot}
\end{figure}

We show in Fig.\ \ref{fig: exp_plot} a scatterplot of $cgNA+$ versus $J$-factor-derived energies for sequences studied experimentally in [\cite{Davis1999}, \cite{Kahn_Crothers_1998}, and \cite{Cloutier_Widom_2005}].  The energies on the vertical axis are the global minima coming from our computations, and those on the horizontal axis are derived from experimental $J$-factors as described above.
We note a strong positive correlation in the scatterplot, with the two lowest Cloutier-Widom energies also being the two lowest $cgNA+min$ energies, a strong correlation for the higher-energy
Cloutier-Widom data, and the intermediate CAP/TATA/TACA results lying in between.  There is undoubtedly a length effect at play (i.e., longer DNA tend to have lower energies both experimentally and in our computations), so for a more pointed exploration of sequence-dependence, as distinct from the length effect, we show in Table \ref{table: expt} the correlation coefficients within each family, finding that even these within-family correlations are quite strong (Pearson coefficients from 0.70 to 0.88).
\begin{table}[H]
\caption{Correlation coefficients for $cgNA+min$ versus experimental energies, within each of the families of DNA sequences from Fig.\ \ref{fig: exp_plot}}
\begin{center}
\begin{tabular}{ |p{8.5cm}|p{3.8cm}|p{3.6cm}|  }
\hline
Sequence group (and range of lengths in bp)  & \# of sequences   & Pearson correlation   \\ \hline
CAP (150-160) & 11   & 0.88    \\ \hline 
TATA (147-163) & 18   & 0.74    \\ \hline 
TACA (147-163)  & 18   & 0.70    \\ \hline 
CAP + TATA + TACA (147-163)  & 47   & 0.70    \\ \hline
Cloutier and Widom  sequences (89-116)   & 43   & 0.84    \\ \hline
Cloutier and Widom  sequences (322, 325)   & 2   & -    \\ \hline
All (89-325) & 92   & 0.96  \\ \hline
\end{tabular}
\label{table: expt}
\end{center}
\end{table}
\subsubsection{Sequence-length study for Cloutier and Widom molecules}
\begin{figure}
\centering
\includegraphics[width=6.0in]{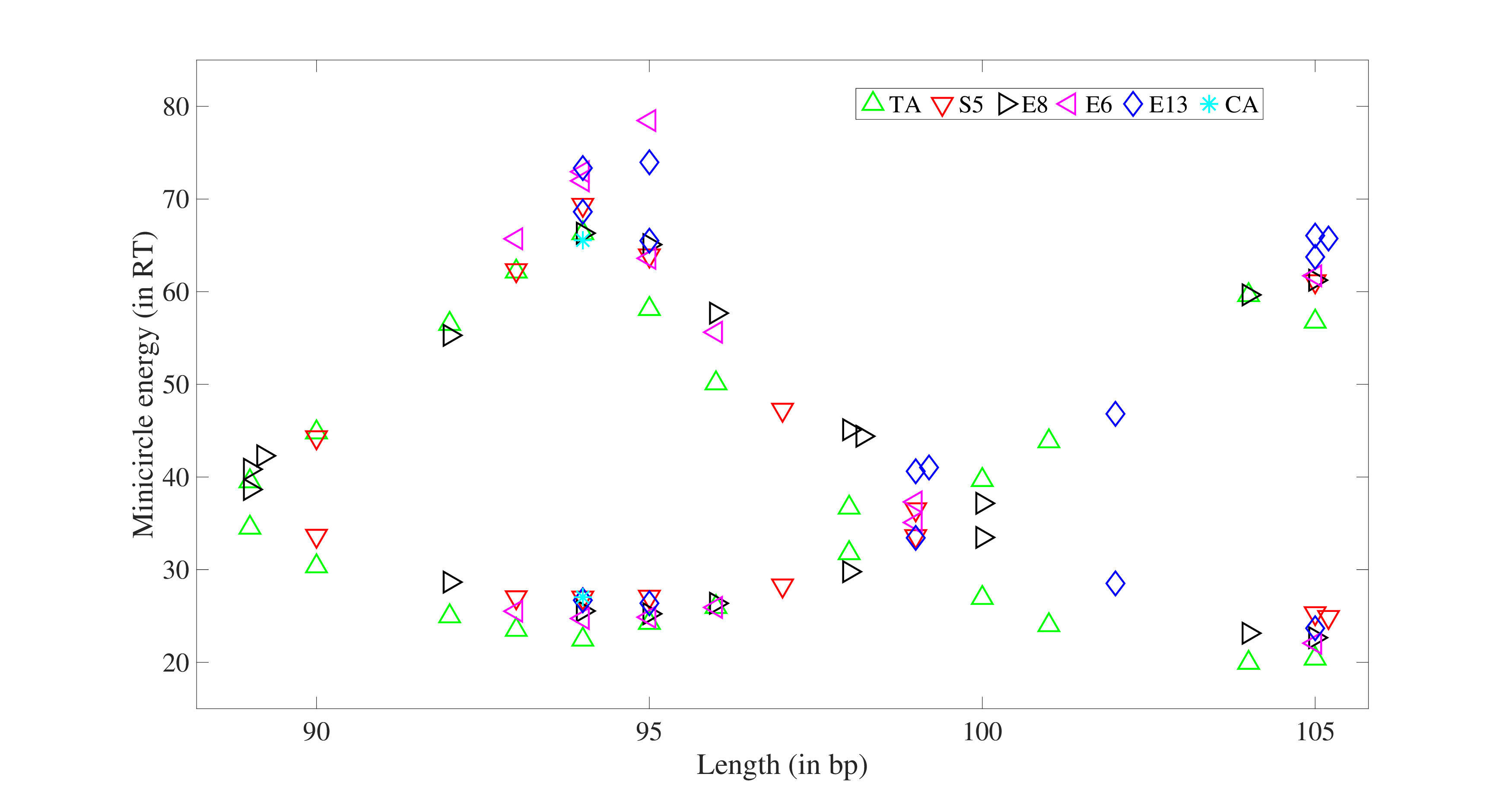}
\caption{Energies of distinct $cgNA+min$ minimisers for the experimentally studied Cloutier and Widom sequences of lengths 89 to 105 bp.  Markers of the same color and shape at a given length denote minimisers with distinct $Lk$ (e.g., the presence of three blue diamonds at length 94 bp indicates that this member of the E13 family has $cgNA+min$ minimisers with three distinct $Lk$).  When multiple minimisers of the {\it same} $Lk$ arise, extra markers are plotted with a small horizontal offset (e.g.\ the two neighboring blue diamonds at length 99 bp). 
}
\label{fig: cloutier_length_dep}
\end{figure}

Since the shorter Cloutier and Widom molecules have received much attention in the literature, we decided to use those sequences to depict how our results vary with DNA length.  We show in Fig.\ \ref{fig: cloutier_length_dep} the energies of all minimisers (of any $Lk$) found by our procedure for the Cloutier and Widom molecules in the 89-105 bp range. For a given family and length, the data in Table\ \ref{table: expt} only includes the lowest-energy marker in Fig.\ \ref{fig: cloutier_length_dep}.  By also including the higher-energy minimisers, we can see a clear pattern in the results, with the data lying close to ``branches'' that cross around lengths of 88 and 99 bp, at which point the previously higher energy minimisers become the global minima.
\subsection{Results for ensembles of randomly generated sequences} \label{subsec:random}
\subsubsection{Distribution of low-energy minimisers}\label{subsubsec: stats_of_ensemble}
\begin{figure}
\centering
\includegraphics[width=5.4in]{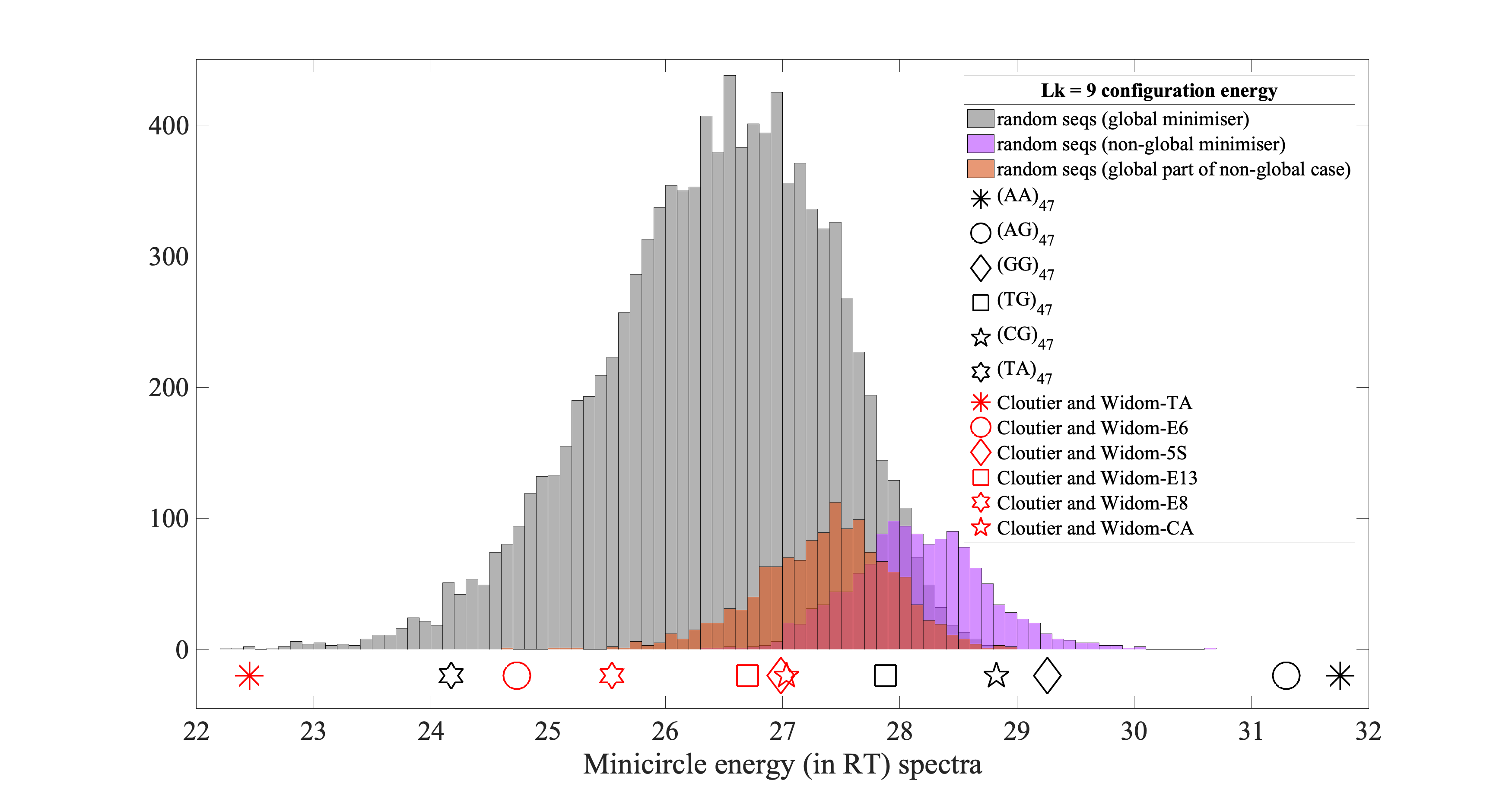}
\caption{Unnormalised histograms of $cgNA+min$ energies of $Lk$ 9 minimisers for 10K random 94 bp sequences. Lowest energies for each sequence are shown in gray.  For sequences with two or more $Lk$ 9 local minimisers, the lower energies are shown in orange and the additional, higher energies are shown in purple.  Markers below the histogram indicate $Lk$ 9 $cgNA+min$ energies for some particular 94 bp sequences: poly(XY) (black) and six Cloutier and Widom sequences (red).  Compared to the 10K random sequences, poly(AA) is a high outlier (above all 10K cases), while the Widom TA sequence is a low outlier (among the four lowest).  
}
\label{fig: low_energy}
\end{figure}
In order to focus on low-energy minicircles, we generated 10,000 sequences of length 94 bp at random (with a 25\% chance of assigning $A$, $C$, $G$, or $T$ to each position) and applied our procedure to find local minimisers of energy.  In Fig.\ \ref{fig: low_energy}, we show the resulting energies for those minimisers with $Lk = 9$; since this is the value of $Lk_0$ for $N = 94$ bp, the resulting energies are relatively low (always below 32 $RT$). The gray histogram shows the global minima in energy for the 10K molecules, indicating a fairly smooth unimodal distribution peaking at $\sim 26.5 RT$ with a slight left skew toward lower energies.  Markers below the histogram indicate energies for some non-random sequences we studied.  Among these are the six poly(XY) sequences, of which all but $CG$ have relatively high energies, with $AA$ and $AG$ being extremely high, exceeding all 10K random sequences.  We also mark energies for six sequences from Cloutier and Widom, of which two (E6 and E8) are relatively low-energy and one (TA) is very low-energy.  That TA sequence is renowned for its high cyclization rate for such a short molecule, so we were interested to find a few random 10K sequences whose energies were a bit lower than TA and thus might merit further study (their sequences can be found in Table \ref{table: low_energy_seqs}).
\begin{table}[H]
\caption{94 bp sequences with the lowest $cgNA+min$ energies (in RT), including Cloutier and Widom's TA (in red).} 
\scriptsize{
\begin{center}
\begin{tabular}{ |p{14.1cm}|p{2.2cm}|}
\hline
Sequence  & minicircle energy    \\ \hline
\tt GCATTTCCTGCCACTGTCGATGCTGCGATGCAGTACATCACCCTCCTAAAACGGTGCCAAAGTGCTACTACGCGCTCGATCCCCCGGATAACAG & 
$22.29$ \\ \hline
\tt AAGGGCCTATCTACTGTTTTAATCATCAATTAACAGCTTATTAAACTGGCGTAACTCCGTTCCTATCGCTTACCGGTTGCGGTACAGCATACCT & 
$22.37$ \\ \hline
\tt 
TACATAAAGTCCGCGTTATGCAAGGGAAATCTGCCAATAAGTTCGAGTTACCCCCTTTAAGGCCTCGAAGATGGTGTTTCAGTGAGAAAAATCT & 
$22.43$ \\ \hline
\tt \addr{GGCCGGGTCGTAGCAAGCTCTAGCACCGCTTAAACGCACGTACGCGCTGTCTACCGCGTTTTAACCGCCAATAGGATTACTTACTAGTCTCTAC} & 
$22.45$ \\ \hline
\tt TCTCACCAAAGTCACGTAGGGGGTCACGTCGCTACTTCACAATTTTCCTACGCTATCCCTGCGCTAAGCGGGTTGAGCGGCGTGAATTCCCAGG & 
$22.45$ \\ \hline
\end{tabular}
\label{table: low_energy_seqs}

\end{center}
}
\end{table}
We similarly show in Table \ref{table: high_energy_seqs} the seven sequences with the highest values for the global minimum energy. All five random sequences in Table \ref{table: high_energy_seqs} exhibit two distinct $Lk$ 9 minimisers whose energies are shown and are very close to each other.  As discussed in Sec.\ \ref{sec: polys}, the poly(XY) sequences have many distinct symmetry-related minima with energies very close to each other, so only a single energy value is shown in Table \ref{table: high_energy_seqs}.
\begin{table}[H]
\caption{94 bp sequences with the highest cgDNA+min energies (in RT).}
\scriptsize{
\begin{center}
\begin{tabular}
{ |p{14.1cm}|p{2.2cm}|}
\hline
Sequence  & minicircle energy \\ \hline
\tt AACTTCAGCCGGACGACTTATTTTCATTGTCTCAGATTCGACAGGCTCAATGTTTTATTCAAGCAAAAGGAAGCCACGGGCGTTCCGCCTGAAC & $28.82, 29.03$ \\ \hline
\tt CGTTTCCCGTTGCCGAGGTCTGATCAGGGGCCGACGAATCTCGTAGCGTCCCCTCAGTCGAAACCTCGAGTGCCAGAGCGATCCGGCCGACCGT & $28.83, 28.86$ \\ \hline
\tt TTGATCGTTACAATTCCGAGTCTTAGGCTGCAAAAGATTTGTTGATTCGTTTACTGGTTTCACGGTGATCAAAGTTGGCCCATCAAAGGGCGAT & $28.91, 29.81$  \\ \hline
\tt TGTACAAACAAAAGTCGCTCCTTGAGGATTCAACAGAACGTCATGAACACTAATGACCGGTGTGTGACACGTTCGCAAAATCTCCTCGTCGATC & $28.95, 29.09$  \\ \hline
\tt CAACGATATGATTCAGACGATCCGGCGAGTCAGACTTGCCTTGTGGGAAAGTCGGGCCCACATCATTCATCAAACAAACCCCGGCAGATTGTTG & $28.95, 28.97$ \\ \hline
$\tt (AG)_{47}$ & $31.29$ \\ \hline
$\tt (AA)_{47}$ & $31.76$ \\ \hline
\end{tabular}
\label{table: high_energy_seqs}
\end{center}
}
\end{table}
\subsubsection{Statistics of energies for random sequences with a range of lengths (92-106 bp)}
\begin{figure}[H]
\centering
\includegraphics[width=2.3in]{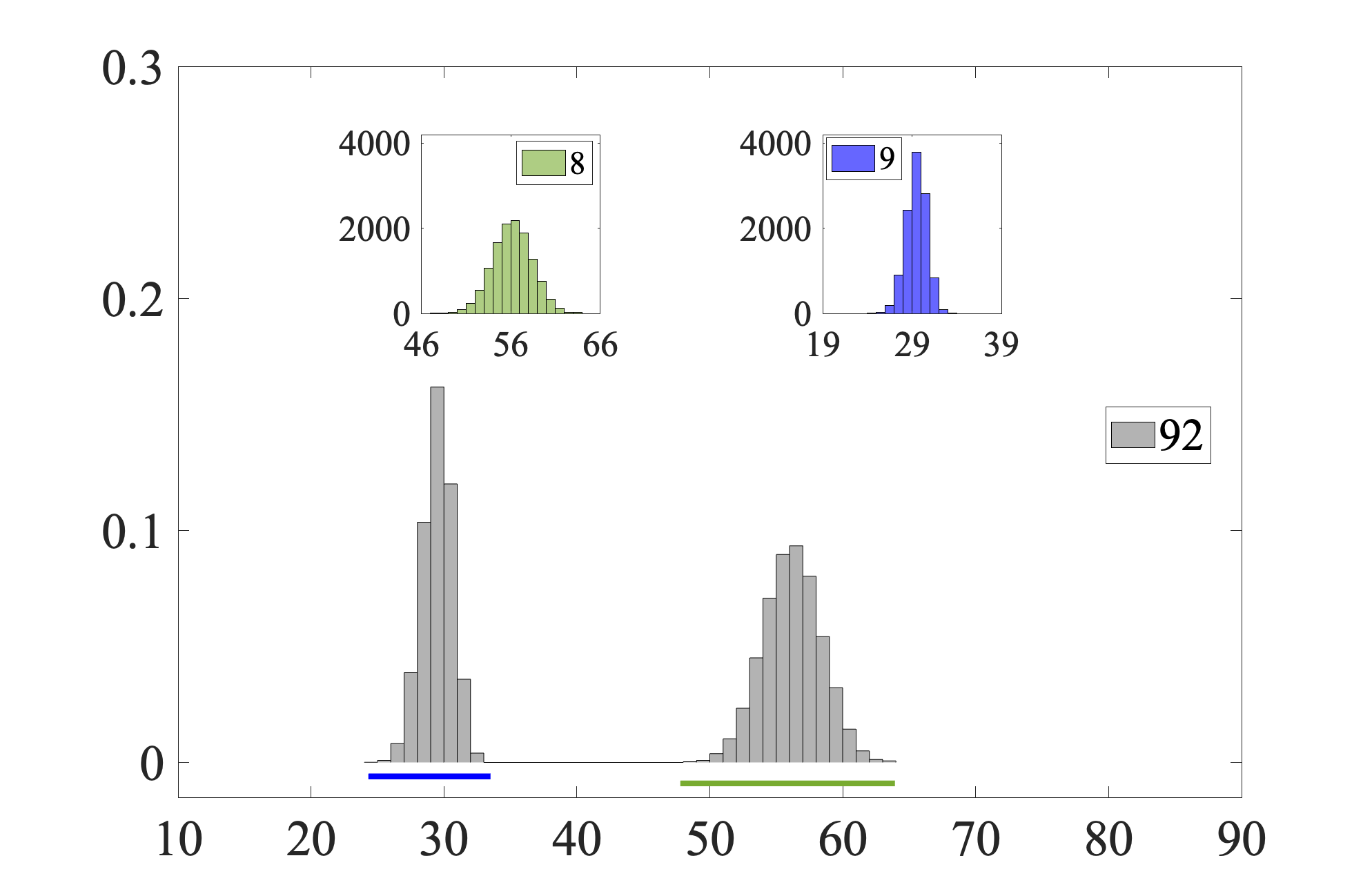} 
\includegraphics[width=2.3in]{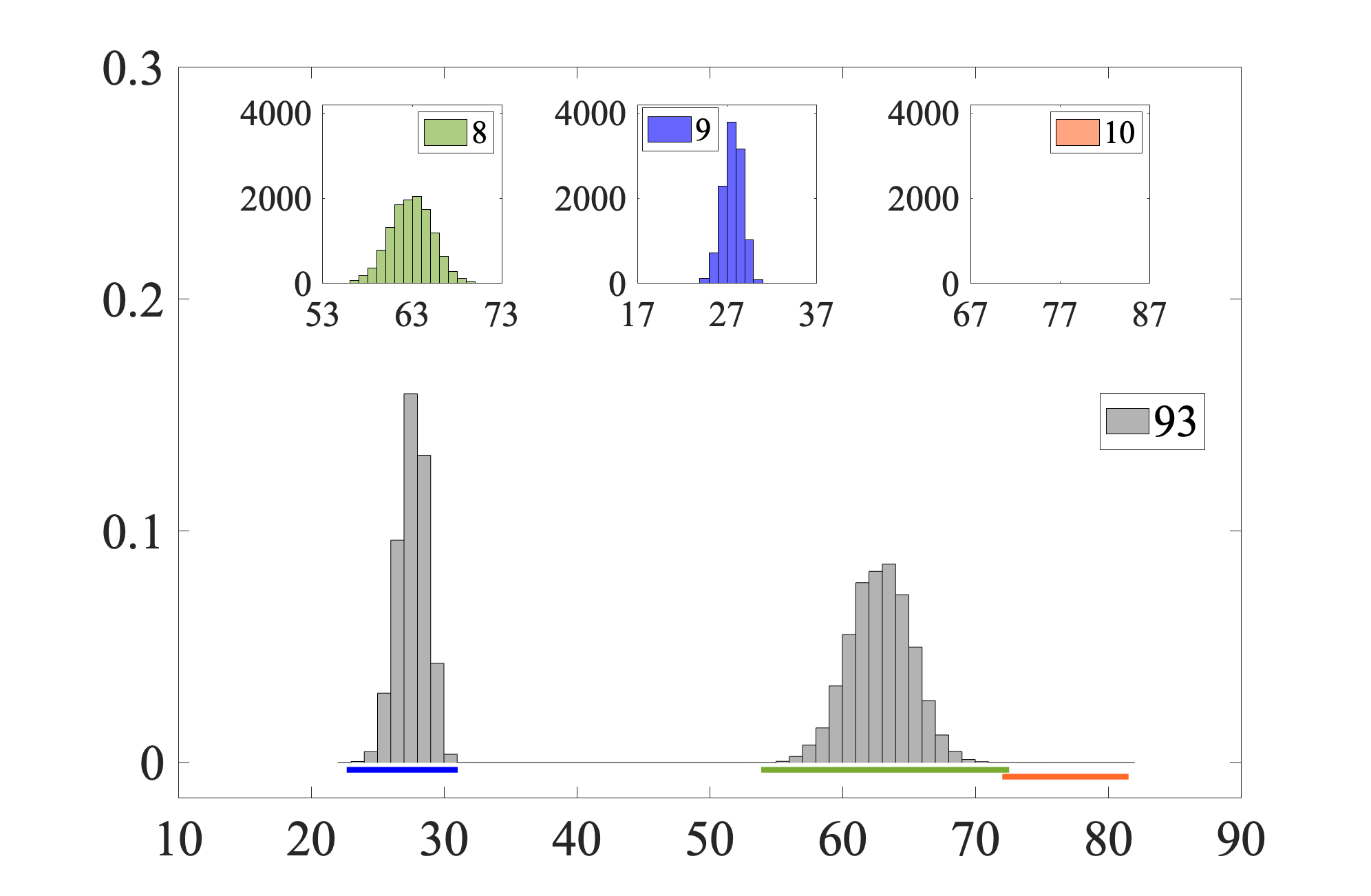} 
\includegraphics[width=2.3in]{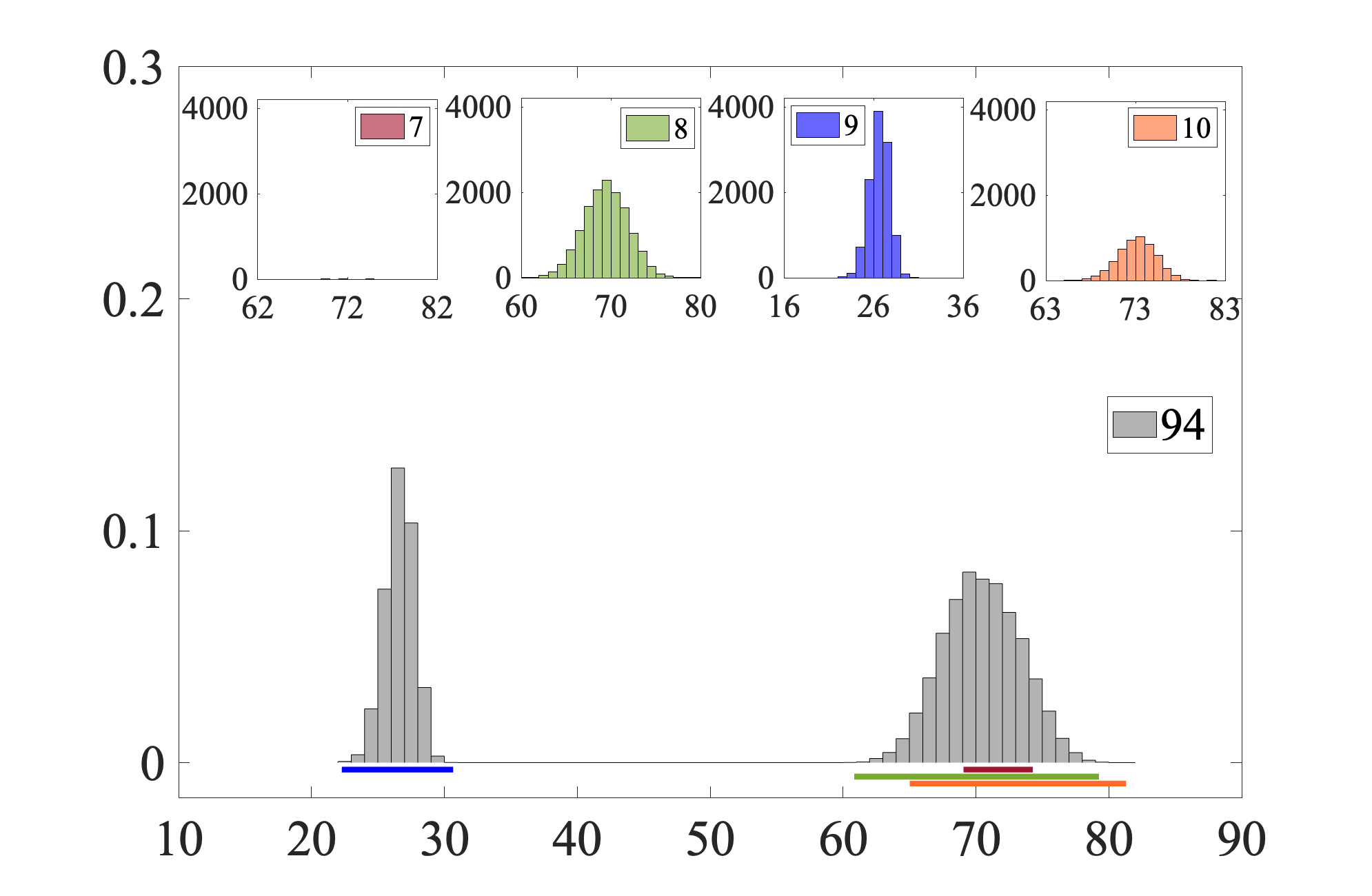}
\includegraphics[width=2.3in]{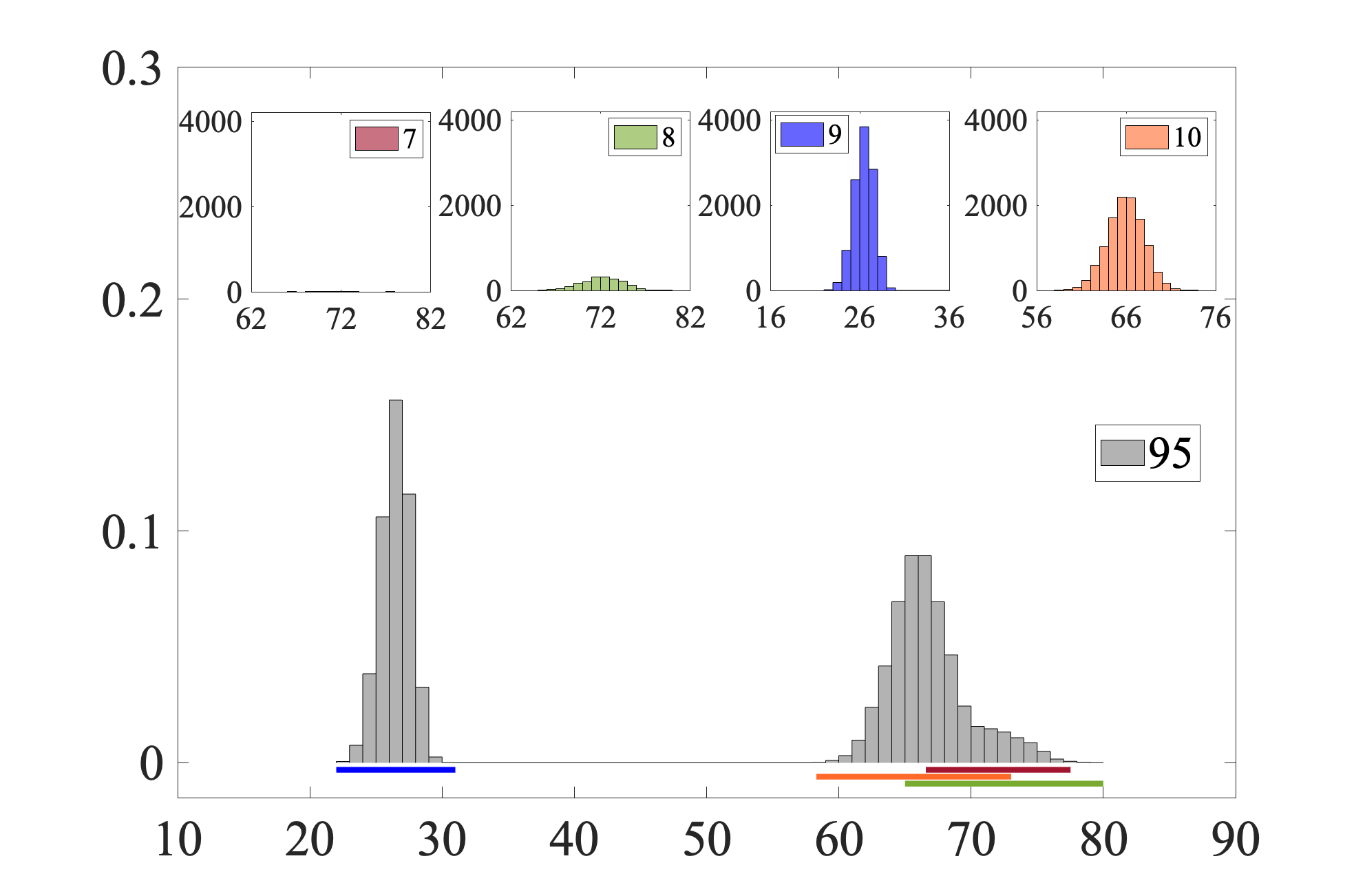}
\includegraphics[width=2.3in]{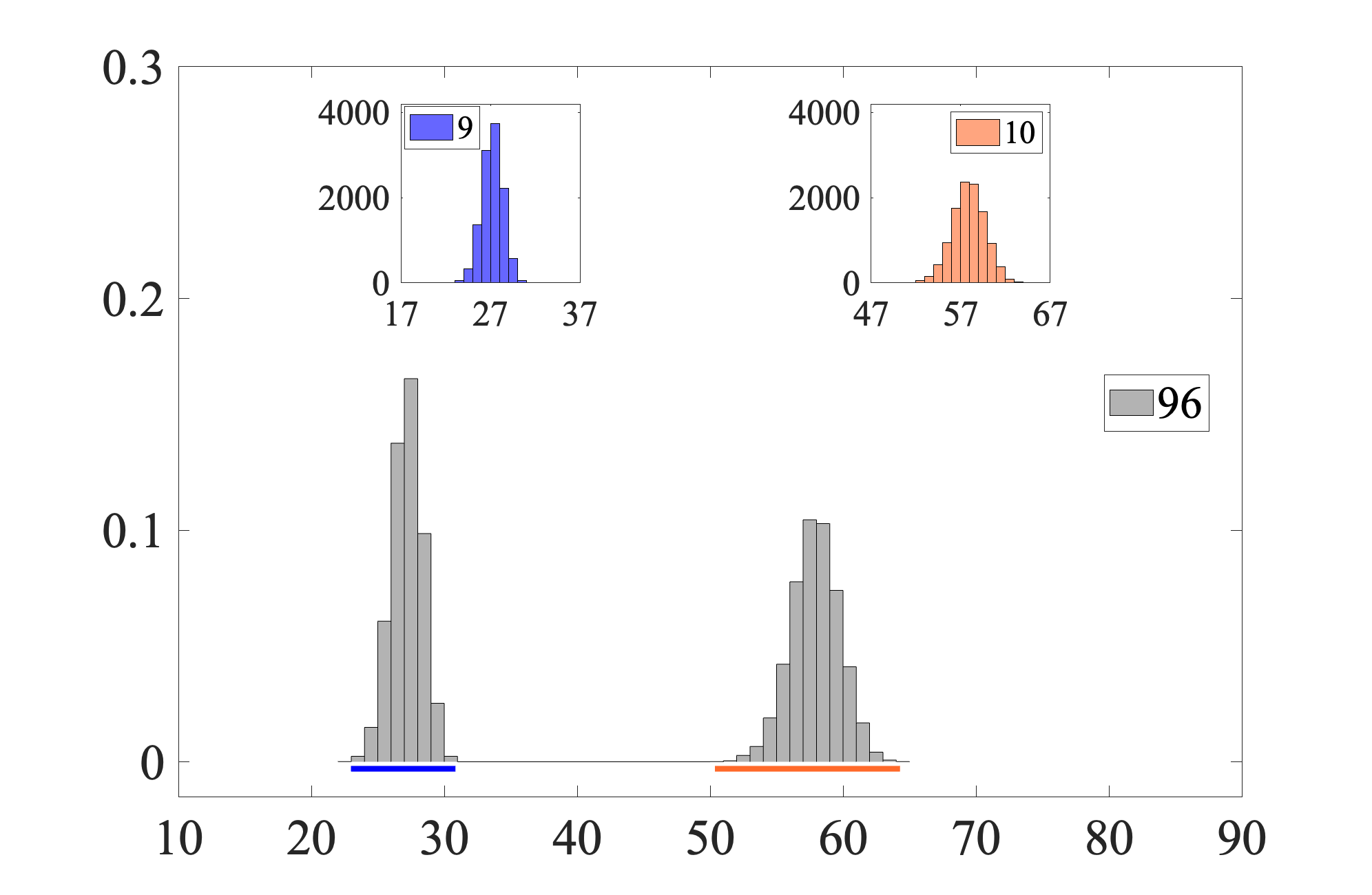}
\includegraphics[width=2.3in]{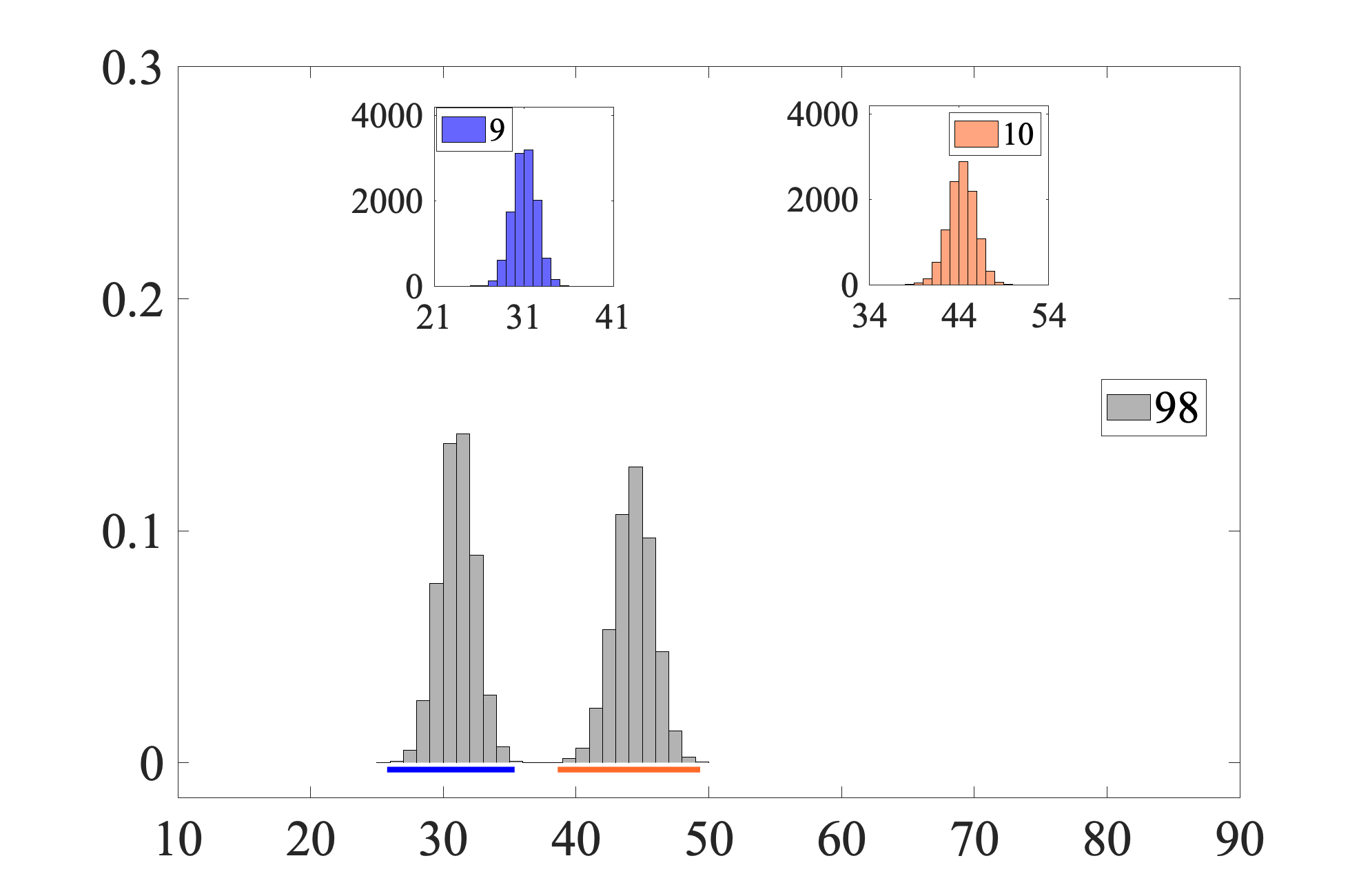} 
\includegraphics[width=2.3in]{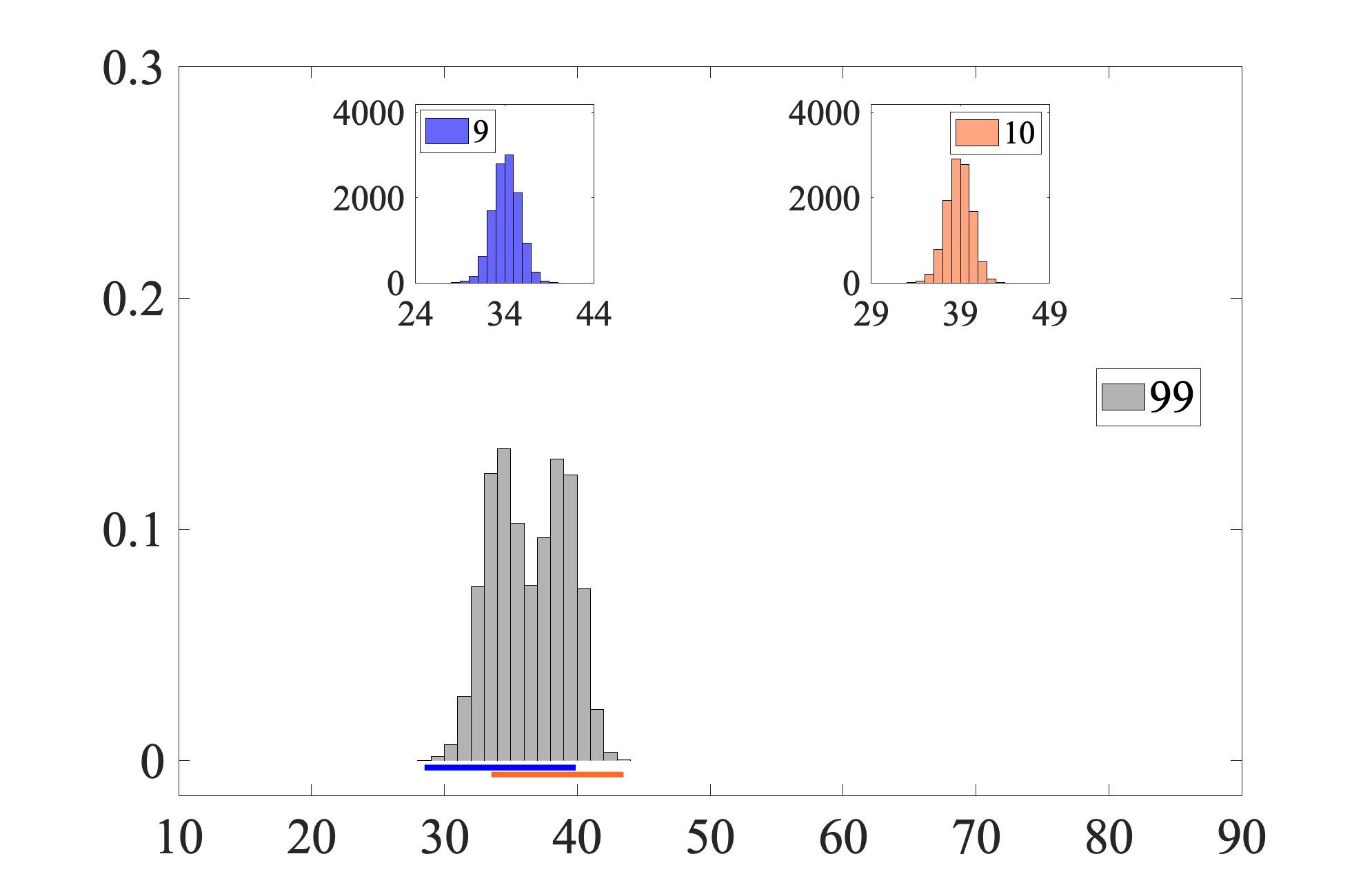} 
\includegraphics[width=2.3in]{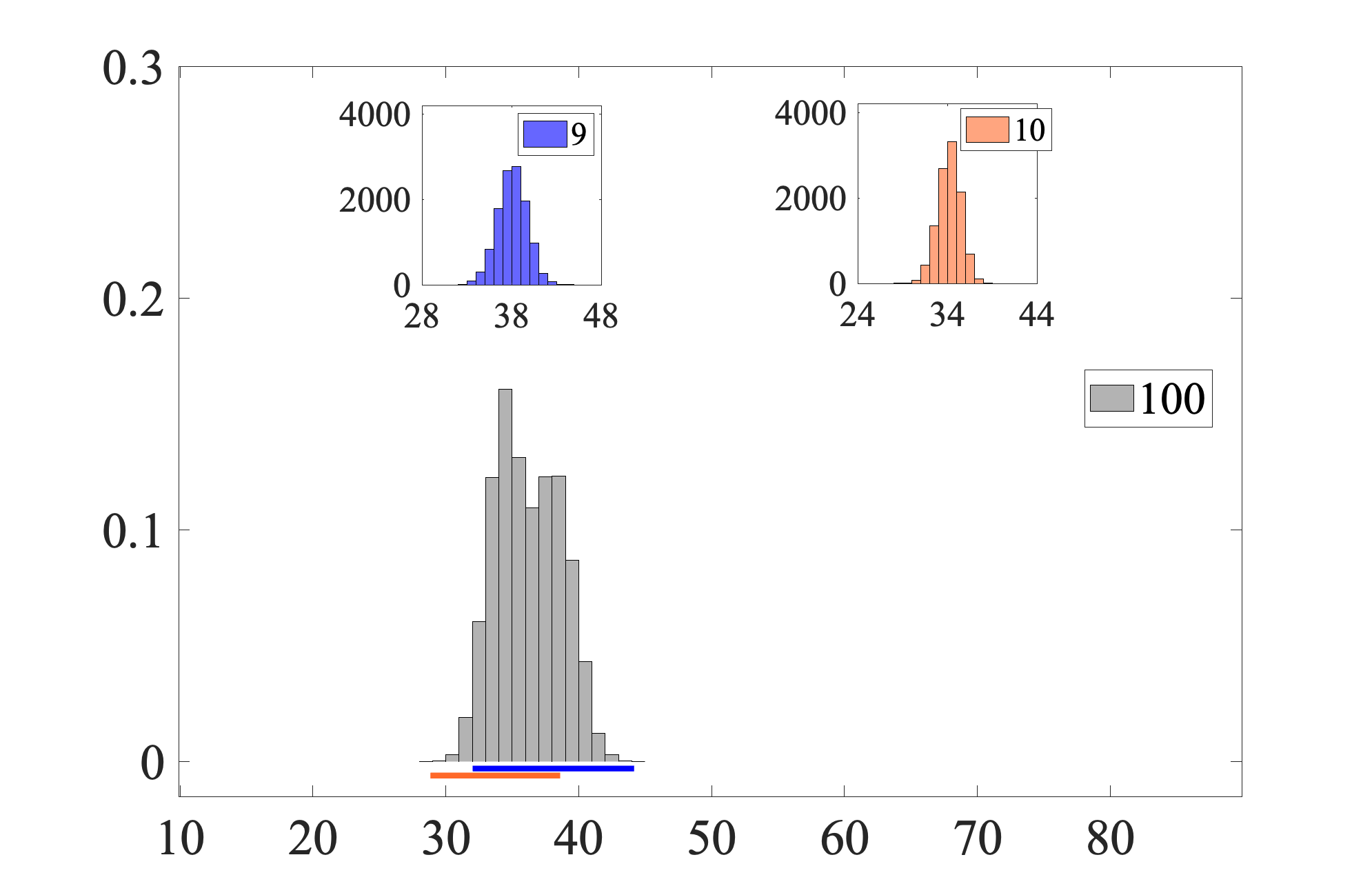}
\includegraphics[width=2.3in]{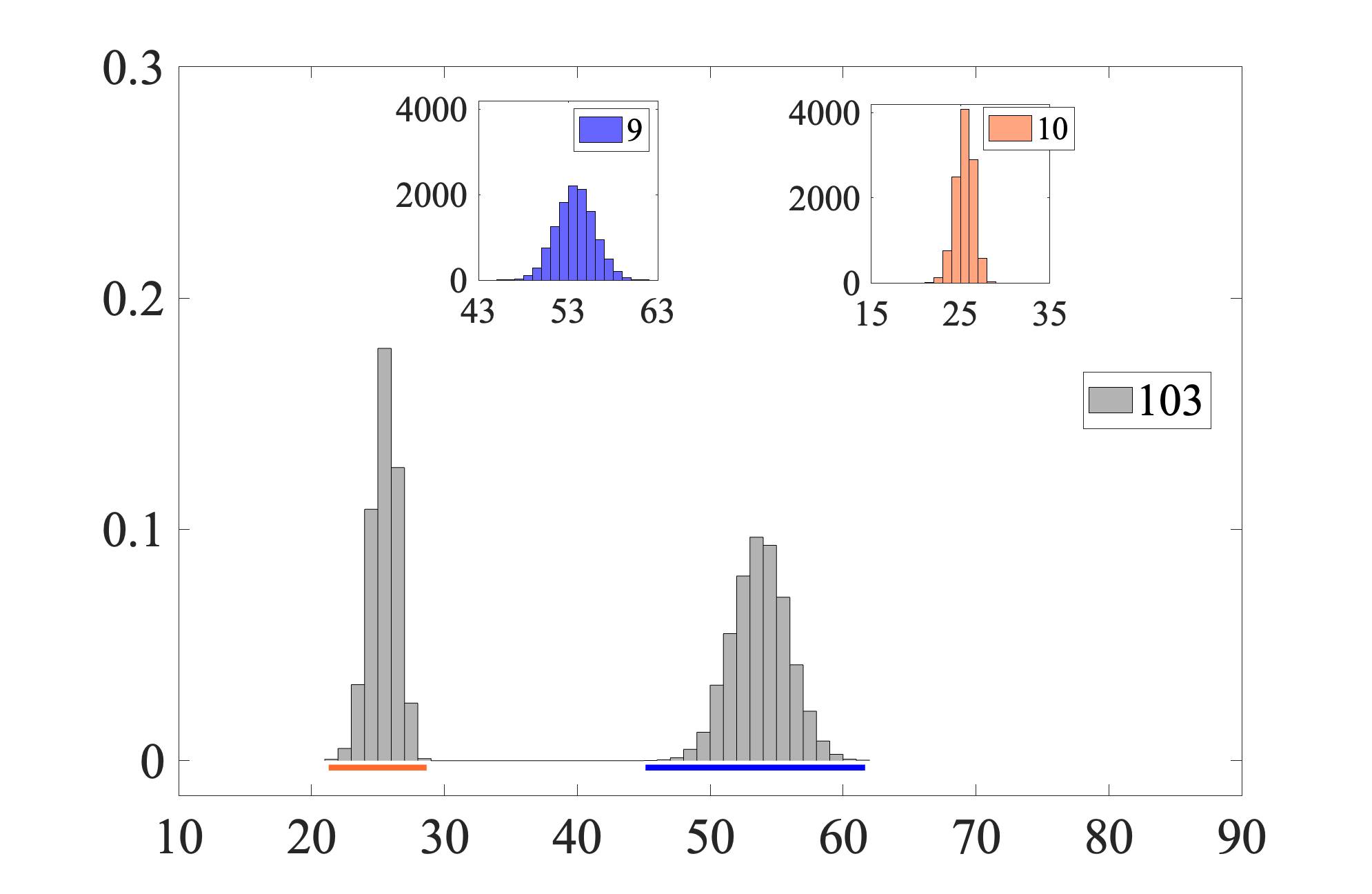} 
\includegraphics[width=2.3in]{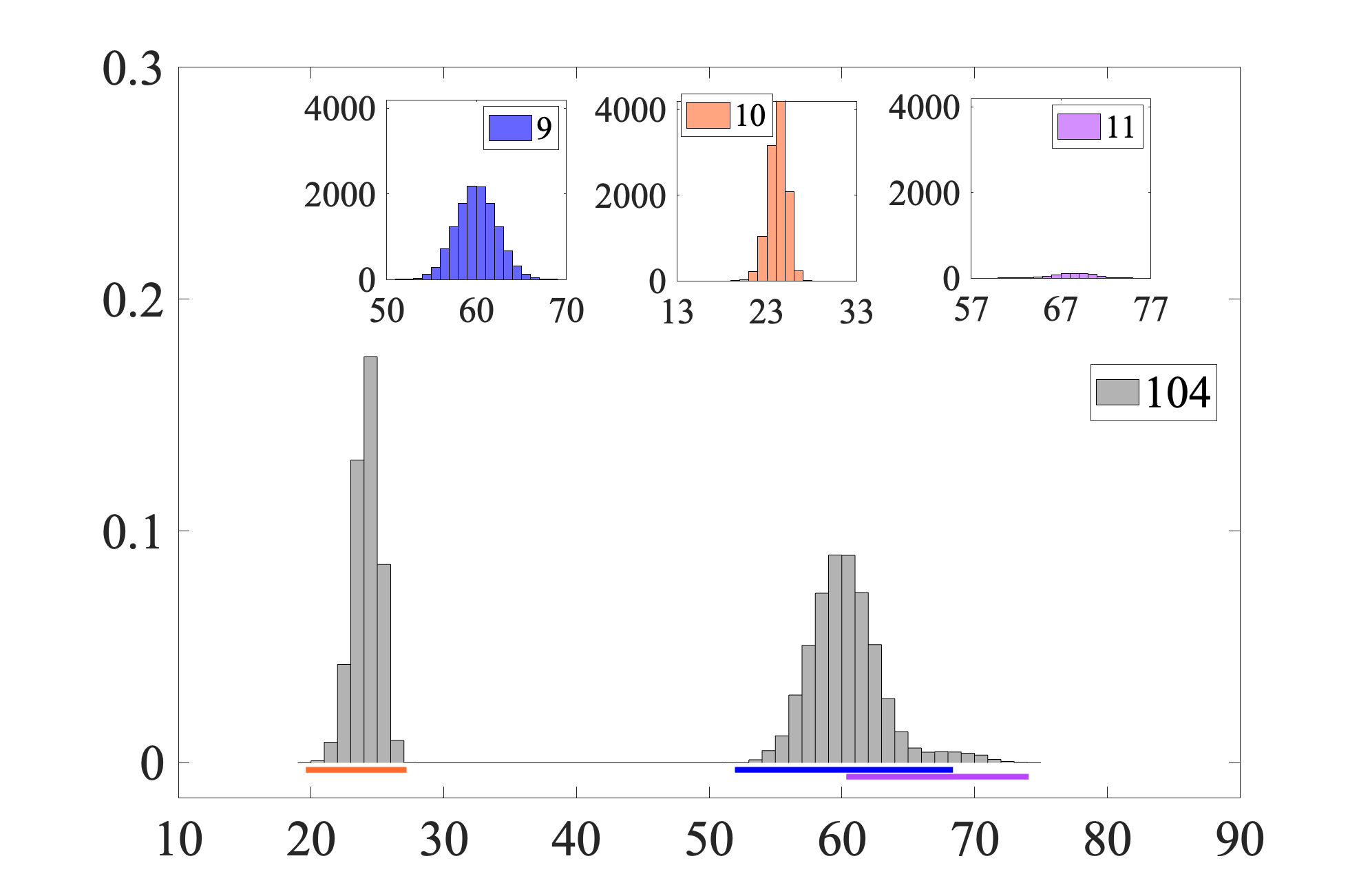} 
\includegraphics[width=2.3in]{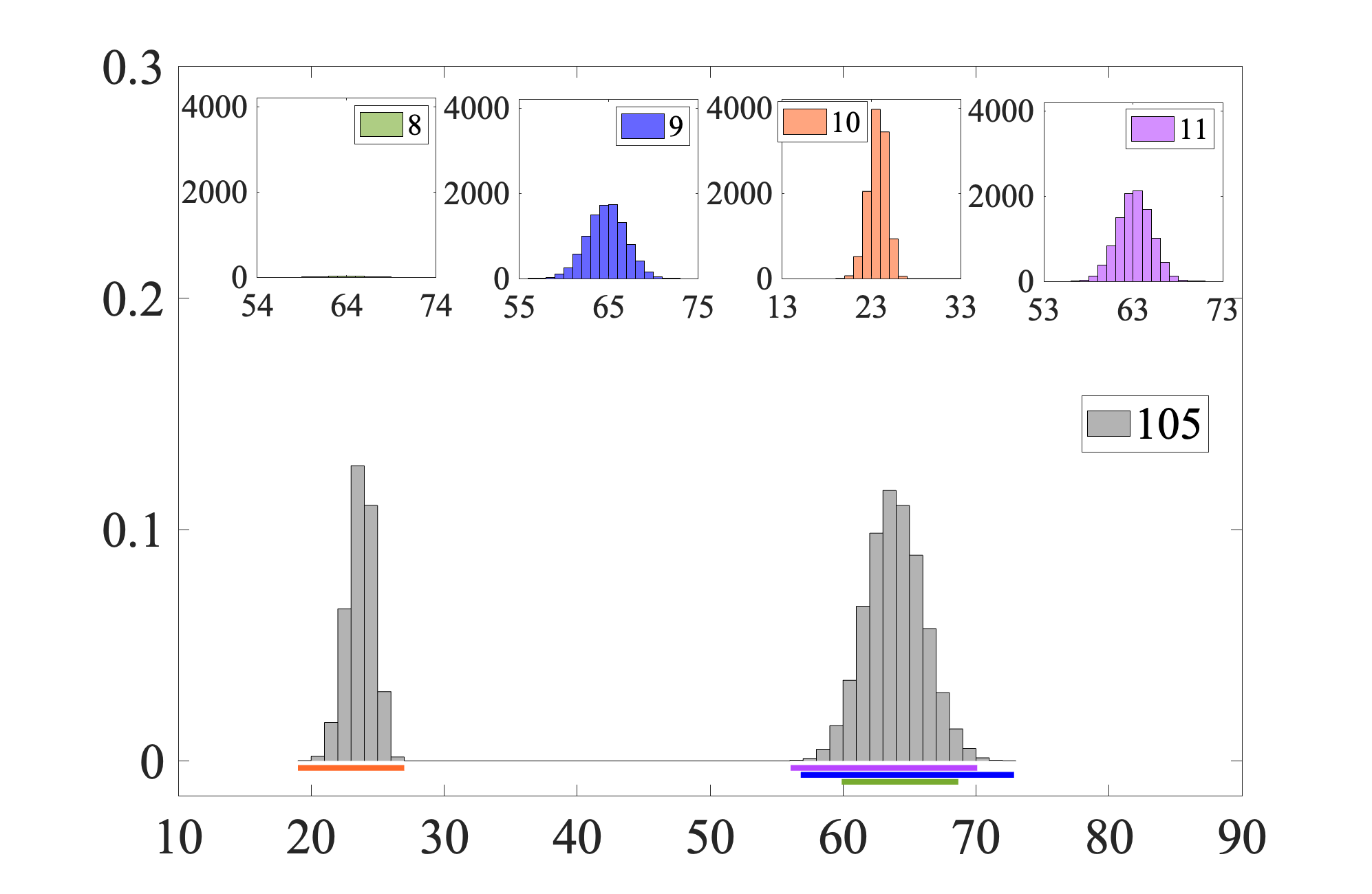} 
\includegraphics[width=2.3in]{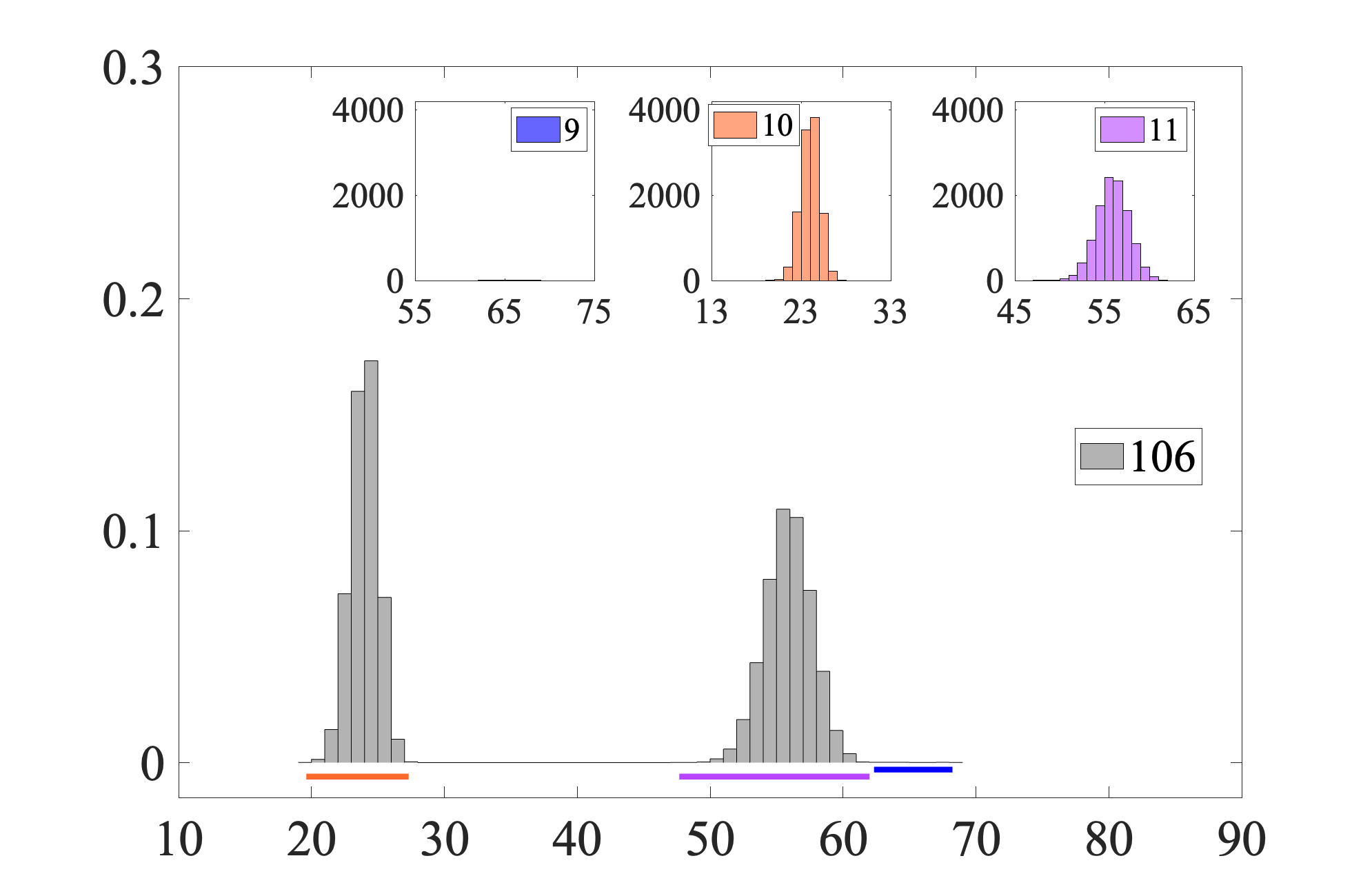}
\caption{Histograms of $cgNA+min$ energies (on x-axis, in RT) as DNA length and $Lk$ vary.  
Each of the twelve main panels shows (in gray) a normalised histogram of all distinct 
$cgNA+min$ energies for 10K random sequences of a prescribed length, with lengths 
ranging panel by panel from 92 to 106 bp.  
Inset sub-panels show unnormalised histograms of the same data, 
but grouped by distinct value of $Lk$ (maroon for $Lk$ 7, green for $Lk$ 8, 
blue for $Lk$ 9, orange for $Lk$ 10, purple for $Lk$ 11).  
The vertical scale in every sub-panel is 4200, and the horizontal scale is $20 RT$, 
but with the centering varying by sub-panel; the location of each sub-panel 
is indicated by colour bar below main axis. 
}\label{fig: hist_different_lengths} 
\end{figure}

In this section, finally we extend our 94 bp study from the previous section to include a range of lengths (92-106 bp) and all the energy-minimisers found by our procedure. First, we show in Fig.\ \ref{fig: hist_different_lengths} histograms of energies of minimisers organized by DNA length and $Lk$.  Each main panel represents one length, progressing from 92 bp in the upper left to 106 bp in the lower right. For each length, we considered 10K random sequences of that length (though a handful out of these 10K failed to produce converged results; see discussion below). Within each panel, the gray histogram shows the energies for all minimisers, and the colored inset histograms show only those energies for particular values of $Lk$, 
color-coded to be consistent from panel to panel.

For many lengths $L$, the main histogram is bimodal, with a low-energy peak coming entirely from $\Lko$, and a high-energy peak coming entirely (or almost so) from the value of $Lk$ that is second-closest to $Lk_0$, call it $Lk_{next}$ (with $\Lko$ being by definition the closest). For example, for $L = 92$ or 93, we have $\Lko = 9$, $Lk_{next} = 8$, and the overall histogram consists of a low-energy peak that matches the blue $Lk=9$ peak and a high-energy peak that matches the green $Lk=8$ peak.  The situation is the same for $L=96$ or 98, only $Lk_{next}$ switches from 8 to 10. Finally, for $L = 103, 104$, we have $\Lko = 10$, $Lk_{next} = 9$, and correspondingly the low-energy peak matches the orange $Lk$ 10 peak and the high-energy peak matches the blue $Lk$ 9 peak. In a few of these cases ($L = 93$ and 104), we see a third value of $Lk$, but the counts are too small to have an impact on the overall gray histogram. When $L$ is close to an integer multiple of 10.5 (such as $L=94$ or 105), we have a very similar situation, with the only change being that there are now two $Lk$ values that contribute substantially the high-energy peak ($Lk = 8,10$ for $L=94$ and $Lk = 9, 11$ for $L = 105$).  Also, in each of those cases, there is a {\it fourth} $Lk$ value, but its counts are too small to impact the overall histogram.

Finally, when $L$ is close to a half-integer multiple of 10.5, the two peaks in the main histogram overlap, though in each case, it is still possible to discern the two peaks, with the lower-energy peak corresponding to $\Lko$ and the higher-energy peak corresponding to $Lk_{next}$. Presumably there could be cases where the two peaks overlap so closely that the overall histogram would show a single peak, but that did not occur in our study (though $L=100$ is close to that situation). 

\begin{figure} 
\centering
\includegraphics[width=6.8in]{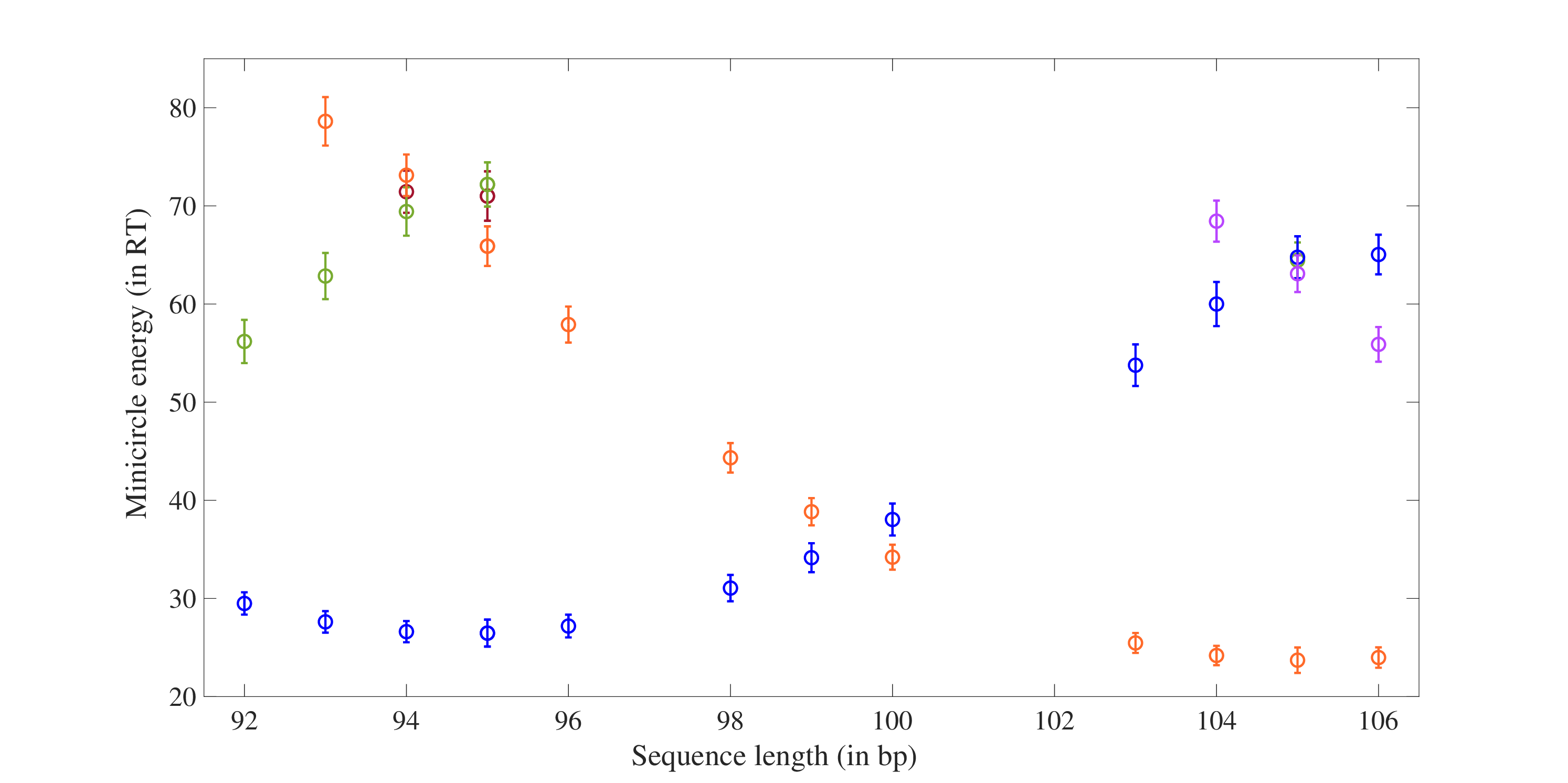} 
\caption{Mean (markers) and standard deviations (error bars) of energies at fixed DNA length and $Lk$;
 colour indicates $Lk$, using the same code as in Fig.\ \ref{fig: hist_different_lengths}.
}
\label{fig: avg_en_different_lengths}
\end{figure}

For another perspective on the data in Fig.\ \ref{fig: hist_different_lengths}, we show in Fig.\ \ref{fig: avg_en_different_lengths} the mean and standard deviation for each sub-panel histogram from Fig.\ \ref{fig: hist_different_lengths}, allowing a visualization of energy (average and variation) by $Lk$ as DNA length varies.  The blue markers ($Lk = 9$) appear to lie on a smooth curve that spans the entire range of lengths considered, and the orange markers ($Lk = 10$) lie on a different curve that crosses the blue curve around $L = 100$ bp (except the orange curve does not exist at $L=92$).  The other markers behave similarly for the ranges of $L$ for which they exist: $92 \le L \le 95$ for $Lk$ 8 (green) and $104 \le L \le 106$ for $Lk$ 11 (purple). 
\begin{table}[H]
\caption{Multiplicities of distinct links and distinct minimisers for random sequences. The notation $L$ denotes the length (in bp) of the sequence, $N$ is the number of sequences that successfully converged (in the sense of yielding at least two distinct minimisers), $M$ is the total number of distinct minimisers found among the $N$ molecules, and $\calN(i,j)$ is the number of molecules that have a total of $j$ distinct minimisers, with $i$ distinct links among those $j$ minimisers.}
\begin{center}
\begin{tabular}{ |p{1.2cm}|p{0.9cm}|p{0.9cm}|p{0.9cm}|p{0.9cm}|p{0.9cm}|p{0.9cm}|p{0.9cm}|p{0.9cm}|p{0.9cm}|p{0.9cm}|p{0.9cm}|p{0.9cm}|p{0.9cm}|  }
\hline
$L$             & 92      & 93      & 94    &  95     & 96    & 98    & 99    & 100   & 103   & 104   & 105   & 106    \\ \hline
$N$            & 9989    & 9988    & 9993  &  9993   & 9989  & 9985  & 9983  & 9973  & 9967  & 9977  & 9968  & 9966   \\ \hline 
$M$             & 23419   & 23837   & 30698 &  24522  & 22605 & 22555 & 22602 & 22571 & 22836 & 24256 & 31165 & 22081  \\ \hline 
\hline
$\calN$(2,2)  & 6804    & 6471    & 2759  &  6213   & 7502  & 7549  & 7493  & 7501  & 7260  & 6203  & 2238  & 7918  \\ \hline
$\calN$(2,3)  & 2930    & 3164    & 1835  &  1925   & 2347  & 2287  & 2344  & 2319  & 2512  & 2913  & 634   & 1933 \\ \hline
$\calN$(2,4)  & 254     & 339     & 314   &  167    & 140   & 149   & 146   & 153   & 195   & 272   & 46    & 99  \\ \hline
$\calN$(2,5)  & 1       & 1       & 13    &  0      & 0     & 0     & 0     & 0     & 0     & 10    & 1     & 0  \\ \hline
\hline
$\calN$(3,3)  & 0       & 10      & 2515  &  1165   & 0     & 0     & 0     & 0     & 0     & 369   & 4176  & 14  \\ \hline
$\calN$(3,4)  & 0       & 3       & 2030  &  458    & 0     & 0     & 0     & 0     & 0     & 185   & 2323  & 2  \\ \hline
$\calN$(3,5)  & 0       & 0       & 477   &  62     & 0     & 0     & 0     & 0     & 0     & 24    & 476   & 0  \\ \hline
$\calN$(3,6)  & 0       & 0       & 43    &  2      & 0     & 0     & 0     & 0     & 0     & 1     & 39    & 0  \\ \hline
$\calN$(3,7)  & 0       & 0       & 5     &  0      & 0     & 0     & 0     & 0     & 0     & 0     & 5     & 0  \\ \hline
\hline
$\calN$(4,4)  & 0       & 0       & 0     &  1      & 0     & 0     & 0     & 0     & 0     & 0     & 23    & 0  \\ \hline
$\calN$(4,5)  & 0       & 0       & 1     &  0      & 0     & 0     & 0     & 0     & 0     & 0     & 6     & 0  \\ \hline
$\calN$(4,6)  & 0       & 0       & 1     &  0      & 0     & 0     & 0     & 0     & 0     & 0     & 0     & 0  \\ \hline
$\calN$(4,7)  & 0       & 0       & 0     &  0      & 0     & 0     & 0     & 0     & 0     & 0     & 1     & 0  \\ 
\hline
\end{tabular}
\label{table: multi_lk_min}
\end{center}
\end{table}
Finally, we show in Table \ref{table: multi_lk_min} how often in our ensembles of random sequences we observed the ``typical'' baseline situation of two minimisers with adjacent $Lk$ and how often we observed something more involved.  For each column of the table, we considered 10,000 random sequences of the length $L$ shown, but a small number (under 40) of our simulations failed to produce two distinct converged minimisers, so those random sequences were omitted from this study; the row labeled $N$ indicates how many random sequences did produce at least two distinct minimisers. [We confirmed (data not shown) for one value of $L$ that by increasing the number of iterations {\tt fminunc} was allowed to take, we would get $N=$ 10,000 and the remainder of the table would not change significantly.] The row labeled $M$ shows the total number of distinct minimisers detected for the $N$ random sequences.   The remainder of the table sorts the $N$ molecules based on both the total number of distinct minimisers ($j$) and the total number of distinct links ($i$). The baseline case of two minimisers of two different $Lk$ appears in the $\calN(2,2)$ row, and for all but $L=94, 105$, this is the majority (``typical'') outcome.  (Given the notation $\calN(i,j)$ used in the table, we have $N = \sum_{i,j} \calN(i,j)$ and $M = \sum_{i,j} j \calN(i,j)$.)

For every $L$, at least 20\% of the random sequences yield more than the baseline case of two minimisers, e.g., it is not uncommon to find two minimisers with one $Lk$ and one with another $Lk$; these are counted by $\calN(2,3)$. For $L$ close to an integer multiple of 10.5, it is not uncommon to find three distinct $Lk$ (row $\calN(3,3)$ if one of each $Lk$, or row $\calN(3,4)$ if two for one $Lk$ and one each for the other two $Lk$). Larger values of $i$ or $j$ are seen at least 1\% of the time ($\calN(i,j)=100$ or more), and rarely (but multiple times in our ensemble) we found unusual situations like 4 distinct $Lk$ or 6 or more distinct minimisers.
\section{Conclusions and discussion}\label{conclusion}
In the framework of the $cgNA+$ model, a computational strategy (``$cgNA+min$'') is presented for finding
energy-minimising cyclized configurations. This methodology in $cgNA+min$ features the use of absolute base pair frame translation and quaternion coordinates, instead of the internal coordinates in which the original model energy is expressed.  With this approach, the only constraints are that each quaternion is length-one; since each constraint involves a single unknown vector and has a simple algebraic expression, this setup results in a problem that can be handled with an unconstrained minimization algorithm (we use Matlab's {\tt fminunc}). We use a high-throughput automated version of $cgNA+min$ that feeds the minimization algorithm $100$ initial configurations per DNA sequence, with the initial guesses spanning an array of values of $Lk$ and register. Based on our results, we are confident that for more than $99\%$ of sequences, $cgNA+min$ finds at least two distinct adjacent link energy-minimising configurations, and case-by-case followup would make the same be true for the remaining $1\%$. Thus, for any sequence, energy minimisers are always found with at least two adjacent values of $Lk$, and we argue from a theoretical perspective why this should be the case (see discussion in Section \ref{subsec:link}). 

In general the link of an initial configuration is not conserved along the sequence of configurations generated by our energy minimising algorithm, so that all that can be guaranteed is that there is at least one minimiser with an odd link, and at least one minimiser with an even link. It is our experience that the odd and even values of the link having minimisers always arise at adjacent integers.  This conjecture seems physically intuitive, but we are unaware of any demonstration of this observation within the context of the $cgNA+$ coarse grain energy minimisation formulation 
adopted here. 
In contrast, the parameter continuation and symmetry breaking methods that have previously been applied to compute DNA minicircle equilibrium configurations within coarser grain, phantom, continuum rod, and birod models of dsDNA do provide the conclusion that there must be minimisers at least two {\em adjacent} links \citep{Manning1996, DichmannLiMaddocks1996, Manning1998, ManningMaddocks1999, Furrer2000, Hoffman2003, Maddocks2004, Glowacki2016, Grandchamp2016}. 
The basic idea is that, as the first and last frames are continuously twisted with respect to one another around a fixed axis, then, due to their double covering properties, the values of the first and last frame quaternions can track relative rotations with respect to one another mod $4\pi$ and not just mod  $2\pi$. And it is this feature that leads to the conclusion that there are always minimisers at two adjacent values of link.  

In this phantom-chain model, with no self-repulsion, for most sequences, these two minimisers with adjacent values of $Lk$ are the only minimisers.  But for other cases, there can be minimisers with more than two values of $Lk$ (we found up to four), particularly for lengths corresponding to complete turns of the associated linear fragment e.g.,  94 bp or 105 bp.  It is also fairly common to observe multiple minimisers that share a value of $Lk$. Finally, we also showed that sequence-dependent energies computed by $cgNA+min$ correlate well with a simplistic approximation to the energy extracted from experimentally measured J-factors. By feeding $cgNA+min$ thousands of sequences, we find sequences with outlier (low or high) minicircle energies within the ensemble of all random sequences. For example, $10$ outlier $94$ bp sequences are presented in Tables \ref{table: low_energy_seqs} and \ref{table: high_energy_seqs}.

We believe that the results presented and the proposed method constitute a substantial improvement in the modelling of DNA mechanics. In particular, the minimum energy configurations computed by $cgNA+min$ are a key component in estimating the cyclisation factor $J$ through the Laplace expansion \citep{Cotta-Ramusino_Maddocks_2010, Manning2024}.
Furthermore, the $cgNA+min$ approach can be used for dsDNA in both standard and epigenetically modified alphabets, for dsRNA, and for DNA-RNA hybrid fragments, since $cgNA+$ parameter sets are available for all those cases. Thus, scanning a large number of sequences of dsNAs minicircles (of different kinds subject to the availability of parameter sets for the $cgNA+$ model) can be achieved, and the generated data may discover exceptional sequences which may encourage further experimental studies of those sequences. Finally, the data generated using $cgNA+min$ can also be used to train machine learning algorithms for other further computations.
\section*{Code availability}
The complete $cgNA+min$ Matlab package is available at \url{https://github.com/singh-raushan/cgNA_plus_min} 
\section*{Acknowledgment}
RS and JHM acknowledge partial support from the Swiss National Science Foundation Grant 200020-182184. RS acknowledges partial support from IIT Madras NFIG B23241556.   
\appendix
\section{Basic properties of rotation matrices}\label{Annexe: Cayley vector}
In this Section we gather for convenience all of the standard material about $3\times 3$ proper rotation matrices, i.e.\ elements of the matrix group $SO(3)$, that we use. The material is more fully described in many places, for example \cite{Courant_Hilbert_2008} p.\ 536. 

The defining property of matrices $R\in SO(3)$ is $R^TR=I_3$ and $\det R = +1$, i.e.\ the matrix inverse is its transpose, which implies $(\det R)^2= 1$, and proper rotations have positive determinant. Such matrices have eigenvalues $\{+1, \exp(\pm {\mathbf i}\, \Theta)\}$, with $0\le \Theta \le \pi$, and therefore trace $\trace(R) = 1+2\cos\Theta \ge -1$. Except for the case $\Theta =0$, which arises only when $R=I_3$, the eigenspace of the eigenvalue $1$ is one dimensional, spanned by the (real) unit vector ${\mathbf n}$ say. Then geometrically $R$ is a rotation through the angle $\Theta$ around the rotation axis ${\mathbf n}$ and the indeterminacy in the sign $\pm {\mathbf n}$ of the unit eigenvector is removed by adding the convention that $\Theta$ is a right handed rotation about ${\mathbf n}$. Axis-angle coordinates on $SO(3)$ (with rotation angle measured in radians) are then points $\Theta {\mathbf n} $ lying in the solid ball of radius $\pi$ in $\R^3$. The mapping between matrices $R$ and coordinates  $\Theta {\mathbf n} $ is invertible, but only after antipodal points on the surface of the ball are identified. This corresponds to the elementary fact that a right handed rotation through the angle $\pi$ about $+{\mathbf n}$ coincides with a right handed rotation through the angle $\pi$ about $-{\mathbf n}$. This identification is also central to two pertinent observations. First, there is no singularity-free global set of three coordinates for the group $SO(3)$. (This observation is standard, but not entirely obvious. It follows because $SO(3)$ is not simply connected in the sense that there are closed curves of matrices in $SO(3)$, containing a rotation through $\pi$, that cannot be continuously shrunk to a point, i.e.\ to a single matrix in $SO(3)$. But $\R^3$ is simply connected, which gives rise to a contradiction if there were a global, singularity-free set of 3D coordinates.) Second, close to the inevitable singularity in any set of 3D coordinates there are matrices $R_1$ and $R_2$ that are arbitrarily close to each other in $SO(3)$ for which the respective coordinates are far from each other. In the case of axis-angle coordinates the discontinuity is explicit, and arises precisely close to rotations through $\pi$. When $\Theta = \pi^-$, i.e.\ rotations through angles just less than $\pi$, the two sets of coordinates $\pi^- {\mathbf n} $ and  $-\pi^- {\mathbf n} $ are far from each other, while the corresponding two matrices are arbitrarily close. Rotations through $\pi$ have the further special property that they are the only (apart from the trivial case of $I_3$) proper rotation matrices that are symmetric i.e.\ $R^T=R$. (This is just because symmetric matrices have real eigenvalues.) And rotations through $\Theta = \pi$ are the only proper rotations at which the (strict) inequality
\begin{equation}
\label{trace inequality}
1+\trace(R) = 2(1+\cos \Theta) = 4 \cos^2 (\Theta/2)> 0, \qquad 0\le \Theta < \pi,
\end{equation}
fails.

In general the Cayley transform and its inverse are mappings between two classes of matrices, analogous to the matrix exponential and its inverse the matrix logarithm. But the Cayley transform is of a more explicit, algebraic character. For the specific case of matrices in $SO(3)$ the Cayley transform is a relation with $3\times 3$ skew symmetric matrices, which we write with the notation
\begin{equation}
\label{Skew}
    w\times \equiv [w\times ] :=
    \begin{bmatrix}
    0 & -w_3 & w_2 \\
    w_3 & 0 & -w_1 \\
    -w_2 & w_1 & 0
    \end{bmatrix}.
\end{equation}
Then
\begin{equation}
\label{Cayley}
R(w) := \left( I_3 + {w\times} \right) \left( I_3 - {w\times} \right)^{-1}
\end{equation}
is a matrix in $SO(3)$ for all $w$, where we call $w=[w_1,w_2,w_3]^T\in \R^3$ the Cayley vector. (The names Gibbs or rotation vector are also both commonly used for $w$, sometimes with a scale factor, as discussed in Appendix \ref{Annexe: quaternion}.)  The sign conventions in (\ref{Skew}) mean that the matrix-vector product $[w\times ]x$ coincides with the vector product $w\times x$ between the Cayley vector $w$ and any other vector $x\in \R^3$. The inverse of $\left( I - {w\times} \right)$ appearing in (\ref{Cayley}) always exists, as $\left( I - {w\times} \right)x=0\implies \norm{x}^2=0$ because $w\times$ is skew. That $R(w)\in SO(3)$ follows from (\ref{Cayley}) via direct evaluation of $R(w)^TR(w)$ and $\det R(w)$ using the facts that $\left( I - {w\times} \right)^T =\left( I + {w\times} \right)$ and $\left( I - {w\times} \right)$ and $\left( I + {w\times} \right)$ commute. The explicit formula
\begin{equation}
\label{neumann_w}
\left( I_3 - {w\times} \right)^{-1} = I_3 + \frac{1}{1+\norm{w}^2} [w\times] + \frac{1}{1+\norm{w}^2} [w\times]^2
\end{equation}
can be derived using a matrix Neumann series expansion for the inverse combined with the Cayley-Hamilton matrix identity $[w\times]^3 = - \norm{w}^2 [w\times]$. (Or (\ref{neumann_w}) can just be verified directly.) Substitution of (\ref{neumann_w}) in (\ref{Cayley}) then yields the completely explicit formula 
\begin{equation}
\label{Cayley_ER}
R(w) = \frac{1-\norm{w}^2}{1+\norm{w}^2}\, I_3 + \frac{2}{1+\norm{w}^2}\, [w\times] + \frac{2}{1+\norm{w}^2}\, w\, w^T,
\end{equation}
 which appears (in a scaled form) in equation \eqref{eqn:P} of the main text. We call (\ref{Cayley_ER}) an Euler-Rodrigues formula because it gives an explicit expression for a rotation matrix in terms of a set of parameters, here the components of the Cayley vector $w$.

 There is a general procedure  to invert Cayley transforms of the algebraic form (\ref{Cayley}), but in the case of $SO(3)$ and skew matrices it is simpler to proceed in an {\it ad hoc} fashion. The elements of $SO(3)$ that can be written with the parametrisation (\ref{Cayley}) (or its equivalent (\ref{Cayley_ER})) can be characterised as follows. Post multiplication of both sides of (\ref{Cayley}) by $\left( I_3 - {w\times} \right)$ and rearranging yields
 \begin{equation}
\label{not_pi}
\left( I_3 + R(w) \right)\, \left( I_3 - {w\times} \right) = 2\,I_3,
\end{equation}
which implies that the matrix $\left( I_3 + R(w) \right)$ is invertible, so that $-1$ is not an eigenvalue of $R(w)$, i.e.\ $R(w)$ is a rotation through less than $\pi$. In particular the (strict) trace inequality (\ref{trace inequality}) holds for all $R(w)$.  In turn taking the trace of the matrix equality (\ref{Cayley_ER}) yields
\begin{equation}
1 + 2\,\cos\Theta = \frac{3 - \norm{w}^2}{1+\norm{w}^2 }
\end{equation}
which has the unique solution
\begin{equation}
\label{norm_w}
\norm{w}^2 = \frac{1-\cos\Theta}{1+\cos\Theta} = \tan^2 (\Theta/2).
\end{equation}
Moreover, subtracting the transpose of (\ref{Cayley_ER}) from itself yields 
 \beq
 w\times = \frac{1}{1+tr(R)} \left( R - R^T \right).
 \label{eqn:cay_def}
 \eeq
Thus we see that $\R^3$ provides a set of coordinates for all proper rotation matrices through angles $0\le \Theta < \pi$. For any vector $w\in \R^3$, (\ref{Cayley_ER}) gives the associated $R(w)\in SO(3)$, and given any $R\in SO(3)$ satisfying (\ref{trace inequality}), (\ref{eqn:cay_def}) provides the associated Cayley vector $w$. The formula (\ref{norm_w}) reveals that $\norm{w}$ approaches infinity along any family of matrices $R\in SO(3)$ along which the rotation angle $\Theta$ approaches $\pi$, and this is the singularity in the coordinate system.

As a preliminary step to explain the quaternion parameterisation of $SO(3)$ we further consider the relation between the Cayley vector $w$ and the axis angle coordinates $\Theta \mathbf{n}$. The explicit Euler-Rodrigues formula (\ref{Cayley_ER}), has the Cayley vector $w$ as an eigenvector with eigenvalue $1$, which for $\Theta \neq 0$ is simple. Thus $w$ is either parallel or antiparallel to the unit rotation axis vector $\mathbf{n}$, and it can be verified that with the specific sign conventions in (\ref{Skew}) and (\ref{Cayley}), $w$ is actually parallel to the choice of $\mathbf{n}$ corresponding to right handed rotation through the angle $\Theta$, so that $w= \tan(\Theta/2)\, \mathbf{n}$, for all $0\le \Theta < \pi$. The vector part of a unit quaternion  has the scaling $(q^1,q^2,q^3)= \sin(\Theta/2)\mathbf{n}$, with the scalar part $q^4 =\cos(\Theta/2)$, which leads to the quaternion Euler-Rodrigues formula (\ref{eqn:quaternion_defn}) in the main text.
 For quaternions, crucially, the rotation angle $\Theta$ can now be taken to lie in the range $0\le \Theta \le 2\pi$, which generates a double covering of $SO(3)$ corresponding to $\pm q$, with the case $q^4=0$ arising for $\Theta = \pi$ being smoothly embedded in the double covering, which is in contrast to the discontinuities that arise in both axis-angle and Cayley vector coordinates at $\Theta = \pi$.  A simple example is instructive. A family of proper rotation matrices making a single complete turn about a common rotation axis have the simple explicit parameterisation
 \beq
 \label{family}
 R(\tau) :=  \begin{bmatrix}
    \cos(2\pi \tau) & \sin(2\pi \tau) & 0 \\
    -\sin(2\pi \tau) & \cos(2\pi \tau) & 0 \\
    0 & 0 & 1
    \end{bmatrix},\qquad 0\le \tau \le 1.
\eeq
The Cayley vector coordinates along this family are $w = (0,0,\tan\pi \tau)$, which is discontinuous on either side of the singularity at $\tau = 1/2$ where the Cayley vector is not defined. Similarly the axis-angle coordinates are 
$\Theta \mathbf{n}= (0,0,2\pi \tau)$ for $0\le\tau \le 1/2$, and $\Theta \mathbf{n}= (0,0,-2\pi (1-\tau))$ for $1/2 \le\tau \le 1$, with the discontinuity at $\tau=1/2$ resolved by identifying the antipodal points. However the unit quaternion coordinates along this family are smooth, $(q^1,q^2,q^3,q^4)(\tau) = (0,0,\sin\pi\tau, \cos\pi\tau)$, but with the (small) price to pay that the end points of this smooth family of coordinates satisfy $q(0) = -q(1)$, which are both coordinates of $R(0)=R(1)= I_3$. It is the double covering feature of the quaternion parameterisation that allows the desirable continuity property of quaternions that for rotation matrices $R_1$ and $R_2$ that are close to each other there is always a choice of corresponding coordinates $q_1$ and $q_2$ that are close to each other. This continuity property is not shared by either Cayley vector or axis-angle coordinates, or any other set of three-dimensional coordinates. 

We remark that we make no use of any algebraic property of quaternions as introduced by Hamilton \citep{Hamilton1866} 
Consequently it is perhaps more appropriate to describe the sets of four scalar parameters $(q^1,q^2,q^3,q^4)$ that we adopt as the Euler, or Euler-Rodrigues, parameters, which were introduced earlier than quaternions. Indeed the remarkable composition rule for quaternions (\ref{composition}) used in the main text was actually derived using Euler parameters by Rodrigues in 1840 \citep{Rodrigues1840}, 26 
years before Hamilton introduced the name quaternion. 
The difference between a more algebraic version of quaternions and Euler-Rodrigues parameters is directly analogous to that between regarding a complex number as a single object, or data type, with its own algebra, or working directly with its real and imaginary parts. We adopt the nomenclature quaternion because it is both shorter, and now, it seems, more widespread. It is however significant to note that Euler parameters are completely different from the commonly adopted three dimensional Euler {\em angle} description of $SO(3)$, which we assiduously avoid because it is harder to avoid their discontinuities, which arise at the inevitable polar singularity, no matter exactly which of the many conventions for Euler angles is adopted.



\section{Curves+ coordinates, $cgNA+$ units and scalings, and the quaternion of the junction frame} \label{Annexe: quaternion}

The Tsukuba convention \citep{Olson2001} standardised an embedding of a frame in an idealised rigid configuration of individual atoms making up each base, along with the names (buckle, propeller, opening, shear, stretch, stagger) for, respectively, rotational and translation coordinates (respecting the ordering of the indices of the axes of these embedded frames) for the relative rigid body displacements between the two bases comprising a base pair, or what is called the intra base pair coordinates in Curves+ \citep{Lavery2009}. Similarly (tilt, roll, twist, shift, slide, rise), or inters, are the conventional names for the relative rotations and translations between the base pair frames that are defined as an average of each consecutive pair of base frames. In this sense the plots of inter and intra coordinates of a minicircle configuration along the dsDNA as provided in, for example, Fig.\ \ref{fig: ex1f3} of the main text, will be familiar to many readers, albeit that there can still be small differences in the detail of how the coordinates are defined in different conventions and associated software. In Curves+ the relative rotations are recorded in axis-angle coordinates, but scaled to degrees rather than the mathematically more standard radians, and the values of tilt, roll and twist are reported as the components of the axis-angle vector on the respective coordinate axes. It should be noted, however, that care needs to be taken in how these components are interpreted. As is well known, in general the product of rotation matrices is noncommutative, so that a rotation through $\theta_1$ around say the tilt axis followed by a rotation through $\theta_2$ around the roll axis, is not the same composite rotation as a rotation through $\theta_2$ about the roll axis followed by a rotation through $\theta_1$ about the tilt axis. The difference is rather small when $\theta_1$ and $\theta_2$ are both small, but nevertheless the strict sense of Curves+ coordinates is that the ratios of the three coordinates fixes the direction of the axis vector, and the norm of the rotation coordinates fixes the magnitude of the single rotation angle about this axis vector measured in degrees. The same observation pertains to the interpretation of the Cayley vector coordinates adopted in $cgNA+$, but with the magnitude of the rotation angle determined as a nonlinear function of the norm of the Cayley vector and in a scaling explained below.

To describe the configuration of a dsDNA fragment it is necessary to choose an orientation along the {\em double} helix. But due to their detailed chemical structure the two backbones have anti-parallel orientations. Consequently there is an arbitrary decision to be made as to whether to adopt the orientation of the dsDNA that matches the $5^\prime\to 3^\prime$ direction in the Crick or in the Watson backbone. In both Curves+ and $cgNA+$ the convention is to adopt the orientation of the dsNA matching the $5^\prime\to 3^\prime$ orientation of the back-bone of the strand along which the sequence is read, which we refer to as the Watson strand, with the inter variables describing the relative rigid body displacement from base pair frame $D_i= (R_i,o_i)$ to base pair frame $D_{i+1}=(R_{i+1},o_{i+1})$. And the intra coordinates are for the relative rigid body deformation from the Crick base to the Watson base within each base pair. Nevertheless it is convenient, particularly for palindromic sequences of dsDNA or dsRNA in which the $5^\prime\to 3^\prime$ Watson read of the sequence is the same as the $5^\prime\to 3^\prime$ Crick read of the sequence, to have a simple transformation rule between the two sets of coordinates corresponding to the two possible choices of orientation of the dsNA. That desiderata is achieved by the base pair frame orientation $R_i$ being defined as an appropriate average of the two base orientations, and the introduction of a junction frame whose absolute orientation is the appropriate average of the two base pair absolute orientations $R_i$ and $R_{i+1}$. There is a general theory of how to take averages of ensembles of matrices in $SO(3)$ \citep{Moakher2002} 
In Curves+ (and $cgNA+$) the orientation of the junction frame between base pairs $i$ and $i+1$  is defined using the simple case of the general theory that arises when averaging only two matrices, explicitly 
\beq
\label{junction}
J_i := R_i P_i, \qquad P_i := \sqrt{R_i^TR_{i+1}},
\eeq
where the square root of a proper rotation matrix means the rotation about the same axis, but through half the angle. It is a defining property of the rotation axis vector of the base-pair step relative rotation matrix $R_i^TR_{i+1}$ (and therefore also of the associated Cayley vector) that its components  are the same when expressed in the frames $R_i$ or $J_i$ or $R_{i+1}$. But this is not true for the components of the absolute inter translation vector $o_{i+1} - o_i$. The Crick-Watson coordinate transformation rule is significantly simpler when the $i$th inter translation coordinates are specifically the components of $o_{i+1} - o_i$ in the $i$th junction frame $J_i$. And this is the reason that those are the coordinates adopted in both Curves+ and $cgNA+$.

 The fact that Cayley vector coordinates have a singularity at $\Theta = \pi$ is not a drawback when applied to the relative junction rotations between consecutive base-pair frame orientations, because the rotation angles $\Theta$ are always smaller than $\pi$. This is true for simulations carried out within the $cgNA+$ model (as done here) because  the coarse-grain model energy (introduced in (\ref{cgNA+pdf}) of the main text) increases to infinity whenever even only one component of one Cayley vector approaches infinity. The Cayley vector coordinates have other advantages that make them a judicious choice for parameterising the relative rotations between the frames of the $cgNA+$ model. For example the Gaussian integrals that arise during evaluation of first and second moments of the pdf (\ref{cgNA+pdf}) are over all of $\R^m$ (for some value of $m$), and so have familiar, explicit evaluation formulas. 
 But for tracking the absolute orientations of base-pair frames around a minicircle, which is necessary to express the closure conditions simply, it is inevitable that rotations close to $\pi$ will arise.  We introduced unit quaternion coordinates $\{q_i\}$ for each absolute base-pair frame orientation $\{R_i\}$, with $R_i=R(q_i)$ 
 (according to the quaternion Euler-Rodrigues formula (\ref{eqn:quaternion_defn}) in the main text) precisely to avoid the associated discontinuities in Cayley vector coordinates.  In order to rewrite the $cgNA+$ energy as a function of these absolute quaternion coordinates it means that we have need of formula (\ref{junction_quat}) of the main text, namely 
 \beq
 \label{app_junction_quat}
 J_i=R(q_i+ q_{i+1}),
 \eeq
 i.e.\ the quaternion of the $i$th junction frame can be taken to be the arithmetic average of the two flanking base-pair frame unit quaternions, or equivalently the projection $(q_i+ q_{i+1})/\norm{q_i+ q_{i+1}}$ onto the sphere of unit quaternions, which bisects the geodesic arc between the two unit quaternions $q_i$ and $q_{i+1}$. There is a literature \citep{Markley2007, Hartley2013} 
 on averaging ensembles of quaternions, which parallels the general theory of averaging in $SO(3)$  \citep{Moakher2002}, 
 but we here simply verify directly the simple formula (\ref{app_junction_quat}) that the quaternion of the $SO(3)$ group average junction matrix $J_i$ of the two rotation matrices $R_i$ and $R_{i+1}$ as defined in (\ref{junction}) is the simple average of the two unit quaternion coordinates $q_i$ and $q_{i+1}$ for $R_i$ and $R_{i+1}$. This conclusion follows directly from two applications of the quaternion composition rule  (\ref{composition_components}) of the main text, which we repeat here
\begin{equation}
\label{app_composition_components}
p^j = r^T B_j q,\qquad j=1,2,3,4.
\end{equation}
As already computed in the main text, applying (\ref{app_composition_components}) with $q=q_i$ and $r=q_{i+1}$ implies that the matrix $R^T_{i+1}R_i$ has quaternion coordinate
$$
(q_{i+1}^TB_1q_i,\ q_{i+1}^TB_2q_i,\ q_{i+1}^TB_3q_i,\ q_{i+1}^Tq_i),
$$
where for the scalar part $q_{i+1}^Tq_i= \cos(\Theta/2)>0$. And assuming (\ref{app_junction_quat}), and then applying (\ref{app_composition_components}) 
with $q=q_i$ and $r=(q_i+ q_{i+1})/\norm{q_i+ q_{i+1}}$ we find that the matrix $P_i$ in the first equality of  (\ref{junction}) has the quaternion coordinate
$$
 (q_{i+1}^TB_1q_i,\ q_{i+1}^TB_2q_i,\ q_{i+1}^TB_3q_i,\  1+ q_{i+1}^Tq_i)/\norm{q_i+ q_{i+1}},
$$
where we have used that $B_1$, $B_2$ and $B_3$ are skew symmetric matrices, and that $q_{i+1}$ and $q_i$ are both unit quaternions.
Thus we see that the vector parts of the quaternion coordinates of $R^T_{i+1}R_i$ and of $P_i$ are parallel, so that the two matrices share the same rotation axis. And for the scalar part of $P_i$ we use half-angle formulas to compute that $(1+ q_{i+1}^Tq_i)= 2 \cos^2(\Theta/4)$, which, taken along with $\norm{q_i+ q_{i+1}}= \sqrt{2(1+q_{i+1}^Tq_i)}$, implies that $P_i$ is a rotation through $\Theta/2$. In other words $P_i$ is indeed the matrix square root in the second equality of (\ref{junction}), and the justification of the quaternion expression (\ref{app_junction_quat}) for $J_i$ is complete. 

 It remains to explain the ubiquitous factor of $10$ associated with the $cgNA+$ Cayley vectors presented in the main text. In fact there are two factors to explain, $10=5\times2$. For any treatment of rigid body displacements, an arbitrary scaling between rotations and translations arises. In $cgNA+$ model parameter sets, the adopted unit of translations is the standard crystallographic one of Angstroms, as also adopted in Curves+. But $5$ Angstroms is a good physical characteristic length scale for all of the relative translations, inter, intra, and phosphate, that arise along the tree structure illustrated in Figure \ref{fig:cgNA+_schematic} of the main text.  That implies that small changes of size $\epsilon$ in angular coordinates measured in $1/5$ radians will give characteristic changes in inter-atomic distances of order $\epsilon$ Angstroms. This nonstandard choice of unit for rotations is not crucial, but it does give rise to a good numerical scaling in all of the diagonal entries (with different dimensions) in the $cgNA+$ stiffness parameter blocks. 
 
 It is also desirable that plots of $cgNA+$ configurations, such as in Fig.\ \ref{fig: ex1f3} of the main text, can be presented as familiar coordinates expressed in familiar units. For the $cgNA+$ translation coordinates that need is met by adopting the translation unit of Angstroms with components expressed in the Tsukuba convention frames. (For phosphates, the coordinates more typically adopted are backbone angles, but these are quite ill-adapted to the Gaussian $cgNA+$ coarse-grain model, so there was no choice but to abandon them in favour of relative translation and rotation coordinates analogous to the treatment of intra and inter base pair coordinates.) The mathematically natural treatment of Cayley vectors $w$ as presented in Appendix \ref{Annexe: Cayley vector} has the scaling $\norm{w}= \tan(\Theta/2)$ so that for small rotation angles $\Theta$ measured in radians $w \approx \Theta/2 \, \mathbf{n}$, where as before $\mathbf{n}$ denotes the (right handed) unit rotation axis vector. Combining this undesirable geometrical factor of one half, with the physically motivated scaling of $1/5$ described in the previous paragraph, we arrive at the $cgNA+$ scaling of Cayley vectors $u=10 w$, where in any nonlinear formula it is the Cayley vector  $w=u/10$ that effectively appears. For the $cgNA+$ scaled Cayley vector $u$ we have the formula $\norm{u} = 10 \tan(\Psi /10)$,where $\Psi$ is the rotation angle expressed in units of $1/5$ radian, and for $\Psi$ small $u \approx \Psi\, \mathbf{n}$. This motivates the plots of components of the $cgNA+$ Cayley vectors for minicircle configurations, as in Fig.\ \ref{fig: ex1f3}, where the scale on the ordinate is degrees, and the conversion from components of the $cgNA+$ Cayley vector $u$ to degrees is made using the linear scaling that $11.5$ degrees is approximately $1/5$ radian, which is only completely justified for $\norm{u} << 1$. Just as for plots of Curves+ coordinates, the precise interpretation of these graphs is that the ratio of sets of three rotational coordinates sets the orientation of the corresponding rotation axis $\mathbf{n}$, with the magnitude of the rotation angle about that axis set (now nonlinearly) by the norm of the Cayley vector $u$.

\section{Gradient and hessian of energy function}
\label{sec:grad_and_hess}

In this section, we compute the gradient and Hessian of the energy function 
 \[ E(z,{\mathbb P}) = \frac{1}{2} \left( \Omega-\hat\Omega \right)^T
   \calK \left( \Omega-\hat\Omega \right) + p_w \sum_{i=2}^N (\| q_i \|^2-1)^2 \]
(Eq.\ \eqref{eq: cgDNAmin zvec energy} in the main text), where $z$ is the vector
of variables $(w_1,o_2,q_2,w_2,o_3,q_3,\cdots,o_N,q_N,w_N)$ and ${\mathbb P}$ is a vector
of parameters that includes the first and last base pair quantities $o_1,q_1,o_{N+1},q_{N+1}$
that are prescribed by the boundary conditions.  Here,
the vector $\Omega = (w_1,y_1,\cdots,w_N,y_N)$ 
of $cgNA+$ variables is understood to be expressed in terms of $z$
and $o_1,q_1,o_{N+1},q_{N+1}$ via the change of variables described in the main text,
i.e., the variables $w_i$ are unchanged and $y_i=(u_i, v_i)$ are expressed
in the new variables using Eqs. \eqref{quat_junctions_to_cayley}
and \eqref{origins_recursion_inv}.  The vector $\hat\Omega$, the matrix
$\calK$, and the scalar $p_w$ are constants (with the first two derived from the DNA sequence
as outlined in the main text).

\begin{figure}
\centering
\scalebox{0.16}{
\begin{tikzpicture}
    \draw [step=1.0,gray, very thin] (0,0) grid (42,42);
    \draw [step=6.0,gray] (0,0) grid (42,42);
    \draw[black, very thick] (0,0) -- (42,0);
    \draw[black, very thick] (0,18) -- (42,18);
    \draw[black, very thick] (0,24) -- (42,24);
    \draw[black, very thick] (0,42) -- (42,42);
    \draw[black, very thick] (0,0) -- (0,42);
    \draw[black, very thick] (18,0) -- (18,42);
    \draw[black, very thick] (24,0) -- (24,42);
    \draw[black, very thick] (42,0) -- (42,42);
    
    \fontsize{70pt}{10pt}\selectfont\
    \node at (9,33)  () {$N_i$};
    \node at (21,33) () {$P_i$};
    \node at (33,33) () {$Q_i$};
    \node at (21,21) () {$M_i$};
    \node at (33,21) () {$O_i$};
    \node at (33,9)  () {$N_{i+1}$};
    \node at (9,21)  () {$P_i^T$};
    \node at (9,9)   () {$Q_i^T$};
    \node at (21,9)  () {$O_i^T$};
    \normalfont
    \fill[orange,opacity=.2] (0,42)  rectangle (18,24);
    \fill[orange,opacity=.2] (24,18)  rectangle (42,0);
    \fill[red,opacity=.2]    (18,24)  rectangle (24,18);
    \fill[blue,opacity=.2]   (24,42) rectangle (42,24);
    \fill[green,opacity=.2]  (18,42)  rectangle (24,24);
    \fill[yellow,opacity=.2]  (24,24) rectangle (42,18);
\end{tikzpicture}
}
\caption{Sub-blocks of the $i$-th $42\times42$ diagonal block $K_i$ of the stiffness matrix $\calK$. The full block shown here is the diagonal sub-block of $\calK$ with indices ranging from $24(i-1)+1$ to $24(i-1)+42$ in both dimensions (except for $i=N$ the block gets subdivided and some parts distributed elsewhere in $\calK$ as described in the main text).  The shaded blocks denote $P_i$, $Q_i$, $M_i$, $O_i$, $N_i$, and $N_{i+1}$, and the three unshaded blocks are transposes of shaded blocks, as indicated in the figure.}\label{fig: cgDNA+ block split}
\end{figure}
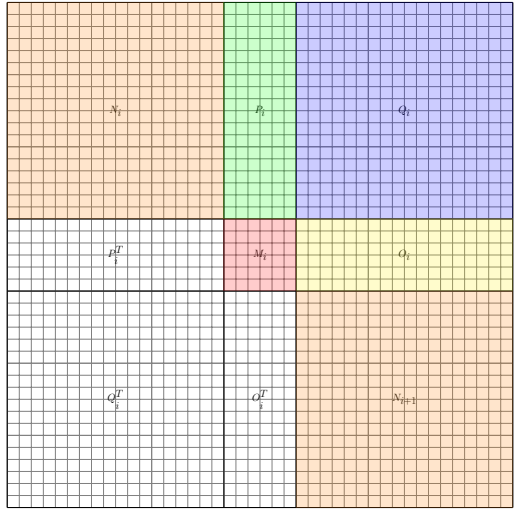

To make our gradient and Hessian more concrete, we begin by decomposing the stiffness matrix $\calK$ into sub-blocks. As defined in the main text,
$\calK$ consists of $42\times 42$ blocks
that have $18 \times 18$ overlaps.  If $K_i$ 
denotes the $i$th $42\times 42$ block,
we give names to particular sub-blocks of $K_i$:
\begin{center}
    \begin{tabular}{c|l|l}
    Notation & Sub-block of $K_i$ & Interaction represented by this sub-block \\
     \noalign{\hrulefill}
    $N_i$ & $K_i$[1:18,1:18] & base pair $i$ with base pair $i$ \\
    $P_i$ & $K_i$[1:18,19:24] & base pair $i$ with inter ($i$-to-$(i+1)$) \\
    $Q_i$ & $K_i$[1:18,25:42] & base pair $i$ with base pair $(i+1)$ \\
    $M_i$ & $K_i$[19:24,19:24] & inter ($i$-to-$(i+1)$) with inter ($i$-to-$(i+1)$) \\
    $O_i$ & $K_i$[19:24,25:52] & inter ($i$-to-$(i+1)$) with base pair $(i+1)$
    \end{tabular}
\end{center}

\noindent where the notation $K_i[a:b,c:d]$ means the sub-block of rows $a$ to $b$ and columns $c$ to $d$ of $K_i$. These sub-blocks are illustrated in Fig. \ref{fig: cgDNA+ block split}.
With these definitions, we can write the first term in the energy more explicitly:
 \begin{eqnarray*} 
 \frac{1}{2} \left( \Omega-\hat\Omega \right)^T \calK \left( \Omega-\hat\Omega \right) &=&
 \frac{1}{2} \sum_{i=1}^N (w_i-\hat w_i)^T N_i (w_i-\hat w_i) + \sum_{i=1}^N (w_i-\hat w_i)^T P_i (y_i-\hat y_i) \\
 && \hskip .2 in + \sum_{i=1}^{N-1} (w_i-\hat w_i)^T Q_i (w_{i+1}-\hat w_{i+1}) 
      + (w_N - \hat w_N)^T Q_N (w_1-\hat w_1) \\ 
 && \hskip .2 in + \sum_{i=1}^{N-1} (y_i-\hat y_i)^T O_i (w_{i+1}-\hat w_{i+1}) 
      + (y_N - \hat y_N)^T O_N (w_1-\hat w_1) \\ 
 && \hskip .2 in + \frac{1}{2} \sum_{i=1}^N (y_i-\hat y_i)^T M_i (y_i-\hat y_i) 
 \end{eqnarray*}
\subsection{Gradient computation}
\label{gradient_subsection}

Since the change of coordinates from $\Omega$ to $z$ leaves the $w_i$ unchanged, the derivatives
of the energy with respect to $w_i$ are straightforward (the energy is quadratic in these variables).
For $2 \le i \le N-1$, 
\begin{equation}
\label{eq: grad intra}
\begin{split}
    \frac{\partial E}{\partial w_i} &= N_i(w_i-\hat{w}_i) + Q_i(w_{i+1}-\hat{w}_{i+1}) + Q_{i-1}^T(w_{i-1}-\hat{w}_{i-1}) 
    + P_i(y_i-\hat{y}_i) + O_{i-1}^T(y_{i-1}-\hat{y}_{i-1}), 
\end{split}
\end{equation}
while the $i=1$ and $i=N$ terms take slightly different forms due to the periodic setup:
\begin{eqnarray}
    \label{eq: grad intra special}
    \frac{\partial E}{\partial w_1} &= N_1(w_1-\hat{w}_1) + Q_1(w_2-\hat{w}_2) + Q_N^T(w_N-\hat{w}_N) 
    + P_1(y_1-\hat{y}_1) + O_N^T(y_N-\hat{y}_N), \nonumber\\
    \frac{\partial E}{\partial w_N} &= N_N(w_N-\hat{w}_N) + Q_N(w_1-\hat{w}_1) + Q_{N-1}^T(w_{N-1}-\hat{w}_{N-1}) \nonumber\\
    & \hskip 2 in + P_N(y_N-\hat{y}_N) + O_{N-1}^T(y_{N-1}-\hat{y}_{N-1}). 
\end{eqnarray}

For derivatives with respect to $o_i, q_i$, we need to use the chain rule, since they are related to the inter coordinates $y_i$ in $\Omega$ via a nonlinear transformation.
To prepare for the $q$ portion of this computation, we start by computing some preliminary derivatives,
\beq
\frac{\partial}{\partial q_{i+1}} \left(\frac{1}{q_{i+1}^T q_i}\right) 
= -  \left(q_{i+1}^T q_i\right)^{-2} q_i, \;\;\;\;\; 
\frac{\partial}{\partial q_i} \left({\frac{1}{q_{i+1}^T q_i}}\right) = -  \left(q_{i+1}^T q_i\right)^{-2} q_{i+1}.
\eeq
Then, from the transformation formula in Eq.\ \eqref{quat_junctions_to_cayley} of the main text, 
\beqs
\frac{\partial u_i^j}{\partial q_{i+1}} &=& \left[{10 q_{i+1}^T B_j q_i}\right] \left[{-\left({q_{i+1}^T q_i}\right)^{-2}}\right] q_i + \left({q_{i+1}^T q_i}\right)^{-1} (10 B_j q_i) \nonumber \\
&=& \left({q_{i+1}^T q_i}\right)^{-1} \left({10 B_j - u_i^j I_4}\right) q_i
\eeqs
(for $I_4$ the $4 \times 4$ identity matrix) and similarly
\beqs
\frac{\partial u_i^j}{\partial q_i} &=& \left[{10 q_{i+1}^T B_j q_i}\right] \left[{-\left({q_{i+1}^T q_i}\right)^{-2}}\right] q_{i+1} + \left({q_{i+1}^T q_i}\right)^{-1} \left[{-10 B_j q_{i+1}}\right] \nonumber \\
&=& \left({q_{i+1}^T q_i}\right)^{-1} \left({-10 B_j - u_i^j I_4}\right)q_{i+1}.
\eeqs
To prepare for the $o$ portion of the gradient, we first define the shorthand notation:
 \[ d_j(q) = \text{column $j$ of} \; R(q) \;\;\;\; j=1,2,3. \]
Then, from the transformation formula in Eq.\ \eqref{origins_recursion_inv} of the main text, 
\beqs
\frac{\partial v_i^j}{\partial o_i} &=& -d_j (q_{i+1} + q_i), \nonumber\\
\frac{\partial v_i^j}{\partial o_{i+1}} &=& d_j (q_{i+1} + q_i), \nonumber\\
\frac{\partial v_i^j}{\partial q_i} &=& \frac{\partial v_i^j}{\partial q_{i+1}} = \left({\frac{\partial d_j}{\partial q} (q_{i+1} + q_i)}\right)^T (o_{i+1}-o_i).
\eeqs
Now (ignoring the penalty term for the moment)
we can compute the $o_i$ and $q_i$ portions of the gradient vector using the chain rule.
\beqs
    \frac{\partial E}{\partial o_i} &=& \sum_{j=1}^3{\left[{
    \frac{\partial E}{\partial v_{i-1}^j} \frac{\partial v_{i-1}^j}{\partial o_i} + \frac{\partial E}{\partial v_i^j} \frac{\partial v_i^j}{\partial o_i} }\right]} 
    = \left({\frac{\partial v_{i-1}}{\partial o_i}}\right)^T \frac{\partial E}{\partial v_{i-1}} + \left({\frac{\partial v_i}{\partial o_i}}\right)^T \frac{\partial E}{\partial v_i} \\ 
    \frac{\partial E}{\partial q_i} &=& \sum_{j=1}^3{\left[{
    \frac{\partial E}{\partial u_{i-1}^j} \frac{\partial u_{i-1}^j}{\partial q_i} + \frac{\partial E}{\partial u_i^j} \frac{\partial u_i^j}{\partial q_i} + 
    \frac{\partial E}{\partial v_{i-1}^j} \frac{\partial v_{i-1}^j}{\partial q_i} + \frac{\partial E}{\partial v_i^j} \frac{\partial v_i^j}{\partial q_i} }\right]} \nonumber \\
    &=& \left({\frac{\partial u_{i-1}}{\partial q_i}}\right)^T \frac{\partial E}{\partial u_{i-1}} + \left({\frac{\partial u_i}{\partial q_i}}\right)^T \frac{\partial E}{\partial u_i} + \left({\frac{\partial v_{i-1}}{\partial q_i}}\right)^T \frac{\partial E}{\partial v_{i-1}} + \left({\frac{\partial v_i}{\partial q_i}}\right)^T \frac{\partial E}{\partial v_i}.
\eeqs
Combining the two results, we obtain
\beqs\label{eq: inter deriv}
    \frac{\partial E}{\partial(o_i,q_i)} &=&
    \begin{bmatrix}
    0 & \left(\partial v_{i-1}/\partial o_i\right)^T \\
    \left(\partial u_{i-1}/\partial q_i\right)^T & \left(\partial v_{i-1}/\partial q_i\right)^T
    \end{bmatrix}
    \begin{bmatrix}
    \partial E/\partial u_{i-1} \\ 
    \partial E/\partial v_{i-1}
    \end{bmatrix} + 
    \begin{bmatrix}
    0 & \left(\partial v_i/\partial o_i\right)^T \\
    \left(\partial u_i/\partial q_i\right)^T & \left(\partial v_i/\partial q_i\right)^T
    \end{bmatrix}
    \begin{bmatrix}
    \partial E/\partial u_i \\
    \partial E/\partial v_i
    \end{bmatrix} \nonumber\\
    &=& 
    \begin{bmatrix}
    0 & \partial u_{i-1}/\partial q_i \\
    \partial v_{i-1}/\partial o_i & \partial v_{i-1}/\partial q_i
    \end{bmatrix}^T \frac{\partial E}{\partial y_{i-1}} + 
    \begin{bmatrix}
    0 & \partial u_i/\partial q_i \\
    \partial v_i/\partial o_i & \partial v_i/\partial q_i
    \end{bmatrix}^T \frac{\partial E}{\partial y_i},
\eeqs
where
\begin{eqnarray}
\frac{\partial E}{\partial y_i} &=& P_i^T(w_i-\hat{w}_i) + M_i(y_i-\hat{y}_i) + O_i(w_{i+1}-\hat{w}_{i+1}), \;\;
 1 \le i \le N-1, \nonumber\\
 \frac{\partial E}{\partial y_N} &=& P_N^T(w_N-\hat w_N) + M_N(y_N-\hat y_N) + O_N(w_1-\hat w_1).
\end{eqnarray}
The two matrices in equation \eqref{eq: inter deriv} will appear in the Hessian matrix computations, so we give them shorthand names:
\beqs\label{eq: name partial matrices}
    \frac{\partial y_{i-1}}{\partial(o_i,q_i)} \equiv
    \begin{bmatrix}
    0 & \partial u_{i-1}/\partial q_i \\
    \partial v_{i-1}/\partial o_i & \partial v_{i-1}/\partial q_i
    \end{bmatrix}, \hspace{5mm}
    \frac{\partial y_i}{\partial(o_i,q_i)} \equiv
    \begin{bmatrix}
    0 & \partial u_i/\partial q_i \\
    \partial v_i/\partial o_i & \partial v_i/\partial q_i
    \end{bmatrix}.
\eeqs
Finally, we compute derivatives of the penalty term $E_{pen} \equiv p_w \sum_{i=2}^N (\| q_i \|^2-1)^2$:
\beq
\frac{\partial E_{pen}}{\partial q_i} = 4 p_w (\norm{q_i}^2-1) q_i, 
\eeq
which would get added to the $q$ portion (bottom 4 rows) of the result in Eq.\ \eqref{eq: inter deriv} in building the gradient
vector.

\subsection{Hessian computation} 
\label{hessian_subsection}

As in the computation of the gradient, it is convenient to separate
the computation of the Hessian into three cases according to whether we are 
differentiating with respect to $w_i$ or $(o_i,q_i)$.  

\paragraph{$w$--$w$ second derivatives}

These are straightforward since the energy is quadratic in the $w_i$:
\begin{eqnarray*}
    \frac{\partial^2 E}{\partial (w_i)^2} &=& N_i, \hspace{3mm} 1\leq i \leq N, \\
    \frac{\partial^2 E}{\partial w_i \partial w_{i+1}} &=& Q_i, \hspace{3mm} 1\leq i \leq N-1, \;\;\;\;
    \frac{\partial^2 E}{\partial w_N \partial w_1} = Q_N, \\
    \frac{\partial^2 E}{\partial w_i \partial w_{i-1}} &=& Q_{i-1}^T, \hspace{3mm} 2\leq i \leq N, \;\;\;\;
    \frac{\partial^2 E}{\partial w_1 \partial w_N} = Q_N^T.
\end{eqnarray*}
(The third line is implied by the second if we replace $i$ by $i-1$ and take the transpose.)

\paragraph{$w$--$(o.q)$ or $(o,q)$--$w$ second derivatives}

We start by rewriting Eqs. \eqref{eq: grad intra} and \eqref{eq: grad intra special} more concisely:
\begin{eqnarray*}
\frac{\partial E}{\partial w_i} &=& 
P_i(y_i-\hat{y}_i) + O_{i-1}^T(y_{i-1}-\hat{y}_{i-1}) + \text{(function of the $w_i$ only)}, 
\;\; 2 \le i \le N, \\
\frac{\partial E}{\partial w_1} &=& 
P_1(y_1-\hat{y}_1) + O_{N}^T(y_{N}-\hat{y}_{N}) + \text{(function of the $w_i$ only)},
\end{eqnarray*}
which implies
\beqs
\frac{\partial^2 E}{\partial w_i \partial(o_{i+1},q_{i+1})} &=& P_i \frac{\partial y_i}{\partial(o_{i+1},q_{i+1})}, 
 \;\; 2 \le i \le N-1, \nonumber\\
\frac{\partial^2 E}{\partial w_i \partial(o_{i-1},q_{i-1})} &=& O_{i-1}^T \frac{\partial y_{i-1}}{\partial(o_{i-1},q_{i-1})}, \;\; 3 \le i \le N, \nonumber\\
\frac{\partial^2 E}{\partial w_i \partial(o_i,q_i)} &=& P_i \frac{\partial y_i}{\partial(o_i,q_i)} + O_{i-1}^T \frac{\partial y_{i-1}}{\partial(o_i,q_i)}, \;\; 2 \le i \le N, \nonumber\\
\frac{\partial^2 E}{\partial w_1 \partial(o_N,q_N)} &=& O_N^T \frac{\partial y_N}{\partial(o_N,q_N)}, \nonumber\\
\frac{\partial^2 E}{\partial w_1 \partial(o_2,q_2)} &=& P_1 \frac{\partial y_1}{\partial(o_2,q_2)},
\eeqs
where the derivatives $\partial y_j/\partial (o_k,q_k)$ on the right-hand side were computed in \eqref{eq: name partial matrices}.  The second partials in the opposite order are obtained by transposing these results.

\paragraph{$(o.q)$--$(o,q)$ second derivatives}
Again we start with some preliminary derivatives:
\beqs
\frac{\partial^2 u_i^j}{\partial (q_{i+1})^2} 
&=&  \left[-\left({q_{i+1}^T q_i}\right)^{-2}\right]\left(10 B_j - u_i^j I_4\right) q_i q_i^T 
- \left({q_{i+1}^T q_i}\right)^{-1} q_i \left({\frac{\partial u_i^j}{\partial q_{i+1}}}\right)^T \nonumber \\
&=& - \left({q_{i+1}^T q_i}\right)^{-2}\left(10 B_j - u_i^j I_4\right)q_i q_i^T - \left({q_{i+1}^T q_i}\right)^{-2} q_i q_i^T \left(-10 B_j -u_i^j I_4\right) \nonumber\\
&=& (q_{i+1}^T q_i)^{-2} \left[{2u_i^j q_i q_i^T - 10(B_j q_i)q_i^T - 10 q_i (B_j q_i)^T}\right], 
\eeqs
\beqs
\frac{\partial^2 u_i^j}{\partial (q_i)^2} &=& \left[{-\left({q_{i+1}^T q_i}\right)^{-2}}\right]\left(-10 B_j - u_i^j I_4\right)q_{i+1} q_{i+1}^T - (q_{i+1}^T q_i)^{-1} q_{i+1} \left({\frac{\partial u_i^j}{\partial q_i}}\right)^T \nonumber\\
&=& - \left({q_{i+1}^T q_i}\right)^{-2}\left(-10 B_j - u_i^j I_4\right)q_{i+1} q_{i+1}^T - (q_{i+1}^T q_i)^{-2} q_{i+1} q_{i+1}^T (10 B_j - u_i^j I_4) \nonumber\\
&=& (q_{i+1}^T q_i)^{-2} \left[{2 u_i^j q_{i+1} q_{i+1}^T + 10(B_j q_{i+1})q_{i+1}^T + 10 q_{i+1}(B_j q_{i+1})^T}\right], \nonumber\\
\eeqs
\beqs
\frac{\partial^2 u_i^j}{\partial q_i \partial q_{i+1}} &=& \left[{-\left({q_{i+1}^T q_i}\right)^{-2}}\right](10 B_j - u_i^j I_4)q_i q_{i+1}^T - (q_{i+1}^T q_i)^{-1} q_i \left({\frac{\partial u_i^j}{\partial q_i}}\right)^T 
+  (q_{i+1}^T q_i)^{-1} (10 B_j - u_i^j I_4) \nonumber\\
&=& -(q_{i+1}^T q_i)^{-2} (10 B_j - u_i^j I_4) q_i q_{i+1}^T - (q_{i+1}^T q_i)^{-2} q_i q_{i+1}^T (10 B_j - u_i^j I_4) 
+  (q_{i+1}^T q_i)^{-1} (10 B_j - u_i^j I_4) \nonumber\\
&=& (q_{i+1}^T q_i)^{-2} \left[{2 u_i^j q_i q_{i+1}^T - 10(B_j q_i)q_{i+1}^T + 10 q_i (B_j q_{i+1})^T}\right] 
+ (q_{i+1}^T q_i)^{-1} (10 B_j - u_i^j I_4), 
\label{eqn:u_sec_deriv}
\eeqs
and, by transposition of \eqref{eqn:u_sec_deriv}:
\beq
\frac{\partial^2 u_i^j}{\partial q_{i+1} \partial q_i}= (q_{i+1}^T q_i)^{-2} \left[{2 u_i^j q_{i+1} q_i^T - 10q_{i+1}(B_j q_i)^T + 10 (B_j q_{i+1})q_i^T}\right] 
+ (q_{i+1}^T q_i)^{-1} (-10 B_j - u_i^j I_4).
\eeq

We also compute some derivatives of $v_i^j$.
\beqs
\frac{\partial^2v_i^j}{\partial o_{i+1} \partial q_{i+1}} &=& \frac{\partial^2v_i^a}{\partial o_{i+1} \partial q_i} = \left[{\frac{\partial d_j}{\partial q}(q_{i+1} + q_i)}\right]^T \nonumber\\
\frac{\partial^2v_i^j}{\partial o_i \partial q_{i+1}} &=& \frac{\partial^2v_i^j}{\partial o_i \partial q_i} = - \left[{\frac{\partial d_j}{\partial q}(q_{i+1} + q_i)}\right]^T \nonumber\\
\frac{\partial^2v_i^j}{\partial (q_{i+1})^2} &=& \frac{\partial^2v_i^j}{\partial (q_i)^2} = \frac{\partial^2v_i^j}{\partial q_{i+1}\partial q_i} = \sum_{k=1}^3{(o_{i+1}^k - o_i^k) \frac{\partial^2 d_{jk}}{\partial q^2}(q_{i+1} + q_i)}.
\eeqs
where the derivatives in the opposite order are obtained through transposition.

These last computations require formulas for some derivatives
of $d_j$ with respect to quaternions.  
From the formulas for the nine entries of $R(q)$, we can verify that for each $k = 1,2,3$,
\begin{eqnarray*}
\frac{\partial d_{1k}}{\partial q} &=& \frac{2 \left[{d_{2k}(B_3q) - d_{3k}(B_2q)}\right]}{\norm{q}^2} \\
\frac{\partial d_{2k}}{\partial q} &=& \frac{2 \left[{d_{3k}(B_1q) - d_{1k}(B_3q)}\right]}{\norm{q}^2}\\
\frac{\partial d_{3k}}{\partial q} &=& \frac{2 \left[{d_{1k}(B_2q) - d_{2k}(B_1q)}\right]}{\norm{q}^2},
\end{eqnarray*}
or, in vector form:
\begin{eqnarray*}
\frac{\partial d_1}{\partial q} &=& \frac{2 \left[{d_2(B_3q)^T - d_3(B_2q)^T}\right]}{\norm{q}^2} \\
\frac{\partial d_2}{\partial q} &=& \frac{2 \left[{d_3(B_1q)^T - d_1(B_3q)^T}\right]}{\norm{q}^2}, \\
\frac{\partial d_3}{\partial q} &=& \frac{2 \left[{d_1(B_2q)^T - d_2(B_1q)^T}\right]}{\norm{q}^2},
\end{eqnarray*}
Note that, for each of the last two sets of three equations, each equation is obtained 
from the previous one by shifting the indices in the pattern $1 \rightarrow 2 \rightarrow 3 \rightarrow 1$.

Since $\frac{\partial}{\partial q}(1/\norm{q}^2) = -(\norm{q}^2)^{-2} (2q) = (1/\norm{q}^2)(-2q/\norm{q}^2)$, we have 
\begin{eqnarray*}
\frac{\partial^2 d_{1k}}{\partial q^2} = \frac{2 \left[{d_{2k} B_3 - d_{3k} B_2}\right]}{\norm{q}^2} 
+ \frac{2 \left[{(B_3q)\left(\partial d_{2k}/\partial q\right)^T 
- (B_2q)\left(\partial d_{3k}/\partial q\right)^T}\right]}{\norm{q}^2} 
- \frac{2 (\partial d_{1k}/\partial q) q^T}{\norm{q}^2}.
\end{eqnarray*}
We can obtain $\partial^2 d_{2k}/\partial q^2$ from this result by shifting the indices according to $1 \rightarrow 2 \rightarrow 3 \rightarrow 1$, and then if we repeat that shift a second time, we can obtain $\partial^2 d_{3k}/\partial q^2$.

Having computed the necessary preliminary derivatives, we proceed to compute blocks of the Hessian matrix.  
From the gradient, and ignoring for now the penalty term, we have
\begin{eqnarray*}
\frac{\partial E}{\partial (o_i,q_i)} 
&=& \left( \partial y_{i-1}/\partial (o_i,q_i) \right)^T
\left[ M_{i-1}(y_{i-1} - \hat{y}_{i-1}) + \left( \text{Function of the $w_i$} \right) \right] \\
&& \hskip 1 in + \left( \partial y_i/\partial (o_i,q_i)\right)^T
\left[ M_i(y_i - \hat{y}_i) + \left( \text{Function of the $w_i$} \right) \right].
\end{eqnarray*}
so that:
\begin{eqnarray*}
\frac{\partial^2 E}{\partial (o_{i-1},q_{i-1}) \partial (o_i,q_i)} &=& \textcolor{red}{\frac{\partial \left[{\partial y_{i-1}/\partial(o_i,q_i)}\right]^T}{\partial (o_{i-1},q_{i-1})}}\frac{\partial E}{\partial y_{i-1}} + \left[{\frac{\partial y_{i-1}}{\partial (o_i,q_i)}}\right]^T M_{i-1} \frac{\partial y_{i-1}}{\partial (o_{i-1},q_{i-1})}, \\
\frac{\partial^2 E}{\partial (o_{i+1},q_{i+1}) \partial (o_i,q_i)} &=& \textcolor{red}{\frac{\partial \left[{\partial y_i/\partial(o_i,q_i)}\right]^T}{\partial (o_{i+1},q_{i+1})}}\frac{\partial E}{\partial y_i} + \left[{\frac{\partial y_i}{\partial (o_i,q_i)}}\right]^T M_i \frac{\partial y_i}{\partial (o_{i+1},q_{i+1})}, \\
\frac{\partial^2 E}{\partial (o_i,q_i) \partial (o_i,q_i)} &=& \textcolor{red}{\frac{\partial \left[{\partial y_{i-1}/\partial(o_i,q_i)}\right]^T}{\partial (o_i,q_i)}}\frac{\partial E}{\partial y_{i-1}} + \left[{\frac{\partial y_{i-1}}{\partial (o_i,q_i)}}\right]^T M_{i-1} \frac{\partial y_{i-1}}{\partial (o_i,q_i)} \\
&& \hskip 1 in + \textcolor{red}{\frac{\partial \left[{\partial y_i/\partial(o_i,q_i)}\right]^T}{\partial (o_i,q_i)}}\frac{\partial E}{\partial y_i} + \left[{\frac{\partial y_i}{\partial (o_i,q_i)}}\right]^T M_i \frac{\partial y_i}{\partial (o_i,q_i)}.
\end{eqnarray*}
The terms in black on the right-hand side were computed as part of 
finding the gradient.  The red terms are $7 \times 7 \times 6$ tensors of the form 
$\partial \left[{\partial y_b/\partial(o_i,q_i)}\right]^T/\partial (o_c,q_c)$ 
with $b = i,i-1$ and $c = i-1,i,i+1$.
If we let $(k.l,m)$ be the indices for the three slots of one of these tensors,
then, for $1 \le m \le 3$, the fixed-$m$ slice of the tensor is the $7 \times 7$ matrix
 \[ \begin{bmatrix} \partial^2 u_b^m/\partial o_c\partial o_i & 
    \partial^2 u_b^m/\partial q_c\partial o_i \\
    \partial^2 u_b^m/\partial o_c\partial q_i & 
    \partial^2 u_b^m/\partial q_c\partial q_i \end{bmatrix} = 
     \begin{bmatrix} 0_{3 \times 3} & 0_{3 \times 4} \\ 
     0_{4\times 3} & \partial^2 u_b^m/\partial q_c\partial q_i \end{bmatrix},
    \]
and the $4 \times 4$ matrices $\partial^2 u_b^m/\partial q_c\partial q_i$ were computed
earlier in this section.  
On the other hand, if $4 \le m \le 6$, the fixed-$m$ slice of the tensor is
the $7 \times 7$ matrix
 \[ \begin{bmatrix} \partial^2 v_b^{m-3}/\partial o_c\partial o_i & 
    \partial^2 v_b^{m-3}/\partial q_c\partial o_i \\
    \partial^2 v_b^{m-3}/\partial o_c\partial q_i & 
    \partial^2 v_b^{m-3}/\partial q_c\partial q_i \end{bmatrix},
    \]
with each of these expressions having been computed earlier in this section.

Once the entries in this $7 \times 7 \times 6$ tensor are computed as just outlined,
the product 
 \[ \frac{\partial \left[{\partial y_b/\partial(o_i,q_i)}\right]^T}{\partial (o_c,q_c)}\frac{\partial E}{\partial y_b} \]
 needed for the Hessian is computed by summing over $m$ (from 1 to 6) and multiplying
 the $m$th slice of the tensor (a $7 \times 7$ matrix) 
 by the $m$th entry of $\partial U/\partial y_c$ (a scalar).

 Finally, we compute the contribution to the Hessian from the penalty term.  
 From the gradient of the penalty term, we compute
   \[ \frac{\partial^2 E_{pen}}{\partial (q_i)^2} = 4 p_w ( \| q_i \|^2 - 1 ) I_4 + 8 p_w q_i (q_i)^T, \]
  which is added to the lower-right $4 \times 4$ block
  of the $7 \times 7$ block $\partial^2 E/\partial (o_i,q_i)\partial (o_i,q_i)$ computed above.

\section{Methodology for generating initial minicircle configurations}
\label{initial}
This section describes a two-step process to generate sequence-dependent initial configurations of covalently closed minicircles with reasonably low energy for any given sequence $\calS$ of length $N$ bp and with a range of prescribed integer linking numbers $Lk$ and initial registers. Having such initial approximations is essential for our optimization algorithm to be able to converge in relatively few iterations to (what is believed to be) all local minima. The basic idea of each step is simple, but some quite detailed calculations arise in the implementation of the ideas.
\\

\noindent \textbf{Helicoidal configurations: }
Our starting observation is that when the inter coordinates along a linear fragment are all uniform ($y = \{u, v \}$) independent of the index of the base pair junction then the origins of the base pair frames  
all lie on a perfect (circular) helix, since 
having uniform inter coordinates means that the relative rotation and translation between any two
base pairs is constant.  
\\
We first compute a helicoidal configuration of a linear periodic fragment by uniformizing. For $y \in \rmath^{6}$ (representing a single set of inter coordinates) and (for $1 \le i \le N$)
$w_i \in \rmath^{18}$ (representing base-pair level coordinates $\{z_i^+,x_i,z_i^-\}$ as in Sec.\ref{subsec:stiffness_matrix}), we let 
\beq\label{eq: periodic_helical_variables}
 \Omega_h 
 \equiv \{w_1, y, w_2, y,..,w_i,y,..,w_N, y \} \in \rmath^{24N} 
\eeq
and seek the particular $\Omega_h$ that minimizes the periodic $cgNA+$ 
energy defined in Eq.\ \eqref{eq: total_energy_periodic}.  To create a vector of unknowns
without redundancy, we define:
\beqs\label{eq: final_periodic_helical_variables}
\Omega_{rh} = \{w_1, w_2,..,w_i,..,w_N, u, v \} \in \rmath^{18N+6},
\eeqs
which is related to $\Omega_h$ by\footnote{Here and throughout, we denote the $a \times a$ identity matrix as $I_a$.}
\beqs
\calD ~ \Omega_{rh} = \Omega_h ~ ~ \text{for} ~ ~ 
\calD = \begin{bmatrix}
I_{18} & 0 &...& 0 & 0\\
0 & 0 & ... & 0 & I_{6} \\ 
0 & I_{18} & ... & 0 & 0\\
0 & 0 & ... & 0 & I_{6} \\ 
. & . & . & .  & . \\
. & . & . & .  & . \\
. & . & . & .  & . \\
0 & 0 & ... & I_{18}  & 0 \\
0 & 0 & ... & 0  & I_{6} \\
\end{bmatrix} \in \rmath^{24N \times (18N+6)}
\eeqs
Plugging $\Omega_h$ into the energy $U_p$ from Eq.\ \eqref{eq: total_energy_periodic}, we have
$U_p(\Omega_h;\calS,\calP) = \frac{1}{2}(\Omega_h - \hat{\Omega})^T \calK (\Omega_h - \hat{\Omega})$
and therefore, in terms of our reduced vector of variables $\Omega_{rh}$, we consider 
\beqs\label{eq: final_helical_energy_periodic}
U_h(\Omega_{rh}; \calS,\calP) &\equiv& \frac{1}{2}(\calD ~ \Omega_{rh} - \hat{\Omega})^T \calK (\calD ~ \Omega_{rh} - \hat{\Omega}) \nonumber\\
&=&  \frac{1}{2}\left[ (\calD ~ \Omega_{rh})^T \calK (\calD ~ \Omega_{rh}) - 2 (\calD ~ \Omega_{rh})^T \calK \hat{\Omega} + \hat{\Omega}^T \calK \hat{\Omega} \right]. 
\eeqs
We seek the vector $\Omega_{rh}$ that minimizes $U_h$ subject to the constraint that the number of helical turns is equal to the prescribed $Lk$, i.e., we take the integer linking number that we seek for our eventual closed fragment and apply it to our open/linear periodic molecule as a number of helical turns.  Accordingly, we define the Lagrangian:
\beqs
\calL(\Omega_{rh},\lambda;\calS,\calP) = U_h(\Omega_{rh}; \calS,\calP) + \lambda ~ C (\Omega_{rh})
\eeqs
where $\lambda$ is a Lagrange multiplier and $C(\Omega_{rh})$ is the constraint function
\beq
C(\Omega_{rh}) \equiv \frac{1}{2} \left(k^2 - \Omega_{rh}^T \calE \Omega_{rh}   \right)
\label{eq: constraint_func}
\eeq
where $k \equiv 10 \tan (Lk ~ \pi/N)$ and 
\beq 
\calE \equiv \begin{bmatrix}
0  & ... & 0 & 0\\
0 & ... & 0 & 0 \\ 
.  & . & .  & . \\
.  & . & .  & . \\
0  & ... & I_{3}  & 0 \\
0  & ... & 0 & 0 \\
\end{bmatrix} \in \rmath^{(18N+6) \times (18N+6)},
\eeq
defined so that $\Omega_{rh}^T \calE \Omega_{rh} = \norm{u}^2$.  To understand this setup, let us first denote $\phi$ as the angle (in radians) between two adjacent base pair frames reconstructed using $\Omega_{rh}$. To achieve the total number of turns $Lk$ (between the first and last frames) in a uniform helicoidal configuration having $N$ base pairs, we need:
\beqs\label{eq: angle_Lk}
\phi N = 2 \pi Lk \implies \frac{\phi}{2} = \frac{Lk ~ \pi}{N}.
\eeqs
We show in the Appendix \ref{Annexe: quaternion} that the Cayley vector $u$ is related to the angle $\phi$ by:
\beqs\label{eq: norm_caley_angle}
\norm{u} = 10 \tan(\frac{\phi}{2}),
\eeqs
and thus we seek $u$ so that $\| u \| = 10 \tan(Lk ~ \pi/N)$, i.e., $C(\Omega_{rh})=0$.
Solving for $Lk$, we find that the constraint can also be expressed in the form
\beq\label{eq: Lk_angle}
Lk = \frac{N}{\pi} ~ \arctan\left(\frac{ \| u \| }{10} \right).
\eeq
We seek solutions $(\Omega_{rh},\lambda)$ 
to $\nabla \calL = 0$ subject to $C = 0$.  From Eqs.\ \eqref{eq: final_helical_energy_periodic}
and \eqref{eq: constraint_func}, we compute:

\beq
\nabla \calL = \nabla U_h (\Omega_{rh}) + \lambda \nabla C (\Omega_{rh})
= \left[\calD^T \calK \calD - \lambda \calE \right]\Omega_{rh} - \calD^T \calK \hat{\Omega}.
\eeq
Our solution approach is to select a grid of possible values of $\lambda$, and for each $\lambda$ in the grid\footnote{We searched for $\lambda$ in grid $[-2500, 2500]$ for all the computations of this article}
find $\Omega_{rh}$ by solving $\left(\calD^T \calK \calD - \lambda \calE \right)\Omega_{rh} = \calD^T \calK \hat{\Omega}$. Then, as a second step, for these computed $\Omega_{rh}$, we extract their subvectors $u$, compute the values of $N \arctan(\|u \|/10)/\pi$, and look for when this expression is close to $Lk$.%

\bigskip

\noindent \textbf{Bending the helix into a toroidal structure (i.e., a minicircle):} For a given $\Omega_{rh}$ yielding a uniform helix with $Lk$ number of helical turns, we define the following:
\beqs
z = \frac{v \cdot u}{\norm{u}}, \quad\quad
r = \frac{\norm{v - z \frac{u}{\norm{u}} } }{2 \sin(\phi /2)}
\eeqs
where $\phi = 2 \arctan(\| u\|/10)$ as per Eq.\ \eqref{eq: norm_caley_angle}.
In the above expressions, $r$ denotes the radius of the helix, and $z$ denotes a height parameter such that the pitch of the helix is $ 2 \pi z$ and the total length of the helix is $L = N z $. 
\\

Next, we compute the absolute origins $o^h$ and quaternions $q^h$ of a helix with the two helical parameters $r$ and $z$:  
\beqs
o_i^h = \begin{bmatrix}
           r \sin \left(i ~ \phi \right) \\
           r \left( \cos \left(i ~ \phi \right) - 1 \right) \\
           i ~ z
         \end{bmatrix} \quad\quad \text{and} \quad\quad
q_i^h = \begin{bmatrix}
           0 \\
           0 \\    
           \sin \left(i ~ \phi/2 \right) \\
           \cos \left(i ~ \phi/2 \right) \\
         \end{bmatrix} 
         \quad (0 \le i \le N)
\eeqs
Since $N\phi = 2 N  \arctan(\| u\|/10) = 2 \pi Lk$, we have $q_N^h = \pm q_0^h$ (and these quaternions describe the same rotation matrix even if the $\pm$ is $-$) and $o_N^h = o_0^h + (0,0,L)$. 
\\

Finally, we deform this helix into a toroidal configuration with toroidal radius
\beqs
R = \frac{L}{2 \pi} = \frac{N z} {2 \pi},
\eeqs
by bending it in the $y$-$z$ plane, so that 
the base pair frames that were on the surface of a vertical cylinder in the helix now live on 
the surface of a torus:

\beq
o_i^t = \begin{bmatrix}
           r \sin \left( i \phi  \right) \\
           R_i \cos(i\alpha) - R \\ 
           R_i \sin \left(i \alpha \right) 
         \end{bmatrix} \quad \text{and} \quad
q_i^t = \begin{bmatrix}
           ~\sin (i \alpha/2) \cos (i \phi/2) \\
         -  \sin (i \alpha/2) \sin (i \phi/2) \\    
           ~\cos (i \alpha/2) \sin (i \phi/2) \\
           ~\cos (i \alpha/2) \cos (i \phi/2) \\
         \end{bmatrix}
         \quad (0 \le i \le N) 
\eeq
for
 \[ \alpha = 2 \pi /N, ~~ R_i = R + r \cos (i ~ \phi). \]
The formula for $q^t_i$ represent the quaternion for $R^t_i := R_{rot} \ R_i^h$, where $R_{rot}$ (with corresponding quaternion $q_{rot} = (\sin(i \alpha/2),0,0,\cos(i \alpha/2))$) represents rotation in the $y$-$z$ plane by $i \alpha$, and rotation matrix $R_i^h$ corresponds to quaternion $q_i^h$.
Notice that $q^t_i$ can be directly computed in terms of $q_i^h$ and $q_{rot}$ using \eqref{composition} in the main text.
The formula for $o_i^t$ can be obtained by rewriting $o_i^t = (0,-r,iz) + (r\cos(i ~ \phi),r\sin(i ~ \phi),0)$ (the first term being the helical axis), changing the first term to the toroidal axis $(0,R(\cos(i ~ \alpha)-1),R\sin(i ~ \alpha))$, and applying the rotation
 \[ \begin{bmatrix} 1 & 0 & 0 \\ 0 & \cos(i \alpha) & -\sin(i \alpha) \\ 0 & \sin(i \alpha) & \cos(i \alpha)
  \end{bmatrix} \]
  to the second term. 
The first and last base pair frames in this toroidal configuration obey 
\beqs
o_0^t = o_N^t = (0,0,0) ~ \text{and} ~  q_0^t = (0,0,0,1),  \;\; q_N^t = \pm ~ q_0^t,
\eeqs
so this toroidal configuration achieves the cyclization boundary conditions (with stipulated value of $Lk$) that we impose on our energy minimisation problem. 
\\

As discussed in \cite{Furrer2000}, the cyclization problem for intrinsically straight DNA molecules has an infinite family of circular energy-minimisers that can be parametrised by a {\it register symmetry} in which the circular configuration is rotated by an angle $\gamma$ about its axis at each point, and then an overall rigid-body rotation by $-\gamma$ is applied to restore the DNA's configuration at its start/final point to obey the boundary conditions. Furrer et al.\ argue that generically the family of minimisers will yield a single minimiser at a single value of the register angle for a DNA with small intrinsic curvature that is close to straight. For non-generic situations or for DNA with more substantial intrinsic curvature, the register symmetry might broken, and this single-register angle phenomenon might persist, or there might be multiple minimisers, each with its own corresponding value of register angle. Accordingly, in order to find the maximum possible minimisers, we apply a range of register angles to our toroidal configuration and use each of the resulting configurations as initial guesses in our energy minimisation. The absolute origins of the base pair frames after applying register $\gamma$ are
 \begin{equation}
     o_i^f = R_\gamma ~  o_i^t
 \end{equation}
for
\beqs
R_{\gamma} = \begin{bmatrix}
\cos\gamma & - \sin\gamma & 0 \\
\sin\gamma & ~  \cos\gamma & 0 \\
0 & 0 & 1 \\
\end{bmatrix},
\eeqs
i.e.,\ rotation of each $o_i^t$ by $\gamma$ about the $z$ axis.  To find the 
register-adjusted quaternions, we define $q_\gamma = (0,0,\sin(\gamma/2),\cos(\gamma/2))$, 
the quaternion corresponding to $R_\gamma$, and then the final register-adjusted rotation matrix is
\begin{equation}
R^f_i (q^f_i) = R_\gamma (q_\gamma) \ R^t_i (q_i^t) \ R_{-\gamma} (q_{-\gamma}) 
\end{equation}
Finally, quaternion $q^f_i$ which corresponds to $R^f_i$ is obtained directly in terms of known $q_\gamma$, $q_i^t$ and $q_{-\gamma}$ using two times the composition equation \eqref{composition} of the main text. 
\\

We illustrate this entire process in Fig.\ \ref{fig: method} for a 94-bp sequence.  The left image (blue) shows the base pair origins corresponding to the periodic groundstate $\hat{\Omega}$.  The next image (black) shows the helix origins $o_i^h$ for stipulated link $Lk=9$. The top right image (red) shows the toroidal origins $o_i^t$ (still for $Lk=9$). Finally, the lower right image shows a ``rosette'' of $Lk = 9$ configurations $o_i^f$ for $20$ different register angles $\gamma = 2\pi k/20$ ($0 \le k \le 19$).
\section{MATLAB optimisation details}
\label{app:opt_details}
Table \ref{fminunc_options} shows the settings we used within MATLAB's unconstrained
minimisation algorithm {\tt fminunc} for each energy minimisation performed for this article.

\begin{table}[H]
\begin{center}
\begin{tabular}{ |p{5.0cm}|p{5.0cm}| }
\hline
Option  & Setting      \\ \hline
Algorithm & trust-region     \\ \hline
GradObj   & on               \\ \hline 
Hessian   & on               \\ \hline 
MaxIter   & 5000             \\ \hline 
TolFun    & 1e-16            \\ \hline 
TolX      & 1e-16            \\ \hline 
\end{tabular}
\caption{Options used when running {\tt fminunc} in MATLAB.}
\label{fminunc_options}
\end{center}
\end{table}

\end{document}